\documentclass[letterpaper,11pt]{article}
\usepackage[margin=1in]{geometry}  
\usepackage{setspace}  
\usepackage{natbib}
 \bibpunct[, ]{(}{)}{,}{a}{}{,}%

\usepackage{amsfonts}
\usepackage{amsthm}
\usepackage{mathrsfs}
\usepackage{amsmath}
\usepackage{amssymb}
\usepackage{latexsym}
\usepackage{indentfirst}
\usepackage{footmisc}
\usepackage{endnotes}
\usepackage{algorithm}
\usepackage{float}
\usepackage{graphicx}
\usepackage{epstopdf}
\usepackage{longtable}
\usepackage{enumerate}
\usepackage{enumitem}
\usepackage{authblk}
\usepackage{bbm}
\usepackage{booktabs}
\usepackage{url}

\usepackage{multirow}
\usepackage{bm}
\usepackage{slashbox}
\usepackage{array}
\usepackage{subfigure}
\usepackage{caption}
\usepackage{color}

\usepackage{pifont}  

\DeclareMathOperator*{\cov}{Cov} \DeclareMathOperator*{\var}{Var}
\DeclareMathOperator*{\E}{E}

\newcommand{\ud}{\mathrm{d}}

\newcommand{\mse}{\mathrm{MSE}}
\newcommand{\pfs}{\mathrm{PFS}}
\newcommand{\pcs}{\mathrm{PCS}}
\newcommand{\rem}{\mathrm{rem}}
\newcommand{\opt}{\mathrm{opt}}

\def\B{\mathbf{B}}

\def\F{\mathcal{F}}

\def\I{\mathbf{I}}

\def\M{\mathbf{M}}

\def\P{\mathbf{P}}
\def\Q{\mathbf{Q}}

\def\V{\mathbf{V}}
\def\X{\mathbf{X}}
\def\Y{\mathbf{Y}}
\def\Z{\mathbf{Z}}

\def\bba{\mathbf{a}}
\def\bbf{\mathbf{f}}

\def\bbv{\mathbf{v}}

\def\x{\mathbf{x}}
\def\z{\mathbf{z}}

\def\uf{\mathrm{f}}

\def\uh{\mathrm{h}}

\def\Acal{\mathcal{A}}

\def\Ecal{\mathcal{E}}

\def\Xcal{\mathcal{X}}

\def\Ucal{\mathcal{U}}

\def\Wcal{\mathcal{W}}

\def\1{\mathbbm{1}}

\def\bSigma{\boldsymbol{\Sigma}}
\def\bbeta{\boldsymbol{\beta}}
\def\bphi{\boldsymbol{\phi}}
\def\bPhi{\boldsymbol{\Phi}}

\def\HH{\mathbb{H}}
\def\NN{\mathbb{N}}
\def\PP{\mathbb{P}}

\def\RR{\mathbb{R}}

\newcommand{\sgn}{\mathrm{sgn}}

\newtheorem{proposition}{PROPOSITION}
\newtheorem{lemma}{LEMMA}
\newtheorem{remark}{REMARK}
\newtheorem{theorem}{THEOREM}

\allowdisplaybreaks

\newcommand{\vertiii}[1]{{\left\vert\kern-0.25ex\left\vert\kern-0.25ex\left\vert #1
		\right\vert\kern-0.25ex\right\vert\kern-0.25ex\right\vert}}

\DeclareMathOperator{\tr}{\mathop{\mathrm{tr}}}

\title{Convergence Analysis of Stochastic Kriging-Assisted Simulation with Random Covariates}

\author[a]{Cheng Li}
\author[b]{Siyang Gao}
\author[c]{Jianzhong Du}
\affil[a]{Department of Statistics and Data Science, National University of Singapore, Singapore 117546, stalic@nus.edu.sg}
\affil[b]{Department of Advanced Design and Systems Engineering and School of Data Science, City University of Hong Kong, Hong Kong, China, siyangao@cityu.edu.hk}
\affil[c]{School of Management, Fudan University, China, jianzhodu2-c@my.cityu.edu.hk}

\date{\vspace{-8ex}}

\begin{document}

\maketitle

\begin{abstract}
We consider performing simulation experiments in the presence of covariates. Here, covariates refer to some input information other than system designs to the simulation model that can also affect the system performance. To make decisions, decision makers need to know the covariate values of the problem. Traditionally in simulation-based decision making, simulation samples are collected after the covariate values are known; in contrast, as a new framework, simulation with covariates starts the simulation before the covariate values are revealed, and collects samples on covariate values that might appear later. Then, when the covariate values are revealed, the collected simulation samples are directly used to predict the desired results. This framework significantly reduces the decision time compared to the traditional way of simulation. In this paper, we follow this framework and suppose there are a finite number of system designs. We adopt the metamodel of stochastic kriging (SK) and use it to predict the system performance of each design and the best design. The goal is to study how fast the prediction errors diminish with the number of covariate points sampled. This is a fundamental problem in simulation with covariates and helps quantify the relationship between the offline simulation efforts and the online prediction accuracy. Particularly, we adopt measures of the maximal integrated mean squared error (IMSE) and integrated probability of false selection (IPFS) for assessing errors of the system performance and the best design predictions. Then, we establish convergence rates for the two measures under mild conditions. Last, these convergence behaviors are illustrated numerically using test examples.

\textbf{Keywords:} simulation with covariates, convergence rate, stochastic kriging, ranking and selection
\end{abstract}


\section{Introduction}
\label{sec:1}

Stochastic simulation is a powerful tool for analyzing large-scale complex systems. In most of the real situations, systems are highly complex, precluding the possibility of applying analytical solutions; in contrast, simulation makes it possible to accurately describe a system through the use of logically complex, and often non-mathematical models. Consequently, detailed dynamics of the system can be faithfully modeled, the system performance can be studied, and the best system design can be selected \citep{chen2011}. Now simulation has been a widely-used operations-research and management-science technique, e.g., in the management of power systems \citep{Benini1998}, production planning \citep{kleijnen1993}, supply chain network \citep{ding2005}, emergency department \citep{Ahmed2009}, etc.

In these applications, the standard process for analyzing the system is to first establish estimators for measures of interest based on the simulation output, and then develop optimization methods to find the best design of the system. This process highlights the two main purposes of a constructed simulation model, for estimating the system performance and optimizing it over a set of system designs. Throughout the paper, we will refer to these two purposes of simulation as the \emph{estimation problem} and the \emph{optimization problem}.

When conducting simulation experiments, a common practice is to first reveal
and fix the covariate values for the problem under consideration, and then repeat experiments on the simulation model with various system designs. Here, covariates refer to some input information other than system designs to the simulation model which will also affect the system performance. In the literature, covariates are also known as the side information or context. For example, in queueing network design, covariates can be the arrival rate of the customers, which influences the queue length and the mean waiting time of the network. In disease treatment, covariates can be the biometric characteristics of the patients, which influence the efficacy of the treatment methods.

However, given the computational expense of simulation experiments, a notable issue with this practice, for both the purposes of estimation and optimization, is that the time for obtaining the desired simulation results can be very long for some real systems. In addition to the huge monetary cost it incurs, it significantly limits the use of simulation for online problems in which system performance and the best system design are expected soon after the covariate values are revealed. This is also one of the key concerns for simulation-related research \citep{law2015}.

To address this issue, \cite{hong2019} and \cite{shen2019} recently proposed a new framework of using simulation. Instead of running simulation after the covariate values are revealed, the new framework does it before that with randomly sampled covariate values that might possibly appear in future problem instances. It establishes an offline simulation dataset that is useful in describing the system. More importantly, this dataset serves for the purpose of prediction. When the covariate values of a certain problem are known, machine learning and data mining tools can be adopted to build predictive models and predict the performance of each design (the estimation problem) and the best design (the optimization problem) in real time\footnote{If certain adaptive methods are used to collect the covariate points, the predictive models need to be built iteratively, instead of once after all the covariate points are collected.}. For example, a doctor can learn the efficacy of the potential treatment methods and recommend a personalized treatment for a diabetic patient immediately upon his/her arrival by checking the simulation results under the same biometric characteristics (covariate values) of this patient \citep{bertsimas2017}. By doing so, the time for obtaining performance estimation and the best decision can be substantially reduced. It enables simulation to be used in a much broader range of applications for which simulation was hardly a feasible technique before. We call this framework \emph{simulation with covariates}.

The framework of simulation with covariates is quite general and new. A lot of key questions remain largely unexplored. In this research, we focus on the use of this framework in prediction and consider a fundamental problem in it, the quantification of the relationship between the offline simulation efforts and the online prediction accuracy. This quantification provides a good assessment on the quality of the estimated system performance and the best design that can be achieved using the offline dataset. We consider a continuous covariate space and a finite number of system designs. We sample the covariate space using a fixed distribution, conduct the same number of simulation replications on all the designs and sampled covariate points, and construct a predictive model for each design for predicting its performance and selecting the best design. Our main research question is to study the convergence rates of the prediction errors with the number of covariate points ever collected and to facilitate further decision making.

We employ the stochastic kriging (SK) model as the predictive model. SK has is one of the most extensively studied models for simulation output, e.g., in \cite{ankenman2010,chenx2013,qu2014,wang2018}. It is a general-purpose model with less structural assumptions than linear and some nonlinear models, and tends to be more resistant to overfitting than general interpolators \citep{sabuncuoglu2002}.

To evaluate the prediction errors of the estimation and optimization problems, we will use the maximal \emph{integrated mean squared error} (IMSE) and \emph{integrated probability of false selection} (IPFS) respectively. IMSE is the integral of the mean squared error of the SK model over the covariate space. An IMSE is associated to a system design, and describes the average MSE of the estimated system performance of this design over all the possible covariate values. The maximal IMSE corresponds to the largest IMSE from the designs. It serves as a measure for the worst-case error of the estimation problem, whose convergence rate governs the prediction errors for the performance of each design under consideration. IPFS is the integral of the probability of false selection, i.e., the probability of falsely selecting the best design using the SK predictions. It serves as a measure for the error of the optimization problem.

In this study, we use a fixed distribution to sample the covariate space for three reasons. First, for real systems, covariates usually follow a fixed population distribution that can be estimated from historical data. Therefore, the offline dataset generated from this distribution can faithfully describe the distributional characteristics of the system and lead to more accurate estimation over the covariate space. Second, from the experiment design perspective, although more sophisticated sequential designs may have the benefit of using fewer design points in the covariate space, they may not be able to incorporate the distributional information due to the high computational cost in each iteration and may incur higher simulation cost for certain types of response surfaces. In comparison, sampling from a fixed distribution has the advantage of being simple with a fixed prespecified offline simulation cost. The distributional information also helps achieve sufficiently good performance when the number of covariate points sampled is large, and this advantage becomes more obvious when the covariate space has a higher dimension. Third, the setting of fixed-distribution sampling enables us to theoretically derive concrete convergence rates for the two target measures. These convergence rates serve as a good benchmark against which improvement from future design methods with possibly faster convergence rates might be measured (theoretically or numerically).

\subsection{Contributions}  \label{sec:1.1}

Our work makes three main contributions.

First, we establish a formulation for characterizing the performance of simulation with covariates in both the estimation and optimization problems. As one of the first simulation-based real-time decision making frameworks, simulation with covariates resolves the long-standing issue of efficiency for simulation experiments, but has rendered itself unclear about the effectiveness of the decision that is made. Our research builds an SK prediction model for each system design under study and proposes measures for the estimation and optimization problems that evaluate the quality of the prediction over all the possible problem instances that might be encountered. It lays the ground for theoretical analysis of simulation with covariates and other possible simulation frameworks of this kind.

Second, we derive the convergence rates of the two target measures (the maximal IMSE and IPFS) with the number of sampled covariate points $m$ for three common types of SK covariance kernels: finite-rank kernels, exponentially decaying kernels and polynomially decaying kernels. Derivation for the rates of the two measures is based on the upper bounds of the IMSE of a single SK model, and contains additional analysis on the structures of the target measures. Specifically, we show that convergence rates of the two measures are both at the  magnitudes of $1/m$, $(\log m)^{\frac{d}{\kappa_*}}/m$ and $m^{-\frac{2\nu_*}{2\nu_*+d}}$ for the three types of kernels respectively. In these rates, $\kappa_*$ and $\nu_*$ are some kernel parameters, and $d$ is the dimension of covariates. We also show that the convergence rate of IPFS can be improved to exponential with additional mild assumptions on the tail of MSE of each SK model. They provide good insight into the practical performance simulation with covariates can achieve.

Third, based on the polynomial convergence rates of the maximal IMSE, we further propose a simple regression-based procedure to determine the number of distinct covariate points needed to achieve a target precision of the maximal IMSE in Section 5.3 of the Online Supplement. In addition, we numerically illustrate the convergence behaviors of the maximal IMSE and IPFS via several test examples, and show the impact of several factors on their convergence rates, including the problem structure, dimension of the covariate space, number of simulation replications and sampling distribution.

\subsection{Literature Review}

\label{sec:1.2}

There are two streams of literature related to this study.

The first stream is kriging, or Gaussian process regression, which is a popular interpolation method for building metamodels \citep{Stein99,kleijnen2009}. It interpolates the response surface of an unknown function using the realization of a Gaussian random field, and has proven to be a highly effective tool for global metamodeling. In \cite{ankenman2010}, kriging was extended to simulation modeling, in which the observations of the unknown function are no longer deterministic, but are corrupted by random noises. It is known as the stochastic kriging (SK). \cite{chenx2013} and \cite{qu2014} further enhanced SK by utilizing the gradient information when it is available, called stochastic kriging with gradient estimators (SKG). \cite{wang2018} proved the monotonicity of MSE in a sequential setting for both SK and SKG. Theoretical properties of Gaussian process regression and the related kernel ridge regression have been previously studied in \citet{VarZan11}, \citet{Steetal09}, etc. Instead of a single SK model studied in those papers, in this research, we are interested in measures from multiple SK models that are caused by multiple designs.

The second stream is ranking and selection (R\&S), in particular the fixed-budget R\&S. Fixed-budget R\&S is a basic problem in simulation-based optimization, seeking to determine the allocation of a fixed simulation budget in order to correctly select the best simulated system design among a finite set of alternatives. Popular methods in this field include the optimal computing budget allocation (OCBA, \cite{chen2000,chen2008,gao2017a,gao2017b}) and value of information procedure (VIP, \cite{frazier2008,ryzhov2016}). In particular, \cite{gao2019selecting} utilized the OCBA approach to solve the R\&S problem with discrete covariates and derived the asymptotic optimal sampling rule. Similar to fixed-budget R\&S, this research is also set up with a finite number of designs, and samples them with a fixed simulation budget to make decisions. However, this research is different in objective. It aims to analyze the convergence rates of the target measures based on an existing sampling scheme, instead of developing a new sampling scheme as in fixed-budget R\&S.

The rest of the paper is organized as follows. Section \ref{sec:2} presents the formulation of the problem. Sections \ref{sec:3} and \ref{sec:4} provide the main convergence rate results on the maximal IMSE and IPFS.  Numerical examples are presented in Section \ref{sec:6}, followed by conclusions and discussion in Section \ref{sec:7}. A preliminary study of this research appeared in \cite{gao2019rate}. That paper only focused on the exponentially decaying kernels, and presented the convergence rates of the maximal IMSE and IPFS without proof.

\section{Problem Formulation}
\label{sec:2}

In this section, we provide some preliminaries on the SK model and the definitions of the two target measures. For a summary of the key notation we use, please refer to Table 1 of the Online Supplement. Throughout the paper, the subscript $i$ is exclusively used to index the system design, and we will fold it for circumstances with no ambiguity.

\subsection{Stochastic Kriging}
\label{sec:2.1}

We consider a finite number of $k$ system designs. The performance of each design depends on $\X=(X_1,\ldots,X_d)^\top$, a vector of random covariates with support $\Xcal\subseteq \mathbb{R}^d$. For each $i=1,2,\ldots,k$, let $Y_{il} (\X)$ be the $l$-th simulation sample from design $i$ under covariate $\X$, and $y_i (\X)$ be the mean of design $i$, where the mean is taken with respect to the simulation noise. We assume that for any $\X=\x$, $Y_{il} (\x)= y_i (\x)+\epsilon_{il} (\x)$ where $\epsilon_{il} (\x)$'s are mean-zero simulation noises and are independent across different $i$, $l$ and $\x$.

The relationship between the performance $y_i (\x)$ of design $i$ and $\x$ is generally unknown and can only be estimated via stochastic simulations. In this paper, we use the SK model to describe $y_i(\x)$:
\begin{align}\label{krigmodel1}
	& y_i(\x) = \bbf_i(\x)^\top \bbeta_i + M_i(\x),\quad i=1,\ldots,k,
\end{align}
where $\bbf_i(\x)=(\uf_{i1}(\x),\ldots,\uf_{iq}(\x))^\top$ and $\bbeta_i=(\beta_{i1},\ldots,\beta_{iq})^\top$ are a $q\times 1$ vector of known functions of $\x$ and a $q\times 1$ vector of unknown parameters; $M_i(\x)$ is a realization (or sample path) of a mean zero stationary Gaussian process, with the covariance function $\bSigma_{M,i}(\x,\x')=\cov\left[M_i(\x),M_i(\x')\right]$ quantifying the covariance between $M_i(\x)$ and $M_i(\x')$ for any $\x,\x'\in \Xcal$. Model \eqref{krigmodel1} with regressor functions $\bbf_i(\cdot)$ is sometimes called \textit{universal kriging} (\citealt{Stein99}).

In our model setting, we assume that we randomly draw $m$ covariate (design) points $\X^m=\left\{\X_1,\ldots,\X_m\right\}$ of $\X$ from a sampling distribution $\PP_{\X}$. For a given covariate point sample $\x^m=\left\{\x_1,\ldots,\x_m\right\}$, we perform $n_j$ replications at covariate $\x_j$ for each of the $k$ designs. We denote the sample mean for design $i$ and covariate $\x_j$ by $\overline Y_i(\x_j) = n_j^{-1}\sum_{l=1}^{n_j} Y_{il}(\x_j)$, and correspondingly the averaged simulation errors by $\overline \epsilon_i(\x_j) = n_j^{-1}\sum_{l=1}^{n_j} \epsilon_{il}(\x_j)$. For $i=1,\ldots,k$ and $j=1,\ldots,m$, we let $\Y_{ij}=(Y_{i1}(\x_j),\ldots, Y_{in_j}(\x_j))^\top$, and let $\overline \Y_i = \left(\overline Y_i(\x_1),\ldots, \overline Y_i(\x_m)\right)^\top$. For design $i$, let the $m \times q$ design matrix be $\F_i=(\bbf_i(\x_1),\ldots,\bbf_i(\x_m))^\top$. Let $\bSigma_{M,i}(\x^m,\x^m)$ be the $m\times m$ covariance matrix across all covariate points $\x_1,\ldots,\x_m$, i.e., for $s,t\in\{1,\ldots,m\}$, the $(s,t)$ entry of $\bSigma_{M,i}(\x^m,\x^m)$ is $[\bSigma_{M,i}(\x^m,\x^m)]_{st} = \cov\left[y_i(\x_s),y_i(\x_t)\right]$. For any $\x\in \Xcal$, let
$$\bSigma_{M,i}(\x^m,\x)=\left(\cov\left[y_i(\x),y_i(\x_1)\right], \ldots, \cov\left[y_i(\x),y_i(\x_m)\right]\right)^\top.$$
Let $\bSigma_{\epsilon,i}(\x^m)$ be the $m\times m$ covariance matrix of the averaged simulation errors across $m$ covariate points in the design $i$, i.e., for $s,t\in\{1,\ldots,m\}$, the $(s,t)$ entry of $\bSigma_{\epsilon,i}(\x^m)$ is $\{\bSigma_{\epsilon,i}(\x^m)\}_{st} = \cov\left[\overline \epsilon_{i}(\x_s), \overline \epsilon_{i}(\x_t)\right]$. Let $\bSigma_{y,i}=\bSigma_{M,i}(\x^m,\x^m)+\bSigma_{\epsilon,i}(\x^m)$.

To estimate $y_i(\x)$ in \eqref{krigmodel1}, we consider linear predictors in the form of $\alpha_{i,0}(\x_0)+\bm{\alpha}_i(\x_0)\overline \Y_i$, where $\alpha_{i,0}(\x_0)$ and $\bm{\alpha}_i(\x_0)$ are weights
that depend on the test covariate point $\x_0\in \Xcal$. The mean squared error MSE of the predictors at $\x_0$ is given by $\mse_{i}(\x_0)=\E[(y_i(\x_0)-\alpha_{i,0}(\x_0)-\bm{\alpha}_i(\x_0)\overline \Y_i)^2]$, where the expectation is with respect to the randomness in $\overline \Y_i$, i.e., the simulation noise. We call the predictor that minimizes $\mse_{i}(\x_0)$ MSE-optimal linear predictor. \cite{Stein99} (and also \citealt{ankenman2010}, \citealt{chenx2013}) has shown that the MSE-optimal linear predictor has the form
\begin{align}\label{stockrig1}
	\widehat{y}_i(\x_0) &= \bbf_i(\x_0)^\top \widehat\bbeta_{i} + \bSigma_{M,i}(\x^m,\x_0) ^\top \bSigma_{y,i}^{-1} \left(\overline \Y_i - \F_i \widehat\bbeta_{i}\right),
\end{align}
where $\widehat\bbeta_i= \left(\F_i^\top \bSigma_{y,i}^{-1} \F_i\right)^{-1}\F_i^\top \bSigma_{y,i}^{-1} \overline \Y_i$.

In addition, \citet{ankenman2010} has shown that the optimal MSE from Equation \eqref{stockrig1} at $\x_0\in \Xcal$ is:
\begin{align}\label{optimmse}
	\mse_{i,\opt}(\x_0) &= \bSigma_{M,i}(\x_0,\x_0) - \bSigma_{M,i}^\top(\x^m,\x_0) \left[\bSigma_{M,i}(\x^m,\x^m) +  \bSigma_{\epsilon,i}(\x^m)\right]^{-1} \bSigma_{M,i}(\x^m,\x_0) \nonumber \\
	& ~ + \eta_i(\x_0)^\top \left[\F_i^\top \left(\bSigma_{M,i}(\x^m,\x^m)+\bSigma_{\epsilon,i}(\x^m)\right)^{-1}\F_i\right]^{-1} \eta_i(\x_0),
\end{align}
where $\eta_i(\x_0)=\bbf_i(\x_0) - \F_i^\top \left(\bSigma_{M,i}(\x^m,\x^m)+\bSigma_{\epsilon,i}(\x^m)\right)^{-1} \bSigma_{M,i}(\x^m,\x_0)$.

In the following, we define some useful notation. For any finite dimensional vector $\mathbf{v}$, we let $\|\mathbf{v}\|$ be its Euclidean norm. For any generic matrix $A$, we use $A_{ab}$ to denote its $(a,b)$-entry, $cA$ to denote the matrix whose $(a,b)$-entry is $cA_{ab}$ for any constant $c\in \RR$. For any
positive definite matrix $A$, let $\lambda_{\max}(A)$ and $\lambda_{\min}(A)$ be its largest and smallest eigenvalues.  For two sequences of positive numbers $\{a_l\}_{l\geq 1}$ and $\{b_l\}_{l\geq 1}$, $a_l\lesssim b_l$ means that $\limsup_{l\to\infty} a_l/b_l < \infty$, and $a_l\asymp b_l$ means that both $a_l\lesssim b_l$ and $b_l\lesssim a_l$ hold true.

We introduce some concepts from the reproducing kernel Hilbert space (RKHS) theory that will be used in our theorems. Let $\PP_{\X}$ be a probability distribution over $\Xcal$, $L_2(\PP_{\X})$ be the $L_2$ space under $\PP_{\X}$. The inner product in $L_2(\PP_{\X})$ is defined as $\langle f, g\rangle_{L_2(\PP_{\X})} = {\E}_{\X} [f(\X)g(\X)]$ for any $f, g \in L_2(\PP_{\X})$. For any $f\in L_2(\PP_{\X})$, define the linear operator $[T_{\bSigma_M}f](\x) = \int_{\Xcal} \bSigma_M(\x,\x')f(\x')\ud \PP_{\X}(\x')$ for any $\x\in \Xcal$. Since $\bSigma_{M}(\cdot, \cdot)$ is a continuous symmetric non-negative definite kernel on $\Xcal\times \Xcal$, there exists an orthonormal basis $\left\{\phi_l(\x): l=1,2,\ldots\right\}$ with respect to $\PP_{\X}$ consisting of eigenfunctions of the linear operator $T_{\bSigma_M}$, i.e., $\int_{\Xcal} \phi_l^2(\x) d\PP_{\X}(\x) = 1$, $\int_{\Xcal} \phi_l(\x)\phi_{l'}(\x) d\PP_{\X}(\x) = 0$ for $l\neq l'$, and $[T_{\bSigma_M}\phi_l](\x)=\mu_l \phi_l(\x)$ for some eigenvalue $\mu_l\geq 0$, all $l=1,2,\ldots$ and $\x\in \Xcal$. According to Mercer's theorem (e.g. Theorem 4.2 of \citealt{RasWil06}), the kernel $\bSigma_{M}$ (which can be taken as any $\bSigma_{M,i}$ for $i=1,\ldots,k$) has the series expansion $\bSigma_{M}(\x, \x') = \sum_{l=1}^{\infty} \mu_l \phi_l(\x) \phi_l(\x')$ with respect to $\PP_{\X}$ for any $\x,\x'\in \Xcal$, where we assume that the eigenvalues of $\bSigma_{M}$ are sorted into the decreasing order $\mu_1\geq \mu_2 \geq \ldots\geq 0$. The trace of the kernel $\bSigma_{M}$ is defined as $\tr(\bSigma_{M})=\sum_{l=1}^{\infty} \mu_l$.
Any function $f\in  L_2(\PP_{\X})$ has the series expansion $f(\x) = \sum_{l=1}^{\infty} \theta_l \phi_l(\x)$, where $\theta_l = \langle f,\phi_l\rangle_{L_2(\PP_{\X})}$. The reproducing kernel Hilbert space (RKHS) $\HH$ attached to the kernel $\bSigma_{M}$ is the space of all functions $f \in L_2(\PP_{\X})$ such that its $\HH$-norm $\|f\|_{\HH}^2 = \sum_{l=1}^{\infty} \theta_l^2 / \mu_l<\infty$. We refer the readers to \citet{Gu02} and \citet{HsiEub15} for a complete treatment of the RKHS theory.

Based on the decaying rates of eigenvalues, most commonly used covariance functions (kernels) can be categorized into the three types described below: the finite-rank kernels, exponentially decaying kernels, and polynomially decaying kernels. For a comprehensive review of covariance functions, see Chapter 4 of \citet{RasWil06}.
\begin{enumerate}
	\item \textbf{Finite-rank kernels} satisfy $\mu_1\geq \ldots \geq \mu_{l_*} >0$ and $\mu_{l_*+1} = \mu_{l_*+2} =\ldots=0$ for some finite integer $l_*\in \NN$. One example of finite-rank kernels is $\bSigma_M(\x,\x')= (1+\x^\top \x')^D$ for some fixed positive integer $D$ and any $\x,\x'\in \Xcal$. The sample paths generated from this kernel are the class of all polynomial functions up to the degree $D$, and has the finite rank at most equal to $D+1$ (\citealt{RasWil06}). If $D=1$, then $\bSigma_M(\x,\cdot)$ generates the class of linear functions in $\x$.
	\item \textbf{Exponentially decaying kernels} satisfy $\mu_l\asymp \exp(-c l^{\kappa/d})$ for some constants $c>0,\kappa>0$, with $d$ being the dimension of covariate $\x$. The most important example is the squared exponential kernel $\bSigma_M(\x,\x')=\exp\left\{-\varphi\|\x-\x'\|^2\right\}$ for $\varphi>0$ and $\x,\x'\in \Xcal\subseteq \RR^d$. If $d=1$, $\PP_{\X}=N(0,(4a_1)^{-1})$ for some $a_1>0$, then it is known (\citealt{RasWil06} Section 4.3.1) that for $l=0,1,2,\ldots$, the eigenfunctions can be taken as $\phi_l(\x)=(a_2/a_1)^{1/4}\exp\{-(a_2-a_1)\x^2\}H_{l}(\sqrt{2a_2}\x)/\sqrt{2^l l!}$, and the corresponding eigenvalues are $\mu_l = \sqrt{2a_1/(a_1+a_2+\varphi)} \exp\{-l\log (1/a_3)\}$, where $a_2=\sqrt{a_1^2+2a_1\varphi}$, $a_3 = \varphi/(a_1+a_2+\varphi) \in (0,1)$, and $H_l(z) = (-1)^l \exp(x^2)\tfrac{d^l}{dx^l}\exp(-x^2)$ is the $l$th order Hermite polynomial. So $\mu_l\asymp \exp(-c l^\kappa)$ holds with $c=\log (1/a_3)$ and $\kappa=1$. In general, $\mu_l\asymp \exp(-c l^{\kappa/d})$ holds for infinitely smooth stationary kernels on a bounded domain $\Xcal \subseteq \RR^d$ (\citealt{SanSch16}).
	\item \textbf{Polynomially decaying kernels} satisfy $\mu_l \asymp l^{-2\nu/d-1}$ for some constant $\nu>0$ (such that $\tr(\bSigma_M)<\infty$). One example is the kernel $\bSigma_M(\x,\x')=\min\{\x,\x'\}$ for $\x,\x'\in \Xcal=[0,1]$. This kernel generates the first-order Sobolev class that contains all Lipschitz functions on $[0,1]$. If $\PP_{\X}$ is the uniform distribution on $[0,1]$, then it is known that $\mu_l \asymp 1/l^4$ (\citealt{Gu02}). Another very important example is the Mat\'ern kernel $\bSigma_{M,i}(\x,\x')=\tfrac{2^{1-\nu}}{\Gamma(\nu)}\left(\sqrt{2\nu}\varphi\|\x-\x'\|\right)^{\nu} K_{\nu}(\sqrt{2\nu}$ $\varphi\|\x-\x'\|)$, where $K_{\nu}$ is the modifed Bessel function and the smoothness parameter $\nu$ satisfies $\nu >0$. The Mat\'ern kernel is widely used for fitting spatial surfaces with varying roughness from $\nu$. A smaller $\nu$ generates rougher sample paths. If $\Xcal\subseteq \RR^d$ is a bounded set, then the Mat\'ern kernel has eigenvalues decaying as $\mu_l \leq C l^{-2\nu/d-1}$ for some constant $C>0$ (\citealt{SanSch16}).
	
\end{enumerate}

\subsection{Target Measures}
\label{sec:2.2}

For the estimation problem and a given covariate point sample $\x^m=\left\{\x_1,\ldots,\x_m\right\}$, the optimal MSE of the linear predictor \eqref{stockrig1} for design $i$ is $\mse_{i,\opt}(\X_0)$, where the test point $\X_0$ is randomly drawn from the same distribution $\PP_{\X}$ as for $\X^m$. The IMSE for the $i$-th design is the integral of $\mse_{i,\opt}(\X_0)$ with respect to the sampling distribution of $\X_0$
\begin{equation*}
	\text{IMSE}_i={\E}_{\X_0}\left[\mse_{i,\opt} (\X_0)\right],
\end{equation*}
and the maximal IMSE is defined as $\max_{i\in\{1,\ldots,k\}}\text{IMSE}_i$.

Under our consideration, the maximal IMSE can be viewed as a measurement of the prediction error with the worst MSE-optimal linear predictor among the $k$ designs over all possible locations in $\Xcal$. Our goal for the estimation problem is to prove that as the simulation budget increases to infinity, the maximal IMSE decreases at a certain rate to zero, under the correct specification of Model \eqref{krigmodel1} and other necessary mild technical assumptions. In particular, for the ease of presentation, we assume that all points in $\x^m$ receive the same number of simulation runs $n_1=\ldots =n_m =n$, i.e., we do not need to decide the number of simulation replications among different designs and covariate points. We will show that for any given $n$, the maximal IMSE converges to zero at some decreasing rate of $m$, which is the number of distinct points in $\x^m$. Intuitively, this goal is reasonable, because an SK model allows us to interpolate the unknown surface of $y_i(\x)$ at a new location with higher accuracy if $m$ becomes larger. How fast the maximal IMSE converges to zero in terms of $m$ depends mainly on the smoothness of all the unknown true surfaces $y_i(\x)$, $i=1,\ldots,k$. Since we assume that the true surface $y_i(\x)$ is correctly specified as in Model \eqref{krigmodel1}, then equivalently, the convergence rate of the maximal IMSE depends on the properties of the covariance kernel $\bSigma_{M,i}(\cdot,\cdot)$ and the functions $\bbf_i(\cdot)$. Note that the maximal IMSE is still random with respect to the covariate point sample $\X^m$, and our rate result for the maximal IMSE will be obtained in $\PP_{\X^m}-$ probability.

For the optimization problem, given configuration of designs $M_i(\cdot)$'s and a covariate point sample $\x^m= \left\{\x_1,\ldots,\x_m\right\}$, the real best design $i^\circ(\x_0)$ and the estimated best design $\widehat i^\circ (\x_0)$ at test point $\X_0=\x_0$ are
\begin{align}\label{bestdesigndef}
	& y^\circ(\x_0)=\min_{i\in\{1,\ldots,k\}}y_i (\x_0),\qquad i^\circ(\x_0) \in \arg\min_{i\in\{1,\ldots,k\}} y_i (\x_0), \nonumber \\
	& \widehat y^\circ(\x_0)=\min_{i\in\{1,\ldots,k\}} \widehat y_i (\x_0), \qquad \widehat i^\circ (\x_0) \in \arg\min_{i\in\{1,\ldots,k\}} \widehat y_i (\x_0).
\end{align}

Typically in R\&S problems, the correct selection for the best design is defined as $\widehat i^\circ (\x_0)=i^\circ(\x_0)$. However, due to the continuous nature of $\x_0$ in the framework of simulation with covariates, the best design $i^\circ(\x_0)$ might not be unique for certain values of $\x_0$, causing ambiguity in this definition. To solve this issue, in this research, we will focus the event of good selection \citep{ni2017}. Similarly as in the indifference-zone (IZ) formulation for R\&S problems \citep{kim2006}, suppose there is an IZ parameter $\delta_0>0$ showing the minimal difference for the means of designs that we believe is worth detecting. A good selection for $i^\circ(\x_0)$ happens when the mean of the estimated best design $y_{\widehat i^\circ (\x_0)}(\x_0)$ is better than $y^\circ(\x_0)+\delta_0$ for the test point $\x_0\in \Xcal$; equivalently, a false (not good) selection happens when $y_{\widehat i^\circ (\x_0)}(\x_0)$ is no better than $y^\circ(\x_0)+\delta_0$. This definition allows some flexibility for determining the best design when the means of the top two designs are very close or exactly the same under some covariate value. Consequently, probabilities of good selection $\pcs(\x_0)$ and false selection $\pfs(\x_0)$ among the $k$ alternatives at $\x_0$ are given by
\begin{align} \label{pfs_def}
	\pcs(\x_0)&=\PP_{\epsilon}\left(y_{\widehat i^\circ (\x_0)} (\x_0)-y^\circ(\x_0) < \delta_0\right), \nonumber \\
	\pfs(\x_0)&=\PP_{\epsilon}\left(y_{\widehat i^\circ (\x_0)} (\x_0)-y^\circ(\x_0) \geq \delta_0\right),
\end{align}
where $\PP_{\epsilon}$ is the joint probability measure of all simulation error terms $\epsilon_{il}(\x_j)$ for $i=1,\ldots,k$, $j=1,\ldots,m$ and $l=1,\ldots,n$. To ease the burden of notation, we hide the dependence of $\pcs(\x_0)$ and $\pfs(\x_0)$ on the constant IZ parameter $\delta_0$.

Consequently, the integrated PFS is defined as
\begin{equation*}
	\text{IPFS}={\E}_M{\E}_{\X_0}  \left[\pfs(\X_0)\right],
\end{equation*}
where $M$ contains the randomness from all $M_i(\cdot)$'s, $i=1,\ldots,k$, measuring the \textit{extrinsic uncertainty} (\citealt{ankenman2010}). Our goal for the optimization problem is to identify the convergence rate of IPFS with the number of covariate points $m$. Similarly as for the maximal IMSE, IPFS is still random with respect to $\X^m$, and our rate result for IPFS will be obtained in $\PP_{\X^m}-$ probability.

We note two key differences bewteen our setting and existing research in the simulation literature. First, we assume that $\X_0$ is randomly drawn from $\PP_{\X}$, independently of the random sample $\X^m$. Our treatment of both $\X^m$ and $\X_0$ is different from most SK studies (\citealt{ankenman2010}, \citealt{chenx2013}, \citealt{wang2018}), which usually treat $\X^m$ as fixed covariate points and $\X_0$ as uniformly sampled from $\Xcal$. The randomness in $\X^m$ allows us to derive the asymptotic convergence rates of the two target measures for various types of covariance kernels.

Second, although the maximal IMSE and IPFS are expected (integrated) measures, same in appearance to the expected measure $\text{PCS}_{\text{E}}$ in the research of ranking and selection with covariates \citep{shen2019}, the expectations in these two papers are caused by different types of randomness, leading to intrinsical difference in meaning and structure of these measures and the approaches used to analyze them. \cite{shen2019} considered a fixed number of $m$ covariate points, and the expectation in $\text{PCS}_{\text{E}}$ is with respect to the random covariate points, which seeks to assess the average of selection quality over all the possible covariate values (problem instances). In this paper, expectation is with respect to the random test point, which seeks to assess the average of prediction quality over all the possible covariate values (problem instances). This research also faces the randomness of the covariate point sample $\X^m$, and as discussed above, it is handled with the development of convergence rates in $\PP_{\X^m}-$ probability.

\section{Convergence Rates of the Maximal IMSE}
\label{sec:3}

In this section, we study the convergence rate of the first target measure, the maximal IMSE. We make the following assumptions:

\begin{enumerate}[label=A.\arabic*]
	\item \label{c1} For $i=1,\ldots,k$, Model \eqref{krigmodel1} is correctly specified with $M_i(\cdot)$ being a sample path from a known covariance function $\Sigma_{M,i}(\cdot,\cdot)$. For $i=1,\ldots,k$, $j=1,\ldots,m$, $l=1,\ldots,n$, $\epsilon_{il} (\x_j)$'s are random variables with mean zero and variance $\sigma_i^2(\x_j)$, and they are independent across different $i$, $j$, and $l$. The simulation errors $\epsilon_{il} (\x_j)$'s are independent of the Gaussian process $M_i(\x)$ for all $i$, $j$, $l$ and $\x\in \Xcal$. There exist finite constants $\underline \sigma_0^2$ and $\overline \sigma_0^2$ such that $0<\underline \sigma_0^2 \leq \sigma_i^2(\x)\leq \overline \sigma^2_0 $ for all $i$ and $\x \in \Xcal$.
	\item \label{c2} (Trace class kernel) The kernel $\bSigma_{M,i}$ satisfies $\tr\left(\bSigma_{M,i}\right)<\infty$ for $i=1,\ldots,k$.
	\item \label{c3} (Basis functions) Let $\left\{\phi_{i,l}(\x): l=1,2,\ldots\right\}$ be an orthonormal basis with respect to $\PP_{\X}$ consisting of eigenfunctions of the linear operator $T_{\bSigma_{M,i}}$. There are positive constants $\rho_*$ and $r_* \geq 2$ common for all $i=1,\ldots,k$ such that ${\E}_{\X} \{\phi_{i,l}^{2r_*}(\X)\}  \leq \rho_*^{2r_*}$ for every $l=1,2,\ldots, \infty$.
	\item \label{c4} (Regressors) The regression functions satisfy $\uf_{is}\in \HH_i$ for all $i=1,\ldots,k$ and $s=1,\ldots,q$, where $\HH_i$ the RKHS attached to kernel $\bSigma_{M,i}$. Furthermore, $\lambda_{\min}\left({\E}_{\X}[\bbf_i(\X)\bbf_i(\X)^\top]\right)$ is lower bounded by a positive constant for all $i=1,\ldots,k$ if $\X$ follows the distribution $\PP_{\X}$.
\end{enumerate}

\ref{c1} assumes independence of the simulation noise $\epsilon_{il} (\x_j)$ between different designs, covariate points and replications, so we do not consider the common random number technique in the simulation experiments. An implication of this setting is that learning the performance of a design does not enable learning the performance of another design. \ref{c1} also makes a mild assumption on the second moment of the error distribution. For all derivations related to IMSE in this paper, we do not require $\epsilon_{il} (\x_j)$ to be normally distributed. The lower and upper bounds for the error variance are technical, which is trivially satisfied if the errors are homogeneous with a constant variance.

\ref{c2} assumes that the operator associated to the kernel $\bSigma_{M,i}$ is a trace class operator (\citealt{HsiEub15}). This will be verified later for all the three types of kernels described before, in which their eigenvalues typically decrease at least polynomially and are usually summable. \ref{c3} imposes a mild moment condition on the orthonormal basis functions. Sometimes \ref{c3} can be strengthened to the assumption that the $L_{\infty}$ norms of $\phi_{i,l}(\x)$'s are uniformly bounded for all $l=1,2,\ldots$ and all $\x\in \Xcal$. For example, if $\Xcal=[0,1]$ and $\PP_{\X}$ is the uniform distribution on $\Xcal$, then the eigenfunctions of the Mat\'ern covariance kernel with $\nu=1/2$ are the sine functions (Section 3.4.1 of \citealt{VT01}), whose $L_{\infty}$ norms are naturally bounded from above by constant, so that \ref{c3} trivially holds. The quantities $\rho_*$ and $r_*$ do not need to depend on $i$, because if the $i$th design satisfies ${\E}_{\X} \{\phi_{i,l}^{2r_i}(\X)\}  \leq \rho_i^{2r_i}$ for $r_i\geq 2$, one can let $r_*=\min_{i\in\{1,\ldots,k\}} r_i\geq 2$ and $\rho_*=\max\left(\max_{i\in\{1,\ldots,k\}} \rho_i,1\right)$. By Jensen's inequality, ${\E}_{\X} \{\phi_{i,l}^{2r_*}(\X)\} \leq \left[{\E}_{\X} \{\phi_{i,l}^{2r_i}(\X)\} \right]^{r_*/r_i}\leq \rho_i^{2r_i\cdot r_*/r_i} \leq \rho_*^{2r_*}$ and \ref{c3} holds.

\ref{c4} requires that the matrix ${\E}_{\X}[\bbf_i(\X)\bbf_i(\X)^\top]$ is nonsingular. This is a necessary condition for the identifiability of $\bbeta_i$, since a singular ${\E}_{\X}[\bbf_i(\X)\bbf_i(\X)^\top]$ implies that some functions in $\{\uf_{i1}(\x),\ldots,$ $\uf_{iq}(\x)\}$ can be written as a linear combination of others, making it impossible to estimate $\bbeta_i$. In most real applications, $\uf_{is}$'s are highly smooth functions such as monomials; see p.12 of \citet{Stein99} for a cogent argument. In such cases, $\uf_{is}\in \HH_i$ is satisfied in general. For example, if the domain $\Xcal$ is a bounded set and the covariance kernel is a Mat\'ern kernel, then $\HH$ is norm equivalent to a Sobolev space of functions with certain smoothness. Since a monomial $\uf_{is}$ is infinitely differentiable, $\uf_{is}$ lies in $\HH_i$.

We first restrict our discussion to a single SK model and drop the subscript $i$. From \eqref{optimmse}, for a given test point $\x_0$ and an SK model, we can decompose the optimal MSE into two parts:
\begin{align}\label{msedecomp}
	\mse_{\opt}(\x_0) &= \mse_{\opt}^{(M)}(\x_0) + \mse_{\opt}^{(\bbeta)}(\x_0), \nonumber \\
	\mse_{\opt}^{(M)}(\x_0)  &= \bSigma_{M}(\x_0,\x_0) - \bSigma_{M}^\top(\x^m,\x_0) \left[\bSigma_{M}(\x^m,\x^m)+  \bSigma_{\epsilon}\right]^{-1} \bSigma_{M}(\x^m,\x_0), \nonumber \\
	\mse_{\opt}^{(\bbeta)}(\x_0) &= \eta(\x_0)^\top \left[\F^\top \left(\bSigma_{M}(\x^m,\x^m)+\bSigma_{\epsilon}\right)^{-1}\F\right]^{-1} \eta(\x_0),
\end{align}
where $\eta(\x_0) =\bbf(\x_0) - \F^\top \left(\bSigma_{M}(\x^m,\x^m)+\bSigma_{\epsilon}\right)^{-1} \bSigma_{M}(\x^m,\x_0)$.
They are two distinct contributions to the total MSE from estimating $M(\x)$ and $\bbeta$, respectively.

The following two theorems provide upper bounds for the integrated $\mse_{\opt}^{(M)}(\X_0)$ and $\mse_{\opt}^{(\bbeta)}(\x_0)$ in \eqref{msedecomp}. Based on them, we can analyze the convergence behavior of the integrated $\mse_{\opt}(\x_0)$, and consequently the maximal IMSE.

\begin{theorem}\label{mseMthm}
	Under Assumptions \ref{c1}-\ref{c3}, the following relation holds
	\begin{align}\label{varbound1}
		& {\E}_{\X^m}{\E}_{\X_0} \left[\mse_{\opt}^{(M)} (\X_0)\right]\leq  \frac{2\overline \sigma_0^2}{mn} \gamma \left( \frac{\overline \sigma_0^2}{mn} \right) \nonumber \\
		& ~~ + \underset{\zeta \in \NN}{\inf} \,\left[ \left\{\frac{3mn}{\overline \sigma_0^2}\tr(\bSigma_{M})+1\right\}\tr\left(\bSigma_{M}^{(\zeta)}\right) + \tr(\bSigma_{M}) \left\{  300 \rho_*^2 \frac{b(m,\zeta,r_*) \gamma(\tfrac{\overline \sigma_0^2}{mn})}{\sqrt{m}} \right\}^{r_*}  \right],
	\end{align}
	where
	\begin{align*}
		& b(m,\zeta,r_*) = \max \left( \sqrt{\max(r_*, \log \zeta)},~ \frac{\max(r_*,\log \zeta)}{m^{1/2 - 1/r_*}} \right), \\
		& \gamma(a) = \sum_{l=1}^{\infty} \frac{\mu_l}{\mu_l+a} \text{ for any } a>0, ~~ \tr\left(\bSigma_{M}^{(\zeta)}\right) = \sum_{l=\zeta+1}^{\infty} \mu_l  \text{ for any } \zeta \in \NN.
	\end{align*}
\end{theorem}

Theorem \ref{mseMthm} provides an upper bound for the expectation of the IMSE ${\E}_{\X_0} \left[\mse_{\opt}^{(M)}(\X_0)\right]$. The reason we have another expectation ${\E}_{\X^m}$ before this IMSE is that $\X^m$ is a random sample from $\PP_{\X}$ and hence this IMSE is also random in $\X^m$. The upper bound in Theorem \ref{mseMthm} takes a complicated form and some discussion is in order. First of all, the first term in the upper bound \eqref{varbound1} is the dominant term, while the terms inside the infimum are typically of smaller stochastic orders than the first term, as we will show later in the proof of Theorem \ref{jointmsenew1} for three types of kernels. Second, inside the first term in \eqref{varbound1}, the term $\gamma(\frac{\overline \sigma_0^2}{mn})$ is known as the \textit{effective dimensionality} of the kernel $\bSigma_M$ with respect to $L_2(\PP_{\X})$ (\citealt{Zha05}). As we will show later in Theorem \ref{jointmsenew1}, the term $\frac{\overline \sigma_0^2}{mn}\gamma(\frac{\overline \sigma_0^2}{mn})$ is the dominant term that determines the convergence rate of IMSE. Third, the terms inside the infimum sign are stochastic errors due to the randomness in $\X^m$, and under Assumptions \ref{c1}-\ref{c3}, they are of negligible orders by choosing a proper $\zeta \in \NN$.

For two random variables $U_m$ and $V_m$ that are measurable with respect to the sigma-algebra generated by $\X^m$, we use $U_m ~{\lesssim}_{\PP_{\X^m}} V_m$ to denote the relation that $|U_m/V_m|$ is bounded in $\PP_{\X^m}-$ probability.
\begin{theorem}\label{msebetathm}
	Under Assumptions \ref{c1}-\ref{c4},  the following relation holds
	\begin{align}\label{varbound2}
		{\E}_{\X_0}\left[\mse_{\opt}^{(\bbeta)} (\X_0)\right] {\lesssim}_{\PP_{\X^m}} & \frac{8q\tr(\bSigma_M)}{\lambda_{\min}\left({\E}_{\X}[\bbf(\X)\bbf(\X)^\top]\right)}
		\Bigg\{8C_{\uf}^2 \frac{\overline \sigma_0^2}{mn} \nonumber \\
		&+ \inf_{\zeta\in \NN} \Bigg[8C_{\uf}^2 \frac{mn \overline \sigma_0^2}{ \underline \sigma_0^4} \rho_*^4 \tr\left(\bSigma_M \right) \tr\left(\bSigma_M^{(\zeta)}\right) + C_{\uf}^2 \tr\left(\bSigma_M^{(\zeta)}\right)\nonumber \\
		&~+ C_{\uf}^2  \tr\left(\bSigma_M\right)\left\{ 200 \rho_*^2 \frac{ b(m,\zeta,r_*)  \gamma(\tfrac{\overline \sigma_0^2}{mn})}{\sqrt{m}} \right\}^{r_*} \Bigg]\Bigg\},
	\end{align}
	where $C_{\uf}=\max_{1\leq s\leq q}\| \uf_s \|_{\HH}$, $b(m,\zeta,r_*)$ and $\gamma(\cdot)$ are defined in Theorem \ref{mseMthm}.
\end{theorem}

Similar to the upper bound in Theorem \ref{mseMthm}, the terms inside the infimum can be made negligible compared to the leading term of $\frac{\overline \sigma_0^2}{mn}$ by choosing a proper $\zeta \in \NN$. The upper bound in Theorem \ref{msebetathm} is a bound in probability, which means that as $m\to\infty$, the IMSE in \eqref{varbound2} is upper bounded in probability by the right-hand side. It is slightly weaker than the the upper bound on the expectation of IMSE in Theorem \ref{mseMthm}, but suffices for deriving the convergence rate of the maximal IMSE.

The following theorem gives our main rate result on the maximal IMSE.

\begin{theorem}\label{jointmsenew1}
	Suppose that all $k$ designs have the sampling distribution $\PP_{\X}$ for $\X^m$ and $\X_0$. Under Assumptions \ref{c1}-\ref{c4}, the following results hold with $r_*$ given in Assumption \ref{c3}:
	\begin{itemize}
		\item[(i)] \textit{(Finite-rank kernels)} If for every $i=1,\ldots,k$, $\bSigma_{M,i}$ is a finite-rank kernel of rank $l_{*i}$, i.e., its eigenvalues satisfy $\mu_{i,1}\geq \mu_{i,2}\geq \ldots \geq \mu_{i,l_{*i}}>0$ and $\mu_{i,l_{*i}+1}=\mu_{i,l_{*i}+2}=\ldots=0$, then as $m\to\infty$,
		\begin{align}\label{eqadd1}
			\max_{i\in\{1,\ldots,k\}}{\E}_{\X_0}\left[\mse_{i,\opt}(\X_0)\right]  ~{\lesssim}_{\PP_{\X^m}} ~ R^{F}(m,n)\equiv \max\left(\frac{1}{mn}, \frac{1}{m^{\frac{r_*}{2}}}\right).
		\end{align}
		
		\item[(ii)] \textit{(Exponentially decaying kernels)}  If for every $i=1,\ldots,k$, $\bSigma_{M,i}$ is a kernel with eigenvalues satisfying $\mu_{i,l} \leq c_{1i} \exp\left(-c_{2i} l^{\kappa_i/d} \right)$ for some constants $c_{1i}>0$, $c_{2i}>0$, $\kappa_i>0$ and all $l\in \NN$. Let $\kappa_*=\min_{i\in\{1,\ldots,k\}}\kappa_i$. Then, as $m\to\infty$,
		\begin{align}\label{eqadd2}
			\max_{i\in\{1,\ldots,k\}}{\E}_{\X_0}\left[\mse_{i,\opt}(\X_0)\right] ~{\lesssim}_{\PP_{\X^m}} ~ R^{E}(m,n) \equiv \max\left\{\frac{\left(\log(mn)\right)^{\frac{d}{\kappa_*}} }{mn}, \frac{ \left(\log(mn)\right)^{\frac{r_*(\kappa_*+d)}{\kappa_*}}}{m^{\frac{r_*}{2}}}\right\}.
		\end{align}
		
		\item[(iii)] \textit{(Polynomially decaying kernels)}  If for every $i=1,\ldots,k$, $\bSigma_{M,i}$ is a kernel with eigenvalues satisfying $\mu_{i,l} \leq c_{i} l^{-2\nu_i/d-1}$ for some constants $\nu_i>d/2$, $c_{i}>0$ and all $l\in \NN$. Let $\nu_*=\min_{i\in\{1,\ldots,k\}}\nu_i$. Then, as $m\to\infty$,
		\begin{align}\label{eqadd3}
			\max_{i\in\{1,\ldots,k\}}{\E}_{\X_0}\left[\mse_{i,\opt}(\X_0)\right] ~{\lesssim}_{\PP_{\X^m}} ~ R^{P}(m,n) \equiv \max\left\{\frac{1}{(mn)^{\frac{2\nu_*}{2\nu_*+d}}}, \frac{n^{\frac{dr_*}{2\nu_*+d}}(\log (mn))^{r_*} }{m^{\frac{r_*(2\nu_*-d)}{2\nu_*+d}}} \right\}.
		\end{align}
	\end{itemize}
\end{theorem}

\begin{remark}\label{ratermk1}
	\textit{(Simplified convergence rates for fixed $n$)} The convergence rates of the maximal IMSE for the three types of kernels in Theorem \ref{jointmsenew1} appear somehow complicated. However, since we perform the same number of simulation replications $n$ for each pair of covariate point and design, we can simplify the rate results by considering a \textit{fixed} $n$ and an increasing $m$ (to infinity). If $r_* > 2$ in Assumption \ref{c3}, then the larger terms in \eqref{eqadd1} and \eqref{eqadd2} are the first terms in the brackets; if $r_*>\tfrac{2\nu_*}{2\nu_*-d}$ in Case (iii), then the larger term in \eqref{eqadd3} is also the first term. By dropping the fixed constant of $n$, the convergence rates for the three kernels in Theorem \ref{jointmsenew1} can be simplified to: $1/m$ for Case (i), $(\log m)^{\frac{d}{\kappa_*}}/m$ for Case (ii), and $m^{-\frac{2\nu_*}{2\nu_*+d}}$ for Case (iii).
\end{remark}

The convergence rates of the maximal IMSE have been derived based on the upper bounds of ${\E}_{\X^m}{\E}_{\X_0} \left[\mse_{\opt}^{(M)} (\X_0)\right]$ and ${\E}_{\X_0}\left[\mse_{\opt}^{(\bbeta)} (\X_0)\right]$ in Theorems \ref{mseMthm} and \ref{msebetathm}. These rates are generally tight and cannot be improved. In Remark \ref{finitermk} below, we discuss the finite-rank kernels and formally prove in Theorem \ref{jointmsenew2} that the rate function $R^{F}(m,n)$ is optimal, in the sense that it cannot be improved further.

\begin{remark}\label{finitermk}
	\textit{(Example of a finite-rank kernel)} To illustrate the tightness of the bounds in Theorem \ref{jointmsenew1}, we show that the rate $1/(mn)$ in \eqref{eqadd1} can be attained for fixed $n$ as $m\to\infty$. For simplicity, we assume that in Model \eqref{krigmodel1}, $\bbf_i(\x)\equiv 0$ and $\epsilon_{il}(\x)$ is a homogeneous white noise process with mean 0 and a common constant variance $\sigma^2>0$ for $l=1,2,...,n$, $i=1,\ldots,k$, and $\x\in \Xcal$. Thus the model becomes $\overline Y_i(\x_j) = M_i(\x_j) + \overline \epsilon_i(\x_j)$ for $j=1,\ldots,m$ and $i=1,\ldots,k$. Let $\Xcal\subseteq \RR^d$, and let the $i$th covariance kernel be $\bSigma_{M,i}(\x,\x')=a_i(\x^\top \x'+b_i)$ for some known constants $a_i>0$ and $b_i>0$, $i=1,\ldots,k$. We analyze the MSE-optimal linear predictor in \eqref{stockrig1} and the asymptotic behavior of the optimal MSE in \eqref{optimmse}.

	\begin{theorem}\label{jointmsenew2}
		\textit{(Exact rate for a finite-rank kernel)} Suppose that the covariance kernels are $\bSigma_{M,i}(\x,\x')=a_i\big(\x^\top \x'+b_i\big)$ for $\x,\x'\in \Xcal\subseteq \RR^d$, known constants $a_i>0,b_i>0$ and $i=1,\ldots,k$. Under Assumptions \ref{c1}-\ref{c4} and the model setup described above, the MSE-optimal linear predictor in \eqref{stockrig1} and the optimal MSE in \eqref{optimmse} are given by
		\begin{align} \label{rank1blue}
			\widehat y_i(\x_0) & = a_i\widetilde \x_{i,0}^{\top} \Z_i^\top \left(a_i\Z_i \Z_i^\top + \frac{\sigma^2}{n}\I_m \right)^{-1} \overline \Y_i,  \nonumber \\
			\mse_{i,\opt}(\x_0) & = a_i\widetilde \x_{i,0}^{\top} \left(\I_{d+1} + \frac{a_i n}{\sigma^2}\Z_i^\top \Z_i \right)^{-1} \widetilde \x_{i,0},
		\end{align}
		for any $\x_0\in \RR$ and $i=1,\ldots,k$, where $\I_l$ is the $l\times l$ identity matrix, and
		\begin{align*}
			& 	\overline \Y_i=(\overline Y_i(\x_1),\ldots,\overline Y_i(\x_m))^\top \in \RR^m, \\
			& \widetilde \x_{i,0} = \left( \begin{array}{c}
				\sqrt{b_i} \\
				\x_0
			\end{array}\right) \in \RR^{d+1}, \quad  \Z_i = \left(
			\begin{array}{ccc}
				\sqrt{b_i} & \ldots & \sqrt{b_i} \\
				\x_1 & \ldots & \x_m
			\end{array}
			\right)^\top \in \RR^{m\times (d+1)}.
		\end{align*}
		Let $\PP_{\X}$ be any sampling distribution on $\RR^d$ for $\X_1,\ldots,\X_m,\X_0$, and assume that its second moment ${\E}_{\X_0}(\X_0 \X_0^\top)$ exists. Then as $m\to\infty$,
		\begin{align}\label{mseank1limit}
			mn \cdot \max_{i\in \{1,\ldots,k\}}{\E}_{\X_0}\left[\mse_{i,\opt}(\X_0)\right] \to (d+1)\sigma^2, \quad \text{ almost surely in } \PP_{\X^m}.
		\end{align}
	\end{theorem}
	
	Theorem \ref{jointmsenew2} shows that the maximal IMSE of the covariance kernel $\bSigma_{i,M}(\x,\x')=a_i\big(\x^\top \x'+b_i\big)$ decreases asymptotically at the rate $(d+1)\sigma^2/(mn)$. For fixed $n$, this has shown that the rate $1/m$ given in \eqref{eqadd1} for finite-rank kernels is tight and cannot be improved.
\end{remark}

\section{Convergence Rates of IPFS}
\label{sec:4}

We next consider the problem of selecting the best design from the $k$ alternatives, with their mean functions given in Model \eqref{krigmodel1}, and study how fast $\pfs(\X_0)$ converges to 0 (or equivalently, how fast $\pcs(\X_0)$ converges to 1). Similar to the analysis of the maximal IMSE before, the convergence rate here is again in the average sense, by taking expectations of $\pfs(\X_0)$ under three probability measures: (i) the joint Gaussian measure on $M_i(\cdot)$ ($i=1,\ldots,k$), denoted by $\PP_M$ (with the expectation denoted by ${\E}_M$), induced by the $k$ independent Gaussian processes with mean zero and covariance function $\bSigma_{M,i}(\cdot,\cdot)$ for $i=1,\ldots,k$; (ii) the probability measure of the testing point $\PP_{\X_0}$; and (iii) the probability measure of the sample $\PP_{\X^m}$.

In the following, $R(m,n)$ refers to the rate function of the maximal IMSE, which becomes $R^F(m,n)$, $R^E(m,n)$ or $R^P(m,n)$ under the corresponding kernels in Theorem \ref{jointmsenew1}. The following additional assumptions will lead to faster convergence rates of PFS in some particular scenarios.
\vspace{2mm}

\begin{enumerate}[label=A.\arabic*]
	\setcounter{enumi}{4}
	\item \label{c6} The simulation errors $\epsilon_{il} (\x)$'s are independent normal random variables following $N(0,\sigma_{i}^2(\x))$ for all $i=1,\ldots,k$, $l=1,\ldots,n$ and $\x\in \Xcal$.
	\item \label{c7} For any given $\xi\in(0,1/2)$, there exist constants $w_1>0,w_2>0,m_0\geq 1$ that depend on $\xi$, such that for $m\geq m_0$, for any $t>0$,
	\begin{align}
		&\PP_{\X^m} \left\{\PP_{\X_0}\left(\frac{\max_{i\in \{1,\ldots,k\}}\mse_{i,\opt}(\X_0)}{R(m,n)}\geq t\right) \leq w_1 \exp\left(-w_2 t\right)\right\}\geq 1-\xi.
	\end{align}
	\item \label{c8} For any given $\xi\in(0,1/2)$, there exist constants $w_3>0,m_0\geq 1$ that depend on $\xi$, such that for $m\geq m_0$,
	\begin{align} \label{eq:c8}
		&\PP_{\X^m} \left\{\frac{\max_{i\in \{1,\ldots,k\}}\sup_{\x_0\in \Xcal}\mse_{i,\opt}(\x_0)}{R(m,n)} \leq w_3 \right\}\geq 1-\xi.
	\end{align}

\end{enumerate}

Although \ref{c6} is stronger than \ref{c1} by assuming normal observation noises, it is a common assumption in simulation-based optimization problems. We emphasize that the normality assumption in \ref{c6} is only needed for deriving tighter and exponentially small bounds for IPFS in Theorem \ref{pfsthm} below. Without \ref{c6}, we can still establish convergence rates of IPFS directly from the convergence rates of IMSE in Theorem \ref{jointmsenew1}; see Theorem \ref{pfsthm} Part (i). Assumption \ref{c7} requires that the maximum of the $k$ MSE's decays at an exponential rate with a high probability. This is often the case when the MSE is distributed like chi-square with an exponentially decaying right tail. \ref{c8} is an alternative condition stronger than \ref{c7}, requiring that the supremum of MSE over $\Xcal$ to be bounded with a high probability. Both \ref{c7} and \ref{c8} can be rigorously verified for the finite-rank kernel in Remark \ref{finitermk} and Theorem \ref{jointmsenew2}; see Theorem 6 and its proof in the Online Supplement. \ref{c6} together with either \ref{c7} or \ref{c8} will allow tighter bounds for the tail probability of PFS, and hence, sharpened convergence rates of IPFS, as shown in the next theorem.

\begin{theorem}\label{pfsthm}
	Suppose that all the $k$ designs have the sampling distribution $\PP_{\X}$ for $\X^m$ and $\X_0$. Let $\delta_0$ be the IZ parameter in the definition of $\pfs(\X_0)$.
	\begin{itemize}
		\item[(i)] If Assumptions \ref{c1}-\ref{c4} hold, then as $m\to \infty$, $\E_M\E_{\X_0} [\pfs(\X_0)] {\lesssim}_{\PP_{\X^m}} R(m,n) $;
		\item[(ii)] If Assumptions \ref{c1}-\ref{c7} hold, then as $m\to \infty$,
		\begin{align*}
			& {\E}_M{\E}_{\X_0}  \left[\pfs(\X_0)\right] {\lesssim}_{\PP_{\X^m}} \exp\left\{-\frac{1}{2}w_2^{1/2}\delta_0\left[R(m,n)\right]^{-1/2}\right\},
		\end{align*}
		where $w_2$ is given in Assumption \ref{c7};
		\item[(iii)] If Assumptions \ref{c1}-\ref{c6} and \ref{c8} hold, then as $m\to \infty$,
		\begin{align*}
			& {\E}_M{\E}_{\X_0}  \left[\pfs(\X_0)\right] {\lesssim}_{\PP_{\X^m}} \exp\left\{-\frac{1}{4}w_3^{-1}\delta_0^2\left[R(m,n)\right]^{-1}\right\},
		\end{align*}
		where $w_3$ is given in Assumption \ref{c8}.
		
	\end{itemize}
\end{theorem}

The convergence rates of IPFS in Theorem \ref{pfsthm} include the measure $\PP_M$ and its expectation ${\E}_M$, mainly for the convenience of technical treatment, so that our result is general and does not depend on the particular shapes of the $M_i(\cdot)$ functions.

Theorem \ref{pfsthm} provides three convergence rates, from slower to faster, under sequentially stronger sets of assumptions. In Part (i), if we only assume \ref{c1}-\ref{c4} without the normality assumption on error terms, then by a direct application of Markov's inequality, the convergence rate of IPFS is at least as fast as that of the maximal IMSE given in Theorem \ref{jointmsenew1}. If the covariance kernels of the $k$ designs belong to one of the three types of kernels described before, then when $n$ is fixed, we know from Theorem \ref{jointmsenew1} and Remark \ref{ratermk1} that $R(m,n)$ converges to zero at the rate of $1/m$, $(\log m)^{\frac{d}{\kappa_*}}/m$ and $m^{-\frac{2\nu_*}{2\nu_*+d}}$ for the three types of kernels, respectively. As a result, Part (i) of Theorem \ref{pfsthm} implies that these polynomial rates for IMSE also hold for IPFS (and IPGS): when $n$ is fixed, IPFS converges to zero (and the IPGS converges to one) at least polynomially fast in $m$, at least at the rate of $1/m$, $(\log m)^{\frac{d}{\kappa_*}}/m$ and $m^{-\frac{2\nu_*}{2\nu_*+d}}$ for the three types of kernels, respectively.

In Part (ii) of Theorem \ref{pfsthm}, the additional normality assumption of \ref{c6} and Assumption \ref{c7} provide sharpened convergence rates of IPFS than in Part (i), from the polynomial rate in Part (i) to an exponential rate. In particular, following Theorem \ref{jointmsenew1} and Remark \ref{ratermk1}, if $n$ is fixed and $R(m,n)$ converges to zero at the rate of $1/m$, $(\log m)^{\frac{d}{\kappa_*}}/m$ and $m^{-\frac{2\nu_*}{2\nu_*+d}}$ for the three types of kernels, respectively, then Part (ii) of Theorem \ref{pfsthm} implies that the IPFS converges to zero (and the IPGS converges to one) at least exponentially fast in $m$, at least at the rate of $\exp(-c\sqrt{m})$, $\exp(-c\sqrt{m}(\log m)^{-\frac{d}{2\kappa_*}})$ and $\exp(-cm^{\frac{\nu_*}{2\nu_*+d}})$ for the three types of kernels, respectively, where the constant $c=w_2^{1/2}\delta_0/2$.

In Part (iii) of Theorem \ref{pfsthm}, the additional Assumptions \ref{c6} and \ref{c8} provide even more sharpened convergence rates of IPFS than in Part (ii). Following Theorem \ref{jointmsenew1} and Remark \ref{ratermk1}, if $n$ is fixed and $R(m,n)$ converges to zero at the rate of $1/m$, $(\log m)^{\frac{d}{\kappa_*}}/m$ and $m^{-\frac{2\nu_*}{2\nu_*+d}}$ for the three types of kernels, respectively, then Part (iii) of Theorem \ref{pfsthm} implies that the IPFS converges to zero (and the IPGS converges to one) at least exponentially fast in $m$, at least at the rate of $\exp(-cm)$, $\exp(-cm(\log m)^{-\frac{d}{\kappa_*}})$ and $\exp(-cm^{\frac{2\nu_*}{2\nu_*+d}})$ for the three types of kernels, respectively, where the constant $c=w_3^{-1}\delta_0^2/4$. Each of these exponential rates converges to zero faster than the corresponding exponential rate from Part (ii).

\begin{remark}
	Parts (ii) and (iii) of Theorem \ref{pfsthm} show that under additional assumptions on the distribution of simulation noises and tails of $\max_{i\in \{1,\ldots,k\}}\mse_{i,\opt}(\X_0)$ and $\max_{i\in \{1,\ldots,k\}} \sup_{\x_0\in \Xcal}$ $ \mse_{i,\opt}(\x_0)$, the convergence rate of IPFS can be exponentially fast. Note that this is distinguished from the well-established exponential convergence rate of the PFS in R\&S by comparing sample means of different designs (\citealt{dai1996,glynn2004}). In those studies, PFS is reduced by increasing the number of simulation replications for each design instead of increasing the number of covariate points, and its exponential convergence rate takes the form of $\exp(-\varrho n_{tot})$, where $n_{tot}$ is the total number of simulation samples and $\varrho$ is related to some large-deviations rate function.
\end{remark}

\begin{remark}
	\textit{(On the independence across different designs)} In the development of convergence rates of the two target measures, we have assumed in \ref{c1} that the simulation samples are independent across different designs $i$. This assumption is naturally the case when the designs are categorical, e.g., when the designs are the treatment methods for a certain disease. However, when the designs are represented as vectors in a metric space, they usually demonstrate spatial correlation, i.e., designs that are close to each other tend to have similar performance. For this case, our method and analysis can still be applied, but if the model can capture this spatial correlation between designs, it might lead to higher convergence rates for the maximal IMSE and IPFS. A possible way to do it is to build one SK that includes both the covariates and designs as inputs for predicting the system performance. That model is substantially different from ours, and further investigation along this direction is beyond the scope of this paper.
\end{remark}

\begin{remark}
	\textit{(On the choices of $m$ and $n$)} In Theorems \ref{mseMthm}-\ref{pfsthm}, we have assumed that the number of replications $n_i$ for covariate points of design $i$ remains the same across different designs. In practice, it is possible that the decision maker wants to unevenly allocate the simulation samples among the designs to optimize some target measures. In this case, $n_i$'s are no longer identical to each other. It falls in the well-established problem of ranking and selection (R\&S) in simulation. For this purpose, our analysis can still be applied. We will discuss this direction in Section 4 of the Online Supplement.

	When all the covariate points receive the same number of replications $n$, we can see that in all three cases of Theorem \ref{jointmsenew1}, the first term inside the maximum function in the rate expression is always a function of $n_{c}=mn$, while the second term depends on $m$ and $n$ separately. In order to make the maximal IMSE and IPFS decrease as fast as possible, we need to make the second term as small as possible, which means that for all three cases of Theorem \ref{jointmsenew1}, the best choice is to set $n=O(1)$, such that $m$ increases in the same order as $n_{c}$. Intuitively, this is because the maximal IMSE involves averaging MSE over all potential location $\x_0\in \Xcal$, and we should use as many distinct covariate points as possible in order to cover more locations in $\Xcal$. We emphasize that this analysis on the orders of $m$ and $n$ is only in the asymptotic sense based on our theoretical upper bounds.
\end{remark}

\begin{remark}
	\textit{(Determining the value of $m$)} When $n_i$'s are of a constant order, Theorems \ref{jointmsenew1} and \ref{pfsthm} imply that the maximal IMSE and IPFS decrease no slower than a polynomial order of $m$. This theory supports a natural procedure to determine the number of covariate points $m$. First, for given $m$ and $n_i$'s, the maximal IMSE and IPFS can be either calculated by numerical integration, or approximated by simple Monte Carlo estimators; see Section 3 of the Online Supplement. Second, after we fit a sequence of SK models with different sample sizes $m$, we can further fit a linear regression model with the logarithm of the maximal IMSE or IPFS as the response variable and $\log m$ as the predictor. Third, based on this fitted linear model, we reversely solve for the sample size $m^*$ such that the maximal IMSE or IPFS hits a small prespecified target precision. This simple procedure for determining $m$ is often accurate with IMSE and can be slightly conservative with IPFS, since sometimes IPFS can decay exponentially fast in $m$ as shown in Theorem \ref{pfsthm}. We will illustrate the practical implementation of this procedure in Section 5.3 of the Online Supplement.
\end{remark}

\vspace{1mm}

\section{Numerical Experiments}
\label{sec:6}

In this section, we adopt two benchmark functions and an M/M/1 queue example for numerical testing. These experiments can provide concrete presentation for the rates of the maximal IMSE and IPFS, and show the impact of the factors such as the problem structure, covariance kernel, dimension of the covariate space, number of simulation replications and sampling distribution on the convergence rates.  

For all the experiments, we implement four types of covariance kernels ($\|\cdot\|$ denotes the Euclidean distance):
\begin{enumerate}
	\item[(i)] Squared exponential kernel: $\bSigma_M(\x,\x')=\tau^2\exp\{-\varphi\|\x-\x'\|^2\}$, for $\x,\x'\in \Xcal$, $\tau^2>0$, and $\varphi>0$.
	\item[(ii)] Mat\'ern kernel with smoothness $\nu=5/2$: $\bSigma_{M}(\x,\x')=\tau^2(1+\sqrt{5}\varphi\|\x-\x'\| + \frac{5}{3}\varphi^2\|\x-\x'\|^2)\cdot \exp\{-\sqrt{5}\varphi\|\x-\x'\|\}$, for $\x,\x'\in \Xcal$, $\tau^2>0$, and $\varphi>0$.
	\item[(iii)] Mat\'ern kernel with smoothness $\nu=3/2$: $\bSigma_{M}(\x,\x')=\tau^2(1+\sqrt{3}\varphi\|\x-\x'\|)\cdot \exp\{-\sqrt{3}\varphi\|\x-\x'\|\}$, for $\x,\x'\in \Xcal$, $\tau^2>0$, and $\varphi>0$.
	\item[(iv)] Exponential kernel (Mat\'ern kernel with smoothness $\nu=1/2$): $\bSigma_{M}(\x,\x')=\tau^2\exp\{-\varphi\|\x-\x'\|\}$, for $\x,\x'\in \Xcal$, $\tau^2>0$, and $\varphi>0$.
\end{enumerate}

Similar to \citet{ankenman2010}, the covariance matrices $\bSigma_{\epsilon,i}(\x^m)$'s are estimated by $\text{diag}$ $\left\{\widetilde \sigma_i^2(\x_1)/n_1,\ldots, \widetilde \sigma_i^2(\x_m)/n_m\right\}$, where $\widetilde \sigma_i^2(\x_j)$ ($j=1,\ldots,m$) are estimated by the least-squares method based on the sample variances $\widehat \sigma_i^2(\x_j)=(n_j-1)^{-1}\sum_{l=1}^{n_j} [Y_{il}(\x_j)-\overline Y_i(\x_j)]^2$ ($j=1,\ldots,m$). Then given the estimated $\bSigma_{\epsilon,i}(\x^m)$'s, for each of the four kernels, we estimate the parameters $\varphi$ and $\tau^2$ by the maximum likelihood estimation. The squared exponential kernel (i) belongs to the exponentially decaying kernels and the other three kernels (ii)-(iv) belong to the polynomially decaying kernels. The smoothness of sample paths decreases from kernel (i) to kernel (iv), with (i) giving the smoothest sample paths and (iv) giving the roughest sample paths.

In all experiments below, we compute the estimated MSE at a single point $\x_0$ by the formula $\widehat \mse (\x_0) = [\widehat y(\x_0) -y(\x_0)]^2$, where $y(\x_0)$ is the true function value at $\x_0$ and $\widehat y(\x_0)$ is the fitted mean function. To evaluate the IMSE $\E_{\X_0}[\mse_{\opt}(\X_0)]$ over the domain $\Xcal$, we sample $T$ points of $\x_0$ from $\Xcal$ according to the distribution $\PP_{\X}$ and average their estimated MSEs $\widehat \mse (\x_0)$. In our experiments, $T$ is chosen as $10^3$, $10^4$, or $10^5$, depending on the dimension of $\x$. Monte Carlo estimates based on this setting of $T$ are in general accurate enough. Similarly, for each of the $T$ testing locations $\x_0$, we compute the true minimum mean performance $y^\circ(\x_0)$ and the estimated minimum mean performance $\widehat y^\circ(\x_0)$ according to \eqref{bestdesigndef}. Then the IPFS $\E_{\X_0}[\pfs(\X_0)]$ is computed by averaging over the $T$ points drawn from $\PP_{\X}$.

\subsection{Benchmark Functions}

We consider the following common benchmark functions. In all cases, $\x=(x_1,\ldots,x_d)^\top\in \RR^d$ is the covariate, $\z_i \in \RR^d$'s are the ``solutions" that index the different designs, and $\epsilon(\x)$ is an independent noise normally distributed as $N(0,(\sqrt{2})^2)$.

1. De Jong's function:
\begin{equation}\label{eq:bench1}
	Y(\x)=M(\x)+\epsilon(\x) = \sum_{l=1}^d (x_l-z_l)^2+\epsilon(\x).
\end{equation}
For function $M(\x)$, the global minimum $\x^*$ is obtained at $x_l=z_l$, $l=1,2,...,d$ with $M(\x^*)=0$. We consider 10 discrete designs with the $i$-th design $\z^i=(\underbrace{i,...,i}_{d})$, $i=1,2,...,10$.

2. Griewank's function:
\begin{equation}\label{eq:bench4}
	Y(\x)=M(\x)+\epsilon(\x)=\frac{1}{4000}\sum_{l=1}^d (x_l-z_l)^2-\prod_{l=1}^d \cos\left(\frac{x_l-z_l}{\sqrt{l}}\right)+1+\epsilon(\x).
\end{equation}
For function $M(\x)$, the global minimum $\x^*$ is obtained at $x_l=z_l$, $l=1,2,...,d$ with $M(\x^*)=0$. We consider 10 discrete designs with the $i$-th design $\z^i=(\underbrace{i,...,i}_{d})$, $i=1,2,...,10$.

Note that the performance of these functions depends on both the covariate $\x$ and design (solution) $\z$. We denote $y(\x)$ to highlight the input $\x$ to the SK model.

In this numerical test, we consider the De Jong's functions with $d=1$ and 3 and the Griewank's functions with $d=1$ and $10$. To better understand the two test functions, we have provided plots of them in Section 5.1 of the Online Supplement. The De Jong's functions are relatively smooth. The Griewank's functions are highly nonlinear with many oscillations, which brings difficulty to SK modeling when the number of covariate points $m$ is small.

We consider three sampling distributions for $\X^m$: uniform, truncated normal and normal distributions. The covariate space is $\Xcal = [1,10]^d$ when $d=1$, is $\Xcal = [1,4]^d$ when $d=3,10$ for the uniform and truncated normal sampling, and is $\Xcal = \RR^d$ for the normal sampling. For the truncated normal distribution, the mean and variance on each dimension are $(5.5,7^2)$ when $d=1$ and $(2.5,3^2)$ when $d=3,10$. The normal distribution on each dimension is $N(5.5,(\sqrt{3})^2)$ when $d=1$ and $N(2.5,1^2)$ when $d=3,10$.

\begin{figure}[htbp]
	\begin{center}
		\includegraphics[width=0.99\textwidth]{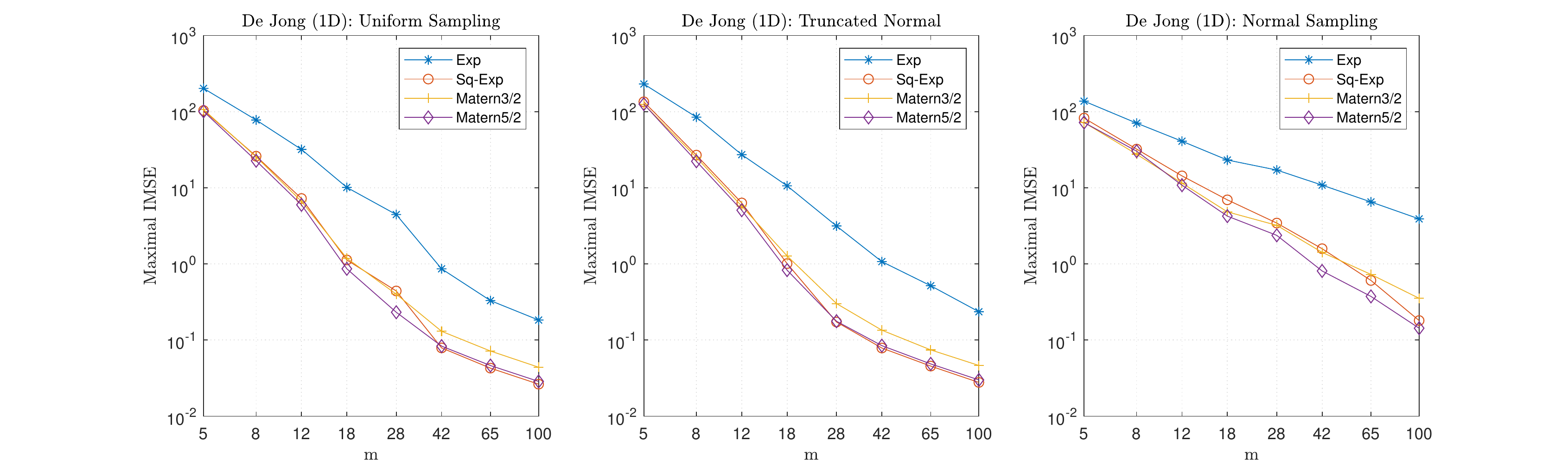}
		\includegraphics[width=0.99\textwidth]{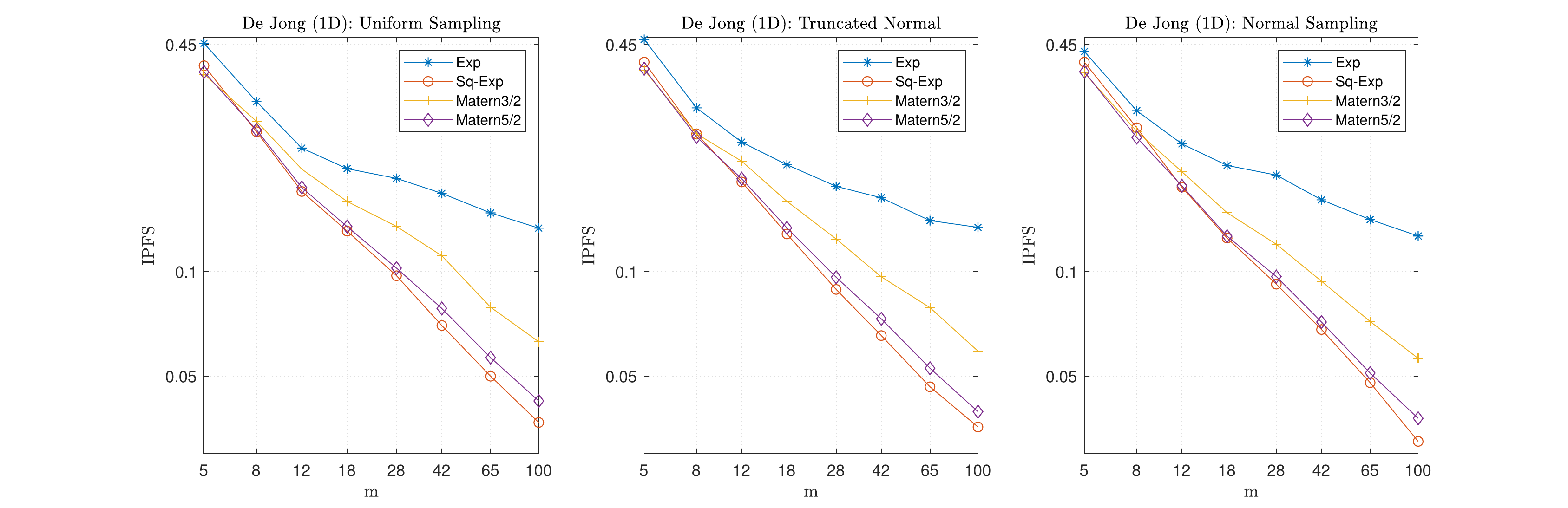}
	\end{center}
	\caption{1-d De Jong's functions: maximal IMSE and IPFS under different covariance kernels and sampling distributions.}
	\label{fig:test1d1}
\end{figure}

\begin{figure}[htbp]
	\begin{center}
		\includegraphics[width=0.92\textwidth]{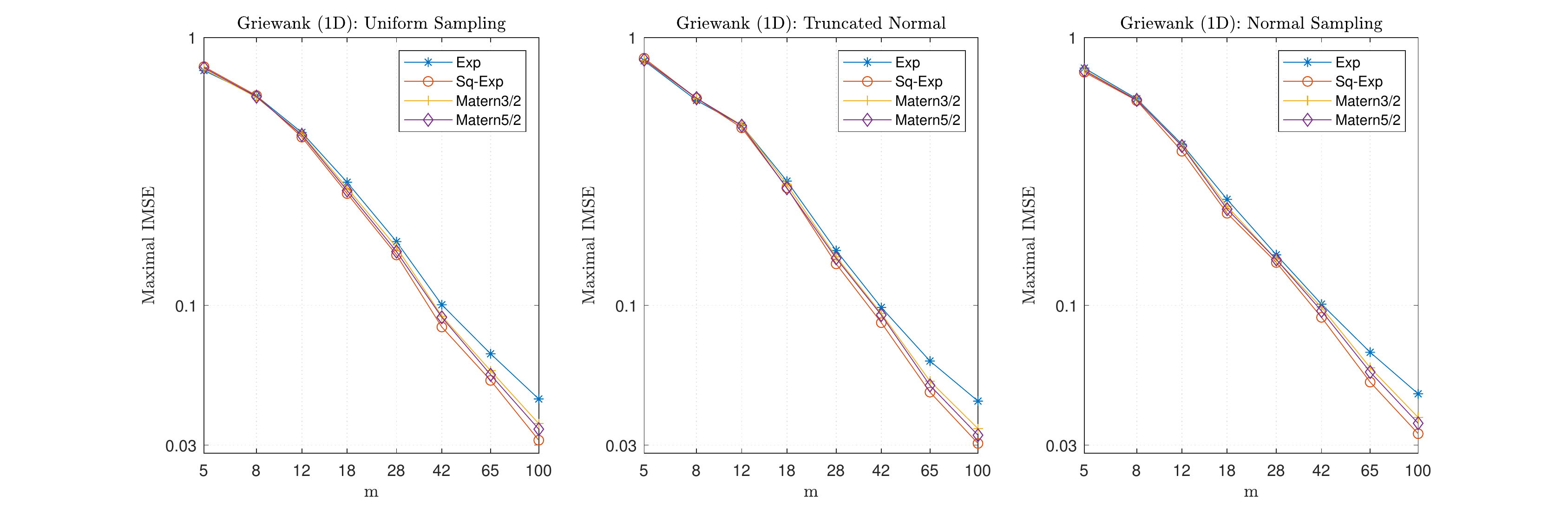}
		\includegraphics[width=0.92\textwidth]{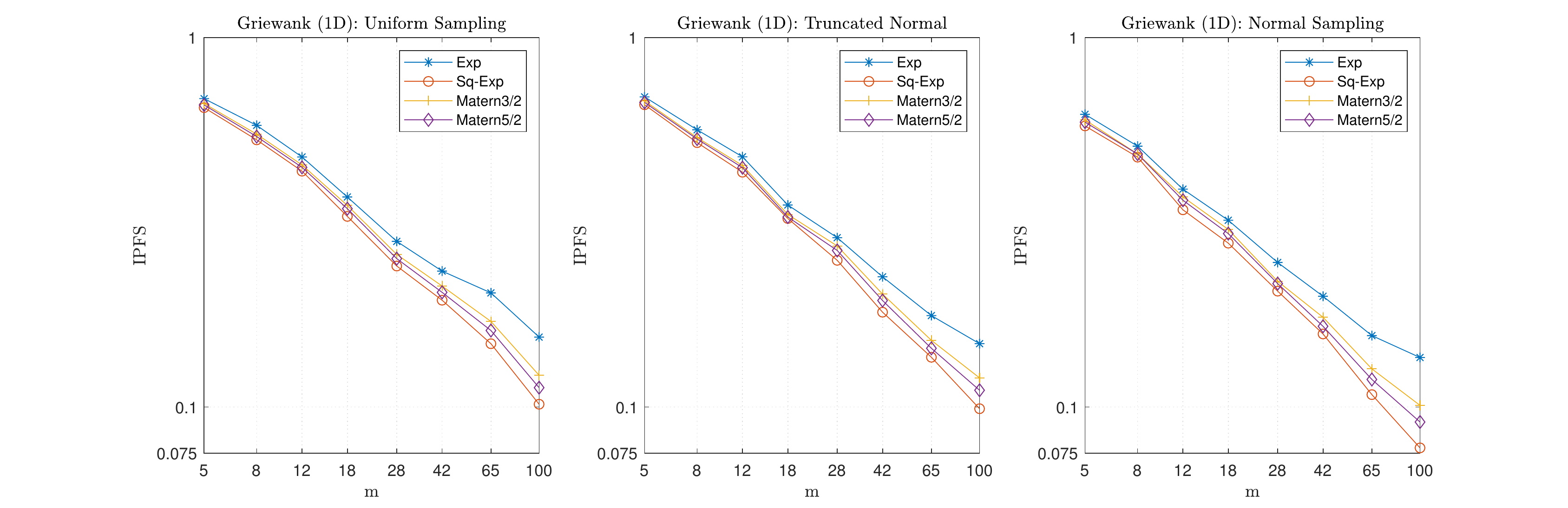}
	\end{center}
	\caption{1-d Griewank's functions: maximal IMSE and IPFS under different covariance kernels and sampling distributions.}
	\label{fig:test2d1}
\end{figure}

\begin{figure}[htbp]
	\begin{center}
		\includegraphics[width=0.92\textwidth]{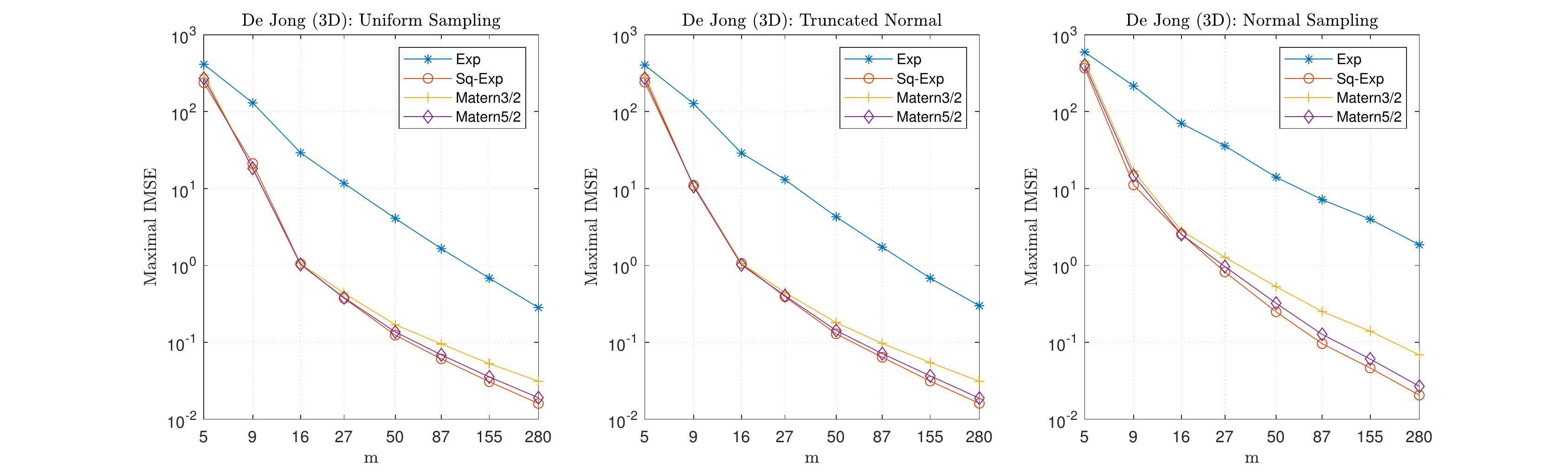}
		\includegraphics[width=0.92\textwidth]{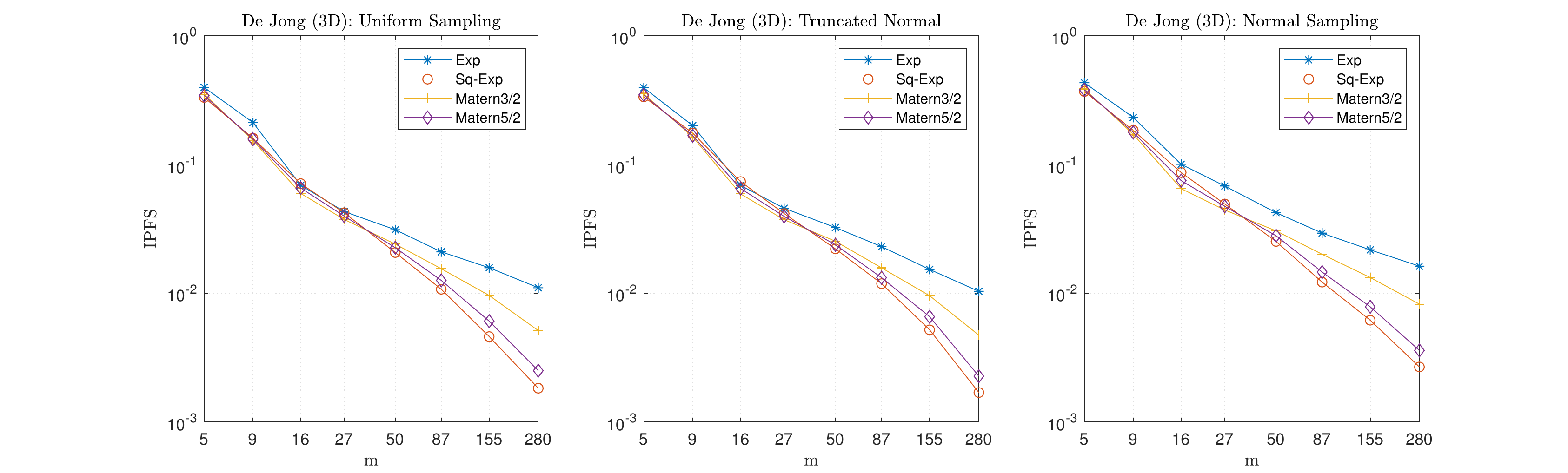}
	\end{center}
	\caption{3-d De Jong's functions: maximal IMSE and IPFS under different covariance kernels and sampling distributions.}
	\label{fig:test1d3}
\end{figure}

\begin{figure}[htbp]
	\begin{center}
		\includegraphics[width=0.92\textwidth]{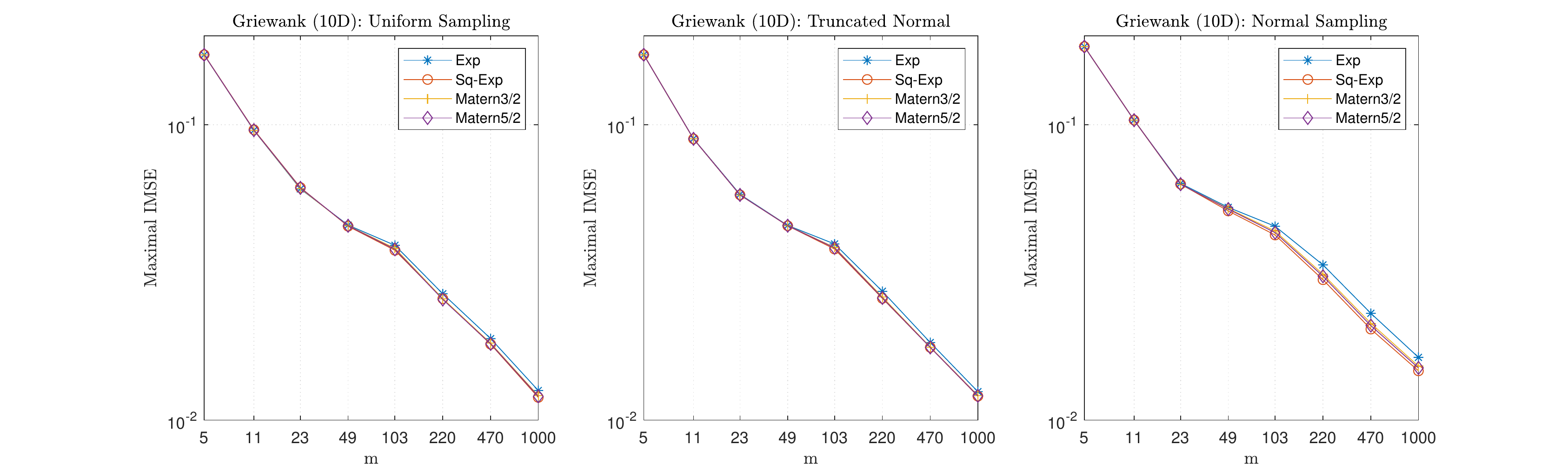}
		\includegraphics[width=0.92\textwidth]{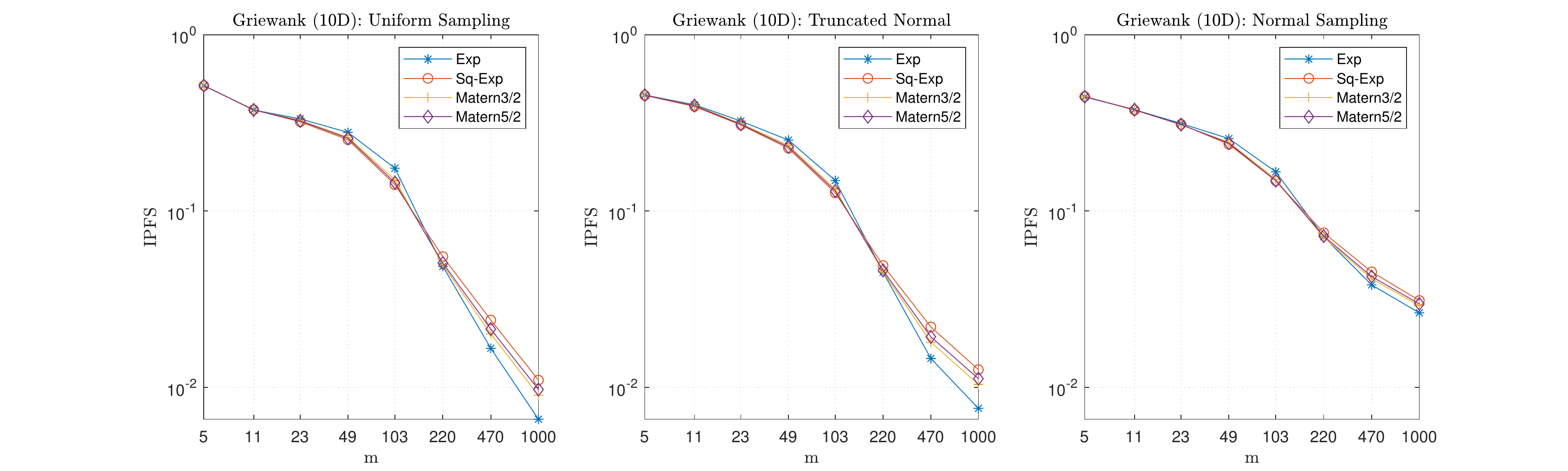}
	\end{center}
	\caption{10-d Griewank's functions: maximal IMSE and IPFS under different covariance kernels and sampling distributions.}
	\label{fig:test2d10}
\end{figure}

We let the number of covariate points $m$ increase geometrically from $m=5$ to $m=100$ in the set $\{5,8,12,18,28,42,65,100\}$, roughly with the common ratio of $1.53$ when $d=1$. When $d=3$, $m$ increases from $m=5$ to $m=280$ in the set $\{5,9,16,27,50,87,155,280\}$, roughly with the common ratio of $1.77$; when $d=10$, $m$ increases from $m=5$ to $m=1000$ in the set $\{ 5,11,23,49,103,220,470,1000 \}$, roughly with the common ratio of $2.13$. We fix the number of replications at each $\x$ for all designs at $n=10$. For the indifference-zone parameter $\delta_0$, we set $\delta_0=0.05$ for the one dimensional De Jong's functions, $\delta_0=0.1$ for the one dimensional Griewank's functions and three dimensional De Jong's functions, and $\delta_0=0.2$ for the ten dimensional Griewank's functions. The maximal IMSE and IPFS in all cases are estimated by the average of 100 macro Monte Carlo replications. The convergence rates of the two measures under different sampling distributions, test functions and covariance kernels are illustrated in Figures \ref{fig:test1d1}-\ref{fig:test2d10}. In the legends, \texttt{SqExp} means the squared exponential kernel, \texttt{Matern 5/2} means the Mat\'ern kernel with $\nu=5/2$, \texttt{Matern 3/2} means the Mat\'ern kernel with $\nu=3/2$, and \texttt{Exp} means the exponential kernel.


In terms of convergence patterns, the maximal IMSE decreases as $m$ increases in all cases, and the decreasing trends are very close to linear when $m$ exceeds 28 with $d=1,3$ and 103 with $d=10$. Since the maximal IMSE and $m$ are plotted on logarithmic scales, it implies that when $m$ is large enough, the maximal IMSE decreases polynomially with $m$. This observation agrees with our rate results in Theorem \ref{jointmsenew1}. The IPFS also decreases as $m$ increases in all cases, and the convergence rates are no slower than those of the maximal IMSE. In some cases, such as the uniform and truncated normal sampling on the 10-dimensional Griewank's function, the decreasing trends of the logarithmic IPFS are superlinear, suggesting that the IPFS might enjoy convergence rates faster than polynomial. These observations agree with the rate results in Theorem \ref{pfsthm}.

Comparing the performances of the four covariance kernels, we can observe that the exponential kernel performs the worst with the largest maximal IMSE and IPFS in all tested cases, and its disadvantage is more obvious on the De Jong's function. This is mainly because the sample paths from the exponential kernel are rough (continuous but not differentiable) while the De Jong's function is very smooth. This mismatch creates bad fitting and predictions, and thus large values of the two target measures. This disadvantage becomes minor on the Griewank's function because the rough sample paths generated from the exponential kernel become appropriate for modeling the oscillations in the Griewank's function. Among the other three kernels, the Mat\'ern kernel with $\nu=5/2$ and the squared exponential kernel often have better performance because their sample paths are smoother.

Among the three sampling distributions, the uniform and truncated sampling have very similar performance. These two distributions are defined on the same supports, i.e., $\Xcal = [1,10]^d$ when $d=1$ and $\Xcal = [1,4]^d$ when $d=3,10$. The truncated normal is set with relatively large variances ($7^2$ when $d=1$ and $3^2$ when $d=3,10$), which results in sufficiently spread out covariate points and hence similar performance to the uniform sampling. The performance of the normal sampling is a little different. This is because the normal sampling is defined on an infinite support, so the space that the MSE and PFS are integrated over is different. However, we can see that the normal sampling is effective in reducing the maximal IMSE and IPFS. The values of the two measures under normal sampling are basically on the same order as those under the uniform and truncated normal sampling.

\subsection{M/M/1 Queue}
\label{subsection:mm1}

The M/M/1 queue is analytical, and thus provides convenience for estimating PFS. In this test, our example is taken from \cite{zhou2015}. Customers arrive at a system according to a Poisson process with rate $x$, and the service time of the server follows an exponential distribution with mean $1/\lambda$. We consider two types of cost, the service cost $c_u \lambda$ with $c_u$ being the per unit cost of the service rate, and the waiting cost, determined by the customers' mean waiting time $\E[\Wcal(\lambda)]$ in the system. In addition, there is an upper bound $\Ucal$ on the total cost. When the system is unstable (i.e., $x/\lambda\geq 1$), it will incur the cost $\Ucal$. Therefore, the total cost $TC$ of this system is
\begin{equation*}
	TC(x,\lambda)=
	\begin{cases}
		\min\{\E[\Wcal(\lambda)]+c_u \lambda,\Ucal\}, \ &\text{ if }x/\lambda<1;\\
		\Ucal, &\text{ otherwise.}
	\end{cases}
\end{equation*}
Note that for the M/M/1 queue, the mean waiting time $\E[\Wcal(\lambda)]$ has an analytical form $1/(\lambda-x)$, and the solution that minimizes the total cost is obtained at $\lambda^*=x+1/\sqrt{c_u}$.

\begin{figure}[htb]
	\begin{center}
		\includegraphics[width=0.9\textwidth]{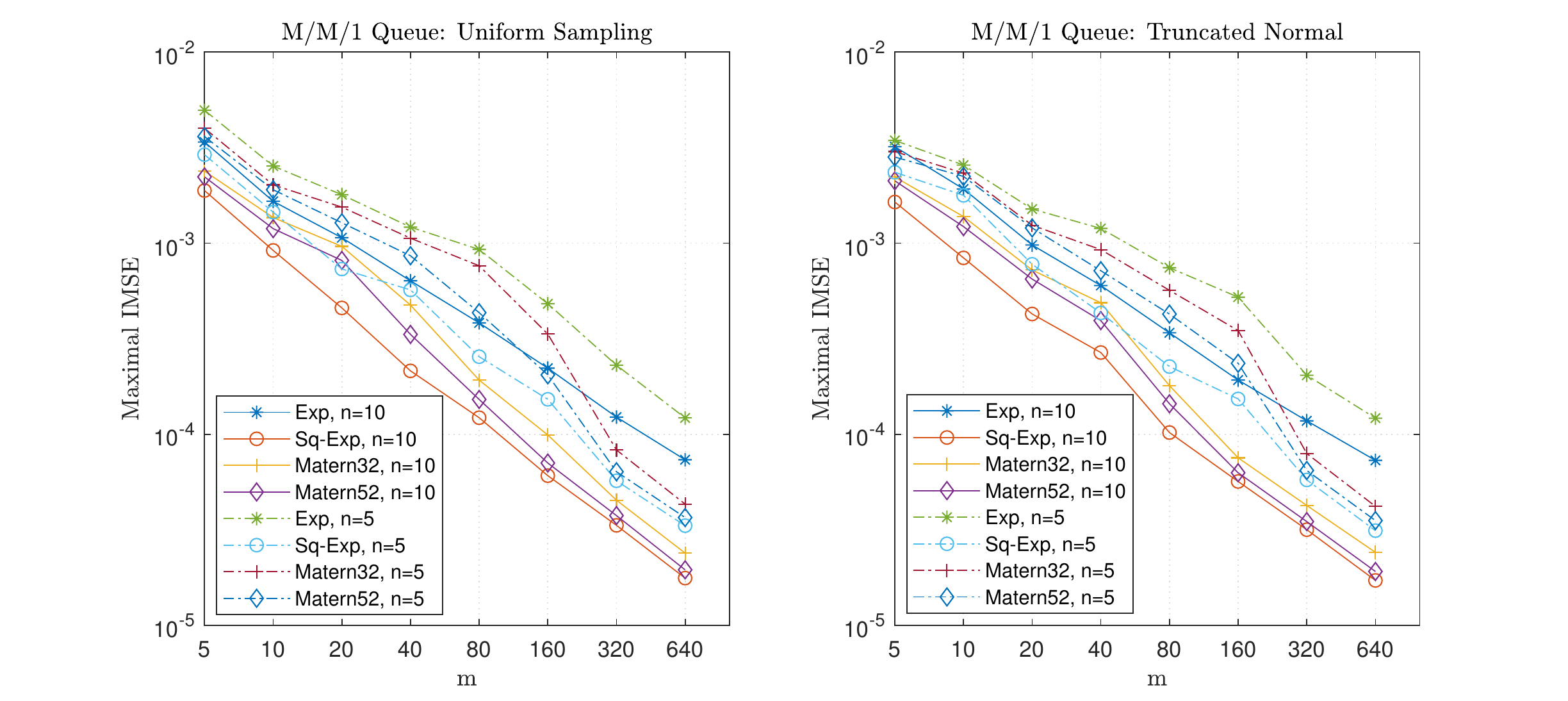}
	\end{center}
	\caption{Maximal IMSE under different covariance kernels, sampling distributions and values of $n$.}
	\label{fig:mm1:mseheter}
\end{figure}

\begin{figure}[htb]
	\begin{center}
		\includegraphics[width=0.9\textwidth]{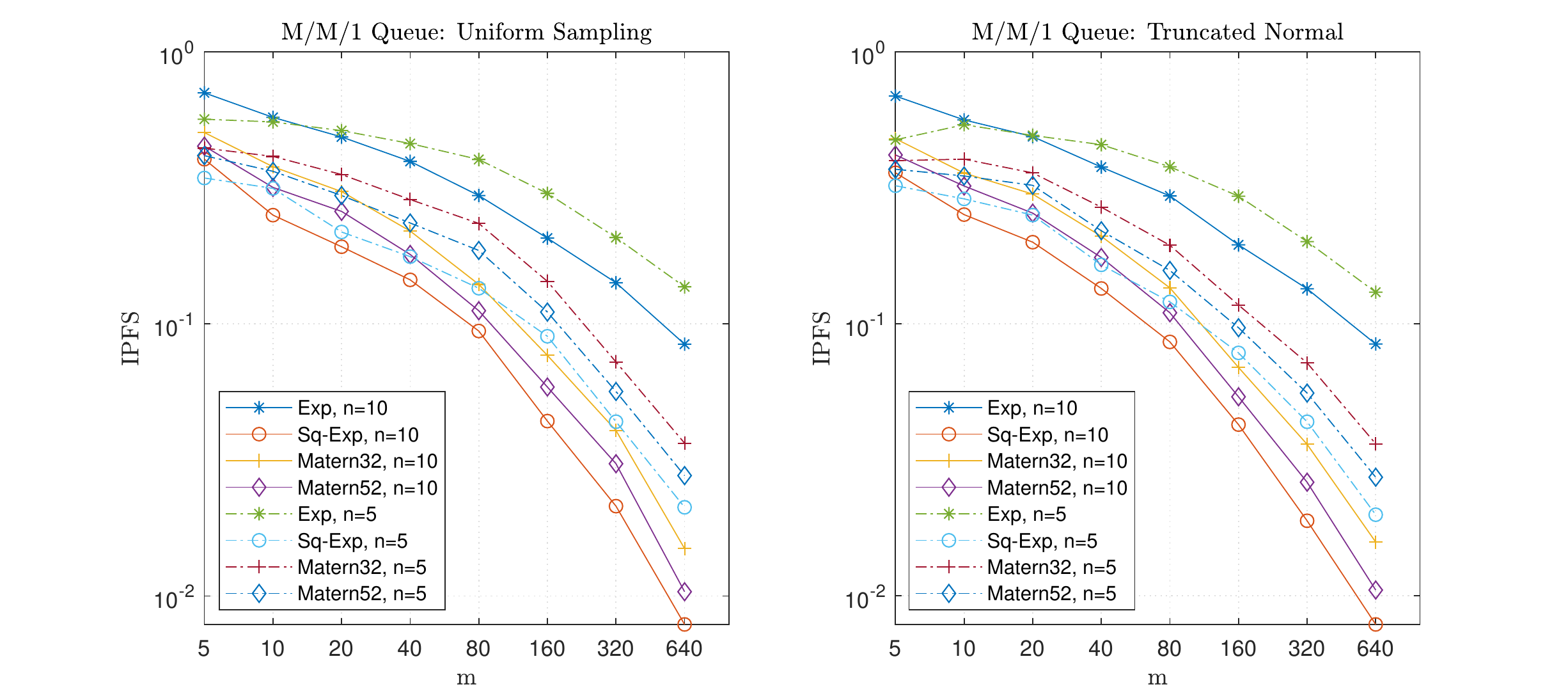}
	\end{center}
	\caption{IPFS under different covariance kernels, sampling distributions and values of $n$.}
	\label{fig:mm1:pfsheter}
\end{figure}

To fit into the framework of simulation with covariates, we consider 10 discrete designs with the $i$-th design $\lambda_i=6+0.3i$, $i=1,2,...,10$, and let $c_u=0.1$ and $\Ucal=2.5$. The covariate $x$ is restricted in an open interval $\Xcal=(0.5,4.5)$. We consider two sampling distributions $\PP_{\X}$ for $\X^m$: uniform on $\Xcal$ and truncated normal on $\Xcal$ with mean 2.5 and variance $3^2$. We let $m$ take values in $\{5,10,20,40,80,160,320,640\}$ and $n$ take values in $\{5,10\}$. The maximal IMSE and IPFS are estimated by the average of 100 macro Monte Carlo replications. The results for the maximal IMSE and the IPFS across the 10 designs are summarized in Figures \ref{fig:mm1:mseheter} and \ref{fig:mm1:pfsheter}.

Figure \ref{fig:mm1:mseheter} shows that on the logarithmic scale, the maximal IMSE across the 10 designs decreases almost linearly as the sample size $\log m$ increases, for all the four kernels and numbers of simulation replications tested. This observation agrees with our theory (Theorem \ref{jointmsenew1}) that the convergence rates of the maximal IMSE are in the polynomial orders of $m$ for the three types of covariance kernels, including all the four kernels we have implemented here. We note that this linear trend can be utilized to help an analyst make the design decision for achieving a target precision of the maximal IMSE. More details are available in Section 5.3 of the Online Supplement.

In Figure \ref{fig:mm1:mseheter}, increasing $n$ from $5$ to $10$ does not significantly reduce the maximal IMSE for all kernels. Among the four kernels, the exponential kernel gives larger maximal IMSE than the other three, again due to the mismatch between its rough sample paths and the smooth target function, since $TC(x,\lambda)$ is always a smooth function in $x$ (infinitely differentiable) for all values of $\lambda_i$. Different sampling distributions on the covariate space do not seem to have a significant impact on the convergence pattern and rate.

Figure \ref{fig:mm1:pfsheter} shows the convergence of IPFS for $\delta_0=0.01$. It can be observed that the relative performance of the IPFS under different kernels, numbers of simulation replications and sampling distributions basically remains the same as that of the maximal IMSE, but the convergence rates of the IPFS are faster, demonstrating a superlinear pattern on the logarithmic scale.

\begin{remark}
	In this research, we have employed the SK models for system performance predictions. It is well-known that the computational complexity of SK (or Gaussian process models) is $O(m^3)$, where $m$ is the number of covariate points. Although with a fixed sampling distribution for the covariate points, we can collect all the covariate points in advance and build the SK models just once, this complexity only makes the computational time practically acceptable when $m$ is no more than a few thousand, or tens of thousand when the offline simulation period is long. When $m$ becomes even larger than that, certain techniques in scalable Gaussian processes \citep{luo2013,hensman2013gaussian,wilson2015kernel} might be considered for improving the computational efficiency.
\end{remark}

\begin{remark}\label{remark:stat}
	In this research, we have adopted a fixed (static) distribution for sampling the covariate space. In the meantime, there has been an increasing interest recently in the development of adaptive design-of-experiment methods (\citealt{Garud2017}). As an initial investigation for the application potential of adaptive methods for the SK construction in simulation with covariates, we numerically compared our static sampling with an intuitive adaptive design procedure (Adaptive MSE Procedure). The results are provided in Section 5.2 of the Online Supplement. We observed that the static sampling considered in this research has similar empirical performance to the Adaptive MSE Procedure in general, and tends to be superior when (i) the dimension of the covariate space is high; (ii) the covariate distribution deviates from uniform; and (iii) the target function has strong oscillation.
\end{remark}

\section{Conclusions and Discussion}
\label{sec:7}

Simulation with covariates is a recently proposed framework for conducting simulation experiments \citep{hong2019,shen2019}. It is comprised of the offline simulation and online prediction periods, and is able to substantially reduce the decision time. We provide theoretical analysis for the predictive performance of the stochastic kriging model under this framework. We focus on two critical measures for the prediction errors, the maximal IMSE and IPFS, and study their convergence rates, in order to understand the relationship between the offline simulation efforts and the online prediction accuracy.

For the maximal IMSE, we show that the convergence rates are $1/m$, $(\log m)^{\frac{d}{\kappa_*}}/m$ and $m^{-\frac{2\nu_*}{2\nu_*+d}}$ for the finite-rank kernels, exponentially decaying kernels and polynomially decaying kernels respectively, where $m$ is the number of sampled covariate points, $\kappa_*$ and $\nu_*$ are some kernel parameters, and $d$ is the dimension of covariates. For the IPFS, we show that the convergence rates are at least as fast as the maximal IMSE, and can be enhanced to exponential rates under some conditions.

Since the rates derived for the maximal IMSE and IPFS are simple and concrete, and are the first to characterize the convergence rates of the prediction errors in simulation with covariates to the best of our knowledge, they serve as a good benchmark against which improvement in rates might be theoretically or numerically measured from future prediction methods built on possibly different assumptions, prediction models, covariance kernels and covariate point collection strategies. In addition, the theoretical analysis in this research has the chance to be extended to facilitate new developments in simulation with covariates, e.g., when adaptive design procedures are used to explore the covariate space.

\appendix

\section{Notation}
\label{sec:appendixa}

We summarize the key notation used in this paper in the following table.
\vspace{-1cm}

\begin{center}
	\begin{longtable}{p{3.6cm}p{2.3cm}p{10.5cm}}
		\renewcommand{\arraystretch}{1.1}\\
		\label{tab:1}\\
		\caption{Table of notation.}\\
		\hline
		\hline
		Symbol & $i$ folded\footnote{For simplicity of notation, in this research, we have folded the design index $i$ in circumstances with no ambiguity. This column shows the symbol if its subscript $i$ has been folded in the paper.} & Meaning \\
		\hline
		$\|\cdot\|$ && Euclidean norm of a vector \\
		$\vertiii{\cdot}$  && operator norm of a matrix, defined as $\sup_{\|\mathbf{v}\|=1}\|\cdot \mathbf{v}\|$\\
		$\|\cdot\|_2^2$ && $L_2$ norm of a function\\
		$a_l\lesssim b_l$ && mean that $\limsup_{l\to\infty} a_l/b_l < \infty$ \\
		$a_l\asymp b_l$ && mean that $a_l\lesssim b_l$ and $b_l\lesssim a_l$ \\
		$k$ && number of system designs  \\
		$d$ && dimension of the covariate space \\
		$m$ && number of covariate points\\
		$n$ && number of replications for each pair of covariate point and design \\
		$q$ && dimension of the regressors $\bbf$ and the regression coefficient $\bbeta$ \\
		$\Xcal$ && support of covariate points \\
		$y_i(\cdot)$ & $y(\cdot)$ & mean of design $i$ \\
		$Y_{il}(\cdot)$ && the $l$-th simulation sample from design $i$ \\
		$\epsilon_{il} (\cdot)$ && simulation noise of the $l$-th sample of design $i$ \\
		$\overline \epsilon_i(\cdot)$  && averaged simulation errors, defined as $n^{-1}\sum_{l=1}^{n} \epsilon_{il}(\cdot)$ \\
		$\sigma_i^2(\cdot)$ && variance of $\epsilon_{il} (\cdot)$ \\
		$\overline Y_i(\cdot)$ &$\overline Y(\cdot)$& sample mean of design $i$ \\
		$\overline \Y_i$ & $\overline \Y$& vector of samples means at $m$ covariate points\\
		$\x,\X$ && vector of covariates with support $\Xcal\subseteq \mathbb{R}^d$ \\
		$\x_0,\X_0$ && test covariate point for the SK model \\
		$\x_j,\X_j$ && the $j$-th covariate point \\
		$\x^m,\X^m$ && vector of $m$ covariate points \\
		$\bbf_i(\cdot)$ & $\bbf(\cdot)$ &vector of known basis functions \\
		$\bbeta_i$ & $\bbeta$ &vector of unknown parameters for $\bbf_i(\cdot)$\\
		$M_i(\cdot)$ & $M(\cdot)$ &realization of a mean zero stationary Gaussian process for design $i$\\
		$\bSigma_{M,i}(\x,\x')$ & $\bSigma_{M}(\x,\x')$ &covariance function, defined as $\cov\left[M_i(\x),M_i(\x')\right]$\\
		$\bSigma_{\epsilon,i}(\x^m)$  & $\bSigma_{\epsilon}(\x^m)$ &covariance matrix of the averaged simulation errors in design $i$\\
		$\widehat{y}_i(\cdot)$& $\widehat{y}(\cdot)$ &MSE-optimal linear predictor of the $i$-th SK model\\
		$\mse_{i,\opt}(\cdot)$& $\mse_{\opt}(\cdot)$ &the MSE of predictor $\widehat{y}_i(\cdot)$\\
		$\lambda_{\max}(\cdot),\lambda_{\min}(\cdot)$ &   & the largest and smallest eigenvalues of a matrix\\
		$A_1\prec A_2,A_2\succ A_1$ &  &mean that $A_2-A_1$ is positive definite\\
		$A_1\preceq A_2,A_2\succeq A_1$ &  &mean that $A_2-A_1$ is positive semi-definite\\
		$\1(\cdot)$ &  & indicator function  \\
		$\PP_{\X},\E_{\X}$ &  &a probability distribution/expectation over $\Xcal$\\
		$L_2(\PP_{\X})$  &  &$L_2$ space under $\PP_{\X}$ \\
		$\langle \cdot, \cdot\rangle_{L_2(\PP_{\X})}$ &  &inner product in $L_2(\PP_{\X})$, defined as ${\E}_{\X} (\cdot)$\\
		$[T_{\bSigma_M}f](\x)$ &  &a linear operator of $\x\in \Xcal$, defined as $\int_{\Xcal} \bSigma_M(\x,\x')f(\x')\ud \PP_{\X}(\x')$\\
		$\big\{\phi_{i,l}(\x):l=1,\ldots\big\}$ &$\phi_l(\x)$& the orthonormal basis for $\bSigma_{M,i}$ (from Mercer's theorem)\\
		$\tr(\cdot)$ & &trace of a kernel (matrix) \\
		$\left\{\mu_{i,l}: l=1,\ldots\right\}$ & $\mu_{l}$& eigenvalues of $\bSigma_{M,i}$ \\
		$\HH_i$ &$\HH$ & reproducing kernel Hilbert space attached to $\bSigma_{M,i}$\\
		$\langle \cdot, \cdot \rangle_{\HH}$ &  &$\HH$-inner product\\
		$\rho_*,r_*$ &   & parameters made for $\phi_{i,l}(\x)$ in Assumption A.3\\
		$\kappa_i,\nu_i,\tau_i,\varphi_i$ &$\kappa,\nu,\tau,\varphi$  &kernel parameters of the $i$-th SK model\\
		$\kappa_*$, $\nu_*$ &  & parameters in rate functions, defined as $\kappa_*=\min_{i\in\{1,2,...,k\}}\kappa_i$ and $\nu_*=\min_{i\in\{1,2,...,k\}\nu_i}$\\
		$~{\lesssim}_{\PP_{\X^m}}$ &  &mean bounding in $\PP_{\X^m}-$ probability \\
		$\delta_0$ & &indifference-zone parameter \\
		$\underline \sigma_0^2,\overline \sigma^2_0$ &  &lower and upper bounds for $\sigma_i^2(\x)$ for all $i$ and all $\x \in \Xcal$\\
		$i^\circ(\cdot),\widehat i^\circ (\cdot)$ &  &the real and estimated optimal designs\\
		$R^F(m,n)$ &  &rate function of the maximal IMSE for finite-rank kernels\\
		$R_i^{E}(m,n)$ & $R^{E}(m,n)$ &rate function of the maximal IMSE for exponentially decaying kernels and design $i$\\
		$R_i^{P}(m,n)$ & $R^{P}(m,n)$ &rate function of the maximal IMSE for polynomially decaying kernels and design $i$\\
		$R(m,n)$ &  &rate function of IMSE \\
		$\mu_i^{\Xcal}$&& expected mean performance of design $i$ over the covariate space\\
		$q_{i,\alpha}^{\Xcal}$&& $\alpha$-quantile of the performance of design $i$ over the covariate space\\
		$w_i^{\Xcal}$&& proportion of design $i$ being the best over the covariate space\\
		\hline
		\hline
	\end{longtable}
\end{center}

\section{Technical Proofs and Additional Theoretical Results}

In this section, we first prove Theorems 1 to 5 in the main text. Next, we present a new theorem (Theorem 6) about the restrictiveness of Assumptions A.6 and A.7.

We reinstate some useful notation and relations. For any finite dimensional vector $\mathbf{v}$, we let $\|\mathbf{v}\|$ be its Euclidean norm. For any generic matrix $A$, we use $A_{ab}$ to denote its $(a,b)$-entry, $cA$ to denote the matrix whose $(a,b)$-entry is $cA_{ab}$ for any constant $c\in \RR$, and $\vertiii{A}=\sup_{\|\mathbf{v}\|=1}\|A\mathbf{v}\|$ to denote its matrix operator norm. For any positive definite matrix $A$, let $\lambda_{\max}(A)$ and $\lambda_{\min}(A)$ be its largest and smallest eigenvalues. For two positive definite matrices $A_1,A_2$, $A_1\prec A_2$ and $A_2\succ A_1$ mean that $A_2-A_1$ is positive definite; $A_1\preceq A_2$ and $A_2\succeq A_1$ mean that $A_2-A_1$ is positive semi-definite. For two sequences of positive numbers $\{a_l\}_{l\geq 1}$ and $\{b_l\}_{l\geq 1}$, $a_l\lesssim b_l$ means that $\limsup_{l\to\infty} a_l/b_l < \infty$, and $a_l\asymp b_l$ means that both $a_l\lesssim b_l$ and $b_l\lesssim a_l$ hold true. Let $\1(\cdot)$ be the indicator function and $\I_k$ be the $k\times k$ identity matrix.

Any function $f\in  L_2(\PP_{\X})$ has the series expansion $f(\x) = \sum_{l=1}^{\infty} \theta_l \phi_l(\x)$, where $\theta_l = \langle f,\phi_l\rangle_{L_2(\PP_{\X})}$. The $L_2$ norm of $f$ is given by $\|f\|_2^2 = \sum_{l=1}^{\infty} \theta_l^2$. The reproducing kernel Hilbert space (RKHS) $\HH$ attached to $\bSigma_{M}$ is the space of all functions $f \in L_2(\PP_{\X})$ such that its $\HH$-norm $\|f\|_{\HH}^2 = \sum_{l=1}^{\infty} \theta_l^2 / \mu_l<\infty$. For any two generic functions $\uh_1,\uh_2\in \HH$, let their $L_2(\PP_{\X})$ expansions be $\uh_s(\x)=\sum_{l=1}^{\infty} h_{sl} \phi_l(\x)$ for $s=1,2$. Their $\HH$-inner product is given by $\langle \uh_1, \uh_2 \rangle_{\HH} = \sum_{l=1}^{\infty} h_{1l}h_{2l}/\mu_l$. For any $\uh \in \HH$, the reproducing property of $\HH$ says that for any $\x \in \Xcal$, $\langle \bSigma_M(\x,\cdot), \uh(\cdot)\rangle_{\HH} = \uh(\x)$.

\vspace{8mm}

\noindent \textbf{Proof of Theorem 1}:

According to Mercer's theorem (e.g. Theorem 4.2 of \citealt{RasWil06}), the series expansion of the kernel function $\bSigma_{M}(\x, \x') = \sum_{l=1}^{\infty} \mu_l \phi_l(\x) \phi_l(\x')$ holds almost surely for any $\x,\x'\in \Xcal$, and hence
\begin{align}\label{expan1}
	&\bSigma_{M}(\x_0, \x_0)=\sum_{a=1}^{\infty} \mu_a \phi_a^2(\x_0),\quad
	\bSigma_{M}(\x_j, \x_0)=\sum_{a=1}^{\infty} \mu_a \phi_a(\x_j)\phi_a(\x_0), \text{ for } j=1,\ldots,m, \nonumber \\
	&\bSigma_{M}(\x^m,\x_0)= \left[\bSigma_{M}(\x_1, \x_0),\ldots,\bSigma_{M}(\x_m, \x_0)\right]^\top.
\end{align}
Under the orthonormal property, if $\X\sim \PP_{\X}$, then $\E_{\X}[\phi_a^2(\X)]=1$ and $\E_{\X}[\phi_a(\X)\phi_b(\X)]=0$ for $a\neq b$. Therefore,
\begin{align}\label{msebound2}
	& \quad {\E}_{\X^m} {\E}_{\X_0} \left[\mse_{\opt}^{(M)}(\X_0)\right] \nonumber \\
	&= {\E}_{\X_0}\left[\bSigma_{M}(\X_0,\X_0)\right]  - {\E}_{\X^m} {\E}_{\X_0} \left\{\bSigma_{M}^\top(\X^m,\X_0) \left[\bSigma_{M}(\X^m,\X^m)+  \bSigma_{\epsilon}(\X^m) \right]^{-1} \bSigma_{M}(\X^m,\X_0)\right\} \nonumber \\
	& \stackrel{(i)}{=} \sum_{a=1}^{\infty} \mu_a {\E}_{\X_0}\left[\phi_a^2(\X_0) \right] - \nonumber \\
	& ~~ {\E}_{\X^m} {\E}_{\X_0} \sum_{j=1}^{m}\sum_{j'=1}^{m}\sum_{a=1}^{\infty}\sum_{b=1}^{\infty} \mu_a \mu_b \left\{\left[\bSigma_{M}(\X^m,\X^m)+  \bSigma_{\epsilon}(\X^m) \right]^{-1}\right\}_{jj'} \phi_a(\X_j)\phi_a(\X_0)\phi_b(\X_j')\phi_b(\X_0) \nonumber \\
	& \stackrel{(ii)}{=} \sum_{a=1}^{\infty} \mu_a {\E}_{\X_0}\left[\phi_a^2(\X_0) \right] -  {\E}_{\X^m}\sum_{j=1}^{m}\sum_{j'=1}^{m}\sum_{a=1}^{\infty}\sum_{b=1}^{\infty} \mu_a \mu_b \left\{\left[\bSigma_{M}(\X^m,\X^m)+  \bSigma_{\epsilon}(\X^m) \right]^{-1}\right\}_{jj'}  \nonumber \\
	& \quad \cdot \phi_a(\X_j)\phi_b(\X_{j'}) {\E}_{\X_0}\left[\phi_a(\X_0)\phi_b(\X_0)\right] \nonumber \\
	&= \sum_{a=1}^{\infty} \mu_a - {\E}_{\X^m} \sum_{j=1}^{m}\sum_{j'=1}^{m}\sum_{a=1}^{\infty} \mu_a^2 \left\{\left[\bSigma_{M}(\X^m,\X^m)+  \bSigma_{\epsilon}(\X^m) \right]^{-1}\right\}_{jj'}  \phi_a(\X_j)\phi_b(\X_{j'}) \nonumber \\
	&= \sum_{a=1}^{\zeta} \mu_a - {\E}_{\X^m}\sum_{a=1}^{\zeta} \sum_{j=1}^{m}\sum_{j'=1}^{m}\mu_a^2 \left\{\left[\bSigma_{M}(\X^m,\X^m)+  \bSigma_{\epsilon}(\X^m) \right]^{-1}\right\}_{jj'}  \phi_a(\X_j)\phi_a(\X_{j'}) \nonumber \\
	& ~~ + \tr\left(\bSigma_{M}^{(\zeta)}\right) - {\E}_{\X^m} \sum_{a=\zeta+1}^{\infty} \sum_{j=1}^{m}\sum_{j'=1}^{m}\mu_a^2 \left\{\left[\bSigma_{M}(\X^m,\X^m)+  \bSigma_{\epsilon}(\X^m) \right]^{-1}\right\}_{jj'}  \phi_a(\X_j)\phi_a(\X_{j'}) \nonumber \\
	&\stackrel{(iii)}{\leq} \sum_{a=1}^{\zeta} \left\{\mu_a - {\E}_{\X^m}\sum_{j=1}^{m}\sum_{j'=1}^{m}\mu_a^2 \left\{\left[\bSigma_{M}(\X^m,\X^m)+  \bSigma_{\epsilon}(\X^m) \right]^{-1}\right\}_{jj'}  \phi_a(\X_j)\phi_a(\X_{j'})\right\} + \tr\left(\bSigma_{M}^{(\zeta)}\right).
\end{align}
In the derivation above, we exchange the expectation and the summation in several steps.
\begin{itemize}[leftmargin=5mm]
	\item For Step (i), because $\left\{\sum_{a=1}^{N} \mu_a \phi_a^2(\X_0), ~~ N=1,2,\ldots\right\}$ is a non-decreasing sequence of functions, by the monotone convergence theorem, we have ${\E}_{\X_0}\left[\bSigma_{M}(\X_0,\X_0)\right] = {\E}_{\X_0}\left[\sum_{a=1}^{\infty} \mu_a \phi_a^2(\X_0) \right] = \sum_{a=1}^{\infty} \mu_a {\E}_{\X_0}\left[\phi_a^2(\X_0) \right]$.
	\item For Step (ii), for any $\x^m$, every $j,j'=1,\ldots,m$, and $N_1,N_2=1,2,\ldots$,
	\begin{align}\label{eq:step2.1}
		& \left| \sum_{a=1}^{N_1}\sum_{b=1}^{N_2} \mu_a \mu_b \left\{\left[\bSigma_{M}(\x^m,\x^m)+  \bSigma_{\epsilon}(\X^m) \right]^{-1}\right\}_{jj'}  \phi_a(\x_j)\phi_b(\x_{j'}) \phi_a(\x_0)\phi_b(\x_0) \right| \nonumber \\
		\leq{}&  \sum_{a=1}^{N_1}\sum_{b=1}^{N_2} \mu_a \mu_b \left\{\left[\bSigma_{M}(\x^m,\x^m)+  \bSigma_{\epsilon}(\X^m) \right]^{-1}\right\}_{jj'}  \phi_a(\x_j)\phi_b(\x_{j'}) \phi_a(\x_0)\phi_b(\x_0) \nonumber \\
		&\quad \cdot \sgn\left(\left\{\left[\bSigma_{M}(\x^m,\x^m)+  \bSigma_{\epsilon}(\X^m) \right]^{-1}\right\}_{jj'}  \phi_a(\x_j)\phi_b(\x_{j'}) \phi_a(\x_0)\phi_b(\x_0)\right),
	\end{align}
	where $\sgn(x)=1$ for $x>0$, $\sgn(x)=-1$ for $x<0$, and $\sgn(x)=0$ if $x=0$. By Assumption A.3 and H\"older's inequality,
	\begin{align}\label{eq:step2.2}
		&{\E}_{\X_0} \left\{  \phi_a(\X_0) \phi_b(\X_0) \sgn \left( \left\{\left[\bSigma_{M}(\x^m,\x^m)+  \bSigma_{\epsilon}(\X^m) \right]^{-1}\right\}_{jj'}    \phi_a(\x_j)  \phi_a(\X_0) \phi_b(\x_{j'}) \phi_b(\X_0) \right) \right\} \nonumber \\
		&\leq {\E}_{\X_0} \{ \left|\phi_a(\X_0)\phi_b(\X_0)\right|  \} \leq \left( {\E}_{\X_0} \left\{ \phi_a^2(\X_0) \right\} \right)^{1/2} \left( {\E}_{\X_0} \left\{ \phi_b^2(\X_0) \right\} \right)^{1/2} \nonumber\\
		&\leq \left( {\E}_{\X_0} \left\{ \phi_a^{2r_*}(\X_0) \right\} \right)^{1/(2r_*)} \left( {\E}_{\X_0} \left\{ \phi_b^{2r_*}(\X_0) \right\} \right)^{1/(2r_*)} \leq \rho_*^{2}.
	\end{align}
	We apply the dominated convergence theorem using \eqref{eq:step2.1} and \eqref{eq:step2.2} to obtain that
	\begin{align*}
		& {\E}_{\X_0}\left\{\sum_{j=1}^{m}\sum_{j'=1}^{m} \sum_{a=1}^{\infty}\sum_{b=1}^{\infty} \mu_a \mu_b \left\{\left[\bSigma_{M}(\X^m,\X^m)+  \bSigma_{\epsilon}(\X^m) \right]^{-1}\right\}_{jj'}  \phi_a(\x_j)\phi_b(\x_{j'}) \phi_a(\x_0)\phi_b(\x_0)\right\}  \\
		={}& \sum_{j=1}^{m}\sum_{j'=1}^{m} {\E}_{\X_0}\Bigg\{ \lim_{N_1,N_2\to\infty}\sum_{a=1}^{N_1}\sum_{b=1}^{N_2} \mu_a \mu_b \left\{\left[\bSigma_{M}(\X^m,\X^m)+  \bSigma_{\epsilon}(\X^m) \right]^{-1}\right\}_{jj'} \\
		&\quad \cdot \phi_a(\x_j)\phi_b(\x_{j'}) \phi_a(\x_0)\phi_b(\x_0) \Bigg\}  \\
		={}& \sum_{j=1}^{m}\sum_{j'=1}^{m} \lim_{N_1,N_2\to\infty}\sum_{a=1}^{N_1}\sum_{b=1}^{N_2} \mu_a \mu_b \left\{\left[\bSigma_{M}(\X^m,\X^m)+  \bSigma_{\epsilon}(\X^m) \right]^{-1}\right\}_{jj'}  \\
		&\quad\cdot \phi_a(\x_j)\phi_b(\x_{j'}) {\E}_{\X_0}\left[ \phi_a(\x_0)\phi_b(\x_0) \right] \\
		={}& \sum_{j=1}^{m}\sum_{j'=1}^{m} \sum_{a=1}^{\infty}\sum_{b=1}^{\infty} \mu_a \mu_b \left\{\left[\bSigma_{M}(\X^m,\X^m)+  \bSigma_{\epsilon}(\X^m) \right]^{-1}\right\}_{jj'}  \phi_a(\x_j)\phi_b(\x_{j'}) {\E}_{\X_0}\left[ \phi_a(\x_0)\phi_b(\x_0) \right] ,
	\end{align*}
	which gives the right-hand side of Step (ii).
	\item For Step (iii), we make the left-hand side larger by dropping the negative quadratic term in the summation $\sum_{a=\zeta+1}^{\infty} \sum_{j=1}^{\infty}\sum_{j'=1}^{\infty}$.
\end{itemize}
\vspace{3mm}

To proceed from \eqref{msebound2}, we define some useful quantities:
\begin{align*}
	& \M = \text{diag}\left(\mu_1,\ldots,\mu_{\zeta}\right), \quad \M^{\rem}=\text{diag}\left(\mu_{\zeta+1},\mu_{\zeta+2},\ldots\right), \\
	& \bphi_a = \left[\phi_a(\X_1),\ldots,\phi_a(\X_m)\right]^\top, \text{ for } a=1,2,\ldots, \\
	& \bPhi = \left[\bphi_1, \ldots, \bphi_{\zeta}\right], \quad \bPhi^{\rem} = \left[\bphi_{\zeta+1}, \bphi_{\zeta+2}, \ldots \right],\\
	& \B = \M - \M \bPhi^\top \left[\bSigma_{M}(\X^m,\X^m)+  \bSigma_{\epsilon}(\X^m) \right]^{-1} \bPhi \M,
\end{align*}
such that $\bPhi$ is a $m\times \zeta$ matrix, and $\B$ is a $\zeta\times \zeta$ positive definite matrix. From this definition and \eqref{msebound2}, we have
\begin{align}\label{msebound3}
	& \tr(\B) = \sum_{a=1}^{\zeta} \mu_a - \sum_{a=1}^{\zeta} \sum_{j=1}^{m}\sum_{j'=1}^{m}\mu_a^2 \left\{\left[\bSigma_{M}(\X^m,\X^m)+  \bSigma_{\epsilon}(\X^m) \right]^{-1}\right\}_{jj'}  \phi_a(\X_j)\phi_b(\X_{j'}), \nonumber \\
	& {\E}_{\X_0}{\E}_{\X^m} \left[\mse_{\opt}^{(M)}(\X_0)\right] \leq  {\E}_{\X^m}\tr(\B) + \tr\left(\bSigma_{M}^{(\zeta)}\right).
\end{align}
Let $\bSigma_M^{\rem} =\bSigma_{M}(\X^m,\X^m) - \bPhi \M \bPhi^\top = \bPhi^{\rem} \M^{\rem} \bPhi^{\rem \top}$, which is a $m\times m$ positive semi-definite matrix. Then by the Woodbury formula (Rasmussen and Williams \citeyear{RasWil06}, Appendix A.3), the matrix $\B$ can be written as
\begin{align}\label{Bmat0}
	\B &= \M - \M \bPhi^\top \left[\bSigma_{M}(\X^m,\X^m)+  \bSigma_{\epsilon}(\X^m) \right]^{-1} \bPhi \M \nonumber \\
	&= \left[\M^{-1} + \bPhi^\top \left\{\bSigma_M^{\rem} +  \bSigma_{\epsilon}(\X^m) \right\}^{-1} \bPhi \right]^{-1} .
\end{align}
By Assumption A.1 and the definition of $n$, we have that $\bSigma_{\epsilon}(\x^m)$ is diagonal and $\bSigma_{\epsilon}(\x^m)\preceq \frac{\overline \sigma_0^2}{n}\I_m$ for any value of $\x^m$, where $\I_m$ is the $m\times m$ identity matrix. Therefore, from \eqref{Bmat0}, we can apply the Woodbury formula again to obtain that
\begin{align}\label{Bmat1}
	\B &\preceq \left[\M^{-1} + \bPhi^\top \left\{\bSigma_M^{\rem} + \frac{\overline \sigma_0^2}{n}\I_m  \right\}^{-1} \bPhi \right]^{-1} \nonumber \\
	&= \frac{\overline \sigma_0^2}{mn} \left[ \I_{\zeta} + \frac{\overline \sigma_0^2}{mn}\M^{-1} + \frac{1}{m} \bPhi^\top \left(\frac{n}{\overline \sigma_0^2}\bSigma_M^{\rem} + \I_m \right)^{-1} \bPhi - \I_{\zeta} \right]^{-1} \nonumber \\
	&= \frac{\overline \sigma_0^2}{mn} \Q^{-2} \left\{ \I_{\zeta} + \Q^{-1}\left[ \frac{1}{m} \bPhi^\top \left(\frac{n}{\overline \sigma_0^2}\bSigma_M^{\rem} + \I_m \right)^{-1} \bPhi - \I_{\zeta} \right] \Q^{-1} \right\}^{-1},
\end{align}
where $\Q = \left(\I_{\zeta} + \frac{\overline \sigma_0^2}{mn}\M^{-1}\right)^{1/2}$.

Define the event $\Ecal_2=\left\{\tfrac{n}{\overline \sigma_0^2}\bSigma_M^{\rem} \preceq \delta_2 \I_m \right\}$. Then since $\bSigma_M^{\rem}$ is positive semi-definite, we have the relation that
\begin{align*}
	\left\{\tr \left(\tfrac{n}{\overline \sigma_0^2} \bSigma_M^{\rem}\right)\leq \delta_2 \right\} \subseteq \left\{\lambda_{\max} \left(\tfrac{n}{\overline \sigma_0^2} \bSigma_M^{\rem}\right) \leq \delta_2 \right\} \subseteq \Ecal_2.
\end{align*}
Therefore, by Markov's inequality and the monotone convergence theorem, we have that
\begin{align}\label{traceerr1}
	& \PP_{\X^m} (\Ecal_2^c) \leq {\PP}_{\X^m} \left\{\tr\left(\tfrac{n}{\overline \sigma_0^2} \bSigma_M^{\rem} \right)>\delta_2\right\} \leq \frac{1}{\delta_2} {\E}_{\X^m} \tr\left(\tfrac{n}{\overline \sigma_0^2} \bSigma_M^{\rem} \right)  \nonumber \\
	& = \frac{n}{\overline \sigma_0^2 \delta_2} \sum_{i=1}^m \sum_{a=\zeta+1}^{\infty} \mu_a {\E}_{\X^m} \phi^2_a(\X_i) = \frac{mn}{\overline \sigma_0^2 \delta_2} \tr\left(\bSigma_M^{(\zeta)}\right).
\end{align}
On the other hand, we consider the event defined in Lemma \ref{zhang15d} with $\delta=\delta_1$, i.e.
$$\Ecal_1=\left\{\vertiii{\Q^{-1}\left( \frac{1}{m} \bPhi^\top \bPhi - \I_{\zeta} \right) \Q^{-1}} \leq \delta_1 \right\}.$$
On the event $\Ecal_1\cap \Ecal_2$, we have that
\begin{align}\label{Bmatbound1}
	&\quad \I_{\zeta} + \Q^{-1}  \left\{ \frac{1}{m} \bPhi^\top \left(\tfrac{n}{\overline \sigma_0^2} \bSigma_M^{\rem} + \I_m \right)^{-1} \bPhi - \I_{\zeta} \right\} \Q^{-1} \nonumber \\
	& \stackrel{(i)}{\succeq} \I_{\zeta} + \Q^{-1}  \left\{ \frac{1}{m}\bPhi^\top \left( \delta_2 \I_m + \I_m \right)^{-1} \bPhi^\top - \I_{\zeta} \right\} \Q^{-1} \nonumber \\
	& = \I_{\zeta} - \left(1-\frac{1}{1+\delta_2}\right) \Q^{-2} + \frac{1}{1+\delta_2} \Q^{-1}  \left\{ \frac{1}{m}\bPhi^\top \bPhi - \I_{\zeta} \right\} \Q^{-1} \nonumber \\
	& \stackrel{(ii)}{\succeq} \I_{\zeta} - \left(1-\frac{1}{1+\delta_2}\right) \I_{\zeta} - \frac{1}{1+\delta_2}\cdot \delta_1\I_{\zeta} = \frac{1-\delta_1}{1+\delta_2}\I_{\zeta},
\end{align}
where (i) follows on the event $\Ecal_2$, and (ii) holds on the event $\Ecal_1$ and from the fact $\Q^{-2}\preceq \I_{\zeta}$.

Therefore, by combining \eqref{traceerr1}, \eqref{Bmatbound1}, and the upper bound for $\PP_{\X^m}(\Ecal_1^c)$ given in Lemma \ref{zhang15d} under our assumptions A.1-A.3, we obtain that
\begin{align} \label{Bmatbound2}
	&\quad {\E}_{\X^m} \tr(\B)  \leq {\E}_{\X^m} \left\{\tr(\B) \1(\Ecal_1\cap \Ecal_2)\right\} + {\E}_{\X^m} \left[\tr(\B) \left\{\1(\Ecal_1^c) + \1(\Ecal_2^c)\right\} \right] \nonumber \\
	& \overset{(i)}{\leq} \frac{1+\delta_2}{1-\delta_1} \frac{\overline \sigma_0^2}{mn} \tr\left(\Q^{-2}\right) + \tr(\bSigma_M) \left\{\PP(\Ecal_1^c)+\PP(\Ecal_2^c)\right\} \nonumber \\
	& \overset{(ii)}{\leq} \frac{1+\delta_2}{1-\delta_1} \frac{\overline \sigma_0^2}{mn} \gamma\left( \frac{\overline \sigma_0^2}{mn} \right) + \frac{mn}{\overline \sigma_0^2 \delta_2}  \tr(\bSigma_M)\tr\left(\bSigma_M^{(\zeta)}\right) + \tr(\bSigma_M) \left\{ 100\rho_*^2\frac{ b(m,\zeta,r_*)  \gamma(\tfrac{\overline \sigma_0^2}{mn})}{\delta_1\sqrt{m}} \right\}^{r_*},
\end{align}
where (i) follows from \eqref{Bmatbound1}, and (ii) follows from \eqref{traceerr1}, Lemma \ref{zhang15d}, and the fact that
\begin{align} 
	\tr\left(\Q^{-2}\right) &= \tr\left\{\left(\I_{\zeta} + \frac{\overline \sigma_0^2}{mn}\M^{-1} \right)^{-1}\right\} = \sum_{a=1}^{\zeta} \left(1+\frac{\overline \sigma_0^2}{mn\mu_a} \right)^{-1} = \sum_{a=1}^{\zeta} \frac{\mu_a}{\mu_a +\frac{\overline \sigma_0^2}{mn}} \leq \gamma\left( \frac{\overline \sigma_0^2}{mn}\right).  \nonumber
\end{align}

Finally, we combine \eqref{msebound3} and \eqref{Bmatbound2} to obtain that
\begin{align} 
	&\quad {\E}_{\X^m} {\E}_{\X_0} \left[\mse_{\opt}^{(M)}(\X_0)\right] \leq  {\E}_{\X^m}\tr(\B) + \tr\left(\bSigma_{M}^{(\zeta)}\right) \nonumber \\
	&\leq  \frac{1+\delta_2}{1-\delta_1} \frac{\overline \sigma_0^2}{mn} \gamma\left( \frac{\overline \sigma_0^2}{mn}\right) + \left\{\frac{mn}{\overline \sigma_0^2 \delta_2}  \tr(\bSigma_M) + 1\right\} \tr\left(\bSigma_M^{(\zeta)}\right) + \tr(\bSigma_M) \left\{ 100\rho_*^2\frac{ b(m,\zeta,r_*)  \gamma(\tfrac{\overline \sigma_0^2}{mn})}{\delta_1\sqrt{m}} \right\}^{r_*}. \nonumber
\end{align}
Taking the infimum with respect to $\zeta$ and setting $\delta_1=\delta_2=1/3$ leads to the conclusion. \hfill $\Box$

\vspace{8mm}

\noindent \textbf{Proof of Theorem 2}:

We define some additional notation. For abbreviation, we write $\sigma^2_j = \sigma^2(\x_j)$, $j=1,\ldots,m$. Let $\F=(\bbf(\X_1),\ldots,\bbf(\X_m))^\top = (\uf_1(\X^m),\ldots,\uf_q(\X^m))$ be the partition of $\F$ according to rows and columns, respectively. For the ``bias" defined in (6) of the manuscript, let $\eta(\x)=\left(\eta_1(\x),\ldots,\eta_q(\x)\right)^\top$ for any $\x \in \Xcal$, where $\eta_s(\x) = \uf_s(\x) - \uf_s(\x^m)^\top \left(\bSigma_M(\x^m,\x^m) + \bSigma_{\epsilon}(\x^m)\right)^{-1} \bSigma_M(\x^m,\x)$. Since by Assumption A.4, $\uf_s(\cdot) \in \HH$ for each $s=1,\ldots,q$ and $\bSigma(\x_j,\cdot)\in \HH$ for each $j=1,\ldots,m$, we have that the function $\eta_s(\cdot)$ also lies in $\HH$. In the following, we investigate and provide upper bound for $\|\eta_s\|_2$, $s=1,\ldots,q$. We first expand the function $\uf_s(\x)$ and $\eta_s(\x)$ in terms of the orthonormal basis $\{\phi_l(\x):l=1,2,\ldots\}$:
\begin{align}\label{etaexpand}
	& \uf_s(\x) = \sum_{l=1}^{\infty} \theta_{sl} \phi_l(\x), \qquad \eta_s(\x) = \sum_{l=1}^{\infty} \delta_{sl} \phi_l(\x),
\end{align}
for any $\x\in \Xcal$ and $s=1,\ldots,q$. For a fixed $\zeta \in \NN$, define $\theta_s^\downarrow = (\theta_{s1},\ldots,\theta_{s\zeta})^\top$, $\theta_s^\uparrow = (\theta_{s,\zeta+1},\theta_{s,\zeta+2},\ldots)^\top$, $\delta_s^\downarrow = (\delta_{s1},\ldots,\delta_{s\zeta})^\top$, $\delta_s^\uparrow = (\delta_{s,\zeta+1},\delta_{s,\zeta+2},\ldots)^\top$. We also define the following quantities:
\begin{align*}
	& \M = \text{diag}\left(\mu_1,\ldots,\mu_\zeta\right), \\
	& \bphi_l = \left[\phi_l(\X_1),\ldots,\phi_l(\X_m)\right]^\top, \text{ for } l=1,2,\ldots, \\
	& \bPhi = \left[\bphi_1, \ldots, \bphi_{\zeta} \right], \\
	& \bbv_s = (v_{s1},\ldots,v_{sm})^\top, \quad v_{sj} = \sum_{l=\zeta+1}^{\infty} \delta_{sl} \phi_l(\X_j),\text{ for } j=1,\ldots,m.
\end{align*}
Then based on Assumptions A.1-A.4, we can prove Lemma \ref{etabound} and Lemma \ref{numeratorbound}. On the other hand, from the definition of $\mse_{\opt}^{(\bbeta)} (\x_0)$ in (6) of the manuscript, we have that
\begin{align}\label{betabound1}
	{\E}_{\X_0}\left[\mse_{\opt}^{(\bbeta)} (\X_0)\right] &= {\E}_{\X_0}\left[\eta(\X_0)^\top \left[\F^\top \left(\bSigma_{M}(\X^m,\X^m)+\bSigma_{\epsilon}(\X^m)\right)^{-1}\F\right]^{-1} \eta(\X_0)\right] \nonumber \\
	&\leq \lambda_{\max}\left(\left[\F^\top \left(\bSigma_{M}(\X^m,\X^m)+\bSigma_{\epsilon}(\X^m)\right)^{-1}\F\right]^{-1}\right)\cdot {\E}_{\X_0}\left[\eta(\X_0)^\top \eta(\X_0)\right] \nonumber \\
	&= \left[\lambda_{\min}\left(\F^\top \left(\bSigma_{M}(\X^m,\X^m)+\bSigma_{\epsilon}(\X^m)\right)^{-1}\F\right)\right]^{-1} \cdot {\E}_{\X_0}\left[ \sum_{s=1}^q \eta_s(\X_0)^\top \eta_s(\X_0)\right] \nonumber \\
	&= \left[\lambda_{\min}\left(\F^\top \left(\bSigma_{M}(\X^m,\X^m)+\bSigma_{\epsilon}(\X^m)\right)^{-1}\F\right)\right]^{-1} \cdot \sum_{s=1}^q \left\| \eta_s \right\|_2^2.
\end{align}

For simplicity, we define $\Gamma_m$ to be the quantity inside the bracelets in Theorem 2:
\begin{align*}
	\Gamma_m &= 8C_{\uf}^2 \frac{\overline \sigma_0^2}{mn} + \inf_{\zeta\in \NN} \Bigg[8C_{\uf}^2 \frac{mn \overline \sigma_0^2}{ \underline \sigma_0^4} \rho_*^4 \tr\left(\bSigma_M \right) \tr\left(\bSigma_M^{(\zeta)}\right) + C_{\uf}^2 \tr\left(\bSigma_M^{(\zeta)}\right)\nonumber \\
	&~+ C_{\uf}^2  \tr\left(\bSigma_M\right)\left\{ 200\rho_*^2 \frac{ b(m,\zeta,r_*)  \gamma(\tfrac{\overline \sigma_0^2}{mn})}{\sqrt{m}} \right\}^{r_*} \Bigg].
\end{align*}
From the upper bound of ${\E}_{\X^m} \left\| \eta_s \right\|_2^2$ in Lemma \ref{etabound}, it is clear that ${\E}_{\X^m} \left\| \eta_s \right\|_2^2\leq \Gamma_m$ for all $s=1,\ldots,q$ since we can make the upper bound in Lemma \ref{etabound} larger by replacing each $\| \uf_s \|_{\HH}$ with $C_{\uf}$. From the Markov's inequality, for any $\xi\in (0,1/4)$, \begin{align}\label{betaboundprob1}
	\PP_{\X^m}\left(\sum_{s=1}^q \left\| \eta_s \right\|_2^2 \geq q\Gamma_m/\xi \right) &\leq \frac{\sum_{s=1}^q{\E}_{\X^m} \left\| \eta_s \right\|_2^2}{q\Gamma_m/\xi} \leq \frac{q\Gamma_m}{q\Gamma_m/\xi} = \xi.
\end{align}

Then from Lemma \ref{numeratorbound}, we have that for any $\xi\in (0,1/4)$, for all $m>m_0$ (with $m_0$ dependent on $\xi,\bSigma_M,\bbf,n,\overline\sigma_0^2,\rho_*$),
\begin{align}\label{betaboundprob2}
	& \PP_{\X^m}\left(\left[\lambda_{\min}\left\{\F^\top \left(\bSigma_{M}(\X^m,\X^m)+\bSigma_{\epsilon}(\X^m)\right)^{-1}\F\right\}\right]^{-1} > \frac{8\tr(\bSigma_M)}{\lambda_{\min}\left({\E}_{\X}[\bbf(\X)\bbf(\X)^\top]\right)} \right) < \xi.
\end{align}
We combine \eqref{betabound1}, \eqref{betaboundprob1} and \eqref{betaboundprob2} together to conclude that for any $\xi\in (0,1/4)$, for all $m>m_0$, there exists a constant $c_{\xi}=1/\xi$, such that
\begin{align}\label{betabound2}
	& \PP_{\X^m} \left({\E}_{\X_0}\left[\mse_{\opt}^{(\bbeta)} (\X_0)\right] > c_{\xi}\cdot \frac{8q\tr(\bSigma_M)}{\lambda_{\min}\left({\E}_{\X}[\bbf(\X)\bbf(\X)^\top]\right)}\Gamma_m \right) \nonumber \\
	\leq{}& \PP_{\X^m} \left(\left[\lambda_{\min}\left(\F^\top \left(\bSigma_{M}(\X^m,\X^m)+\bSigma_{\epsilon}(\X^m)\right)^{-1}\F\right)\right]^{-1} \cdot \sum_{s=1}^q \left\| \eta_s \right\|_2^2 > \frac{8\tr(\bSigma_M)}{\lambda_{\min}\left({\E}_{\X}[\bbf(\X)\bbf(\X)^\top]\right)}\cdot \frac{q\Gamma_m}{\xi}\right) \nonumber \\
	\leq{}& \PP_{\X^m}\left(\sum_{s=1}^q \left\| \eta_s \right\|_2^2 \geq q\Gamma_m/\xi \right) \nonumber \\
	& ~~~ + \PP_{\X^m}\left(\left[\lambda_{\min}\left\{\F^\top \left(\bSigma_{M}(\X^m,\X^m)+\bSigma_{\epsilon}(\X^m)\right)^{-1}\F\right\}\right]^{-1} > \frac{8\tr(\bSigma_M)}{\lambda_{\min}\left({\E}_{\X}[\bbf(\X)\bbf(\X)^\top]\right)} \right) \nonumber \\
	<{}& \xi+\xi = 2\xi.
\end{align}
This has proved that ${\E}_{\X_0}\left[\mse_{\opt}^{(\bbeta)} (\X_0)\right] {\lesssim}_{\PP_{\X^m}} \frac{8q\tr(\bSigma_M)}{\lambda_{\min}\left({\E}_{\X}[\bbf(\X)\bbf(\X)^\top]\right)} \Gamma_m$, which is the conclusion of Theorem 2.  \hfill $\Box$

\vspace{10mm}

\begin{lemma}\label{etabound}
	Under Assumptions A.1-A.4, we have that for each $s=1,\ldots,q$,
	\begin{align*}
		{\E}_{\X^m} \left\| \eta_s \right\|_2^2 & \leq \frac{8\| \uf_s \|_{\HH}^2\overline \sigma_0^2}{mn} + \inf_{\zeta\in \NN} \Bigg[ \frac{8\| \uf_s \|_{\HH}^2 mn \overline \sigma_0^2}{ \underline \sigma_0^4} \rho_*^4 \tr\left(\bSigma_M \right) \tr\left(\bSigma_M^{(\zeta)}\right) + \|\uf_s\|_{\HH}^2 \tr\left(\bSigma_M^{(\zeta)}\right)\nonumber \\
		&~~ + \|\uf_s\|_{\HH}^2 \tr\left(\bSigma_M\right)\left\{ 200\rho_*^2 \frac{ b(m,\zeta,r_*)  \gamma(\tfrac{\overline \sigma_0^2}{mn})}{\sqrt{m}} \right\}^{r_*} \Bigg].
	\end{align*}
\end{lemma}

\noindent \textbf{Proof of Lemma \ref{etabound}}:

By Assumption A.1, we have that $\bSigma_{\epsilon}(\x^m)=\text{diag}\left(\sigma_1^2/n,\ldots,\sigma_m^2/n\right)$, where we let $\sigma_j^2=\sigma^2(\x_j)$ for $j=1,\ldots,m$. For any $\x \in \Xcal$ and any $s\in \{1,\ldots,q\}$, we have the following relation:
\begin{align}\label{krreqn1}
	& \quad \sum_{j=1}^m \frac{n}{\sigma_j^2} \eta_s(\x_j) \bSigma_M(\x_j,\x) \nonumber\\
	&= \sum_{j=1}^m \frac{n}{\sigma_j^2} \left\{\uf_s(\x_j) - \uf_s(\x^m)^\top \left(\bSigma_M(\x^m,\x^m) + \bSigma_{\epsilon}(\x^m)\right)^{-1} \bSigma_M(\x^m,\x_j)\right\} \bSigma_M(\x_j,\x)\nonumber \\
	&= \sum_{j=1}^m \frac{n}{\sigma_j^2} \uf_s(\x_j)\bSigma_M(\x_j,\x) - \sum_{j=1}^m \frac{n}{\sigma_j^2}\uf_s(\x^m)^\top \left(\bSigma_M(\x^m,\x^m) + \bSigma_{\epsilon}(\x^m)\right)^{-1} \bSigma_M(\x^m,\x_j) \bSigma_M(\x_j,\x) \nonumber \\
	&= \uf_s(\x^m)^\top \bSigma_{\epsilon}(\x^m)^{-1} \bSigma_M(\x^m,\x) \nonumber \\
	&\quad - \uf_s(\x^m)^\top \left(\bSigma_M(\x^m,\x^m) + \bSigma_{\epsilon}(\x^m)\right)^{-1} \bSigma_M(\x^m,\x^m) \bSigma_{\epsilon}(\x^m)^{-1} \bSigma_M(\x^m,\x) \nonumber \\
	&= \uf_s(\x^m)^\top \left(\bSigma_M(\x^m,\x^m) + \bSigma_{\epsilon}(\x^m)\right)^{-1} \left\{\bSigma_M(\x^m,\x^m) + \bSigma_{\epsilon}(\x^m) -\bSigma_M(\x^m,\x^m)  \right\} \nonumber \\
	&\quad \cdot \bSigma_{\epsilon}(\x^m)^{-1} \bSigma_M(\x^m,\x) \nonumber \\
	&= \uf_s(\x^m)^\top \left(\bSigma_M(\x^m,\x^m) + \bSigma_{\epsilon}(\x^m)\right)^{-1} \bSigma_M(\x^m,\x) \nonumber \\
	&= \uf_s(\x) - \eta_s(\x).
\end{align}
Therefore, we can rewrite \eqref{krreqn1} as
\begin{align}\label{krreqn2}
	& \sum_{j=1}^m \frac{n}{\sigma_j^2} \eta_s(\x_j) \bSigma_M(\x_j,\x) + \eta_s(\x) -  \uf_s(\x) =0,
\end{align}
for any $\x \in \Xcal$ and any $s\in \{1,\ldots,q\}$.

We proceed with \eqref{krreqn2} in two ways. On one hand, we can take the $\HH$-norm of $\uf_s$ in \eqref{krreqn2}. Since $\eta_s\in \HH$ and it has the expansion in \eqref{etaexpand}, we can derive from \eqref{krreqn2} that
\begin{align}\label{etanorm1}
	\uf_s(\x) &= \sum_{j=1}^m \frac{n}{\sigma_j^2} \sum_{a=1}^\infty \delta_{sa} \phi_a(\x_j) \sum_{b=1}^\infty \mu_b \phi_b(\x_j)\phi_b(\x) + \sum_{b=1}^\infty \delta_{sb} \phi_b(\x) \nonumber \\
	&= \sum_{b=1}^\infty \left\{\mu_b \sum_{j=1}^m \frac{n}{\sigma_j^2} \sum_{a=1}^\infty \delta_{sa} \phi_a(\x_j)  \phi_b(\x_j)+ \delta_{sb}\right\} \phi_b(\x), \nonumber \\
	\|\uf_s\|_{\HH}^2 &= \sum_{b=1}^\infty \frac{1}{\mu_b} \left\{\mu_b \sum_{j=1}^m \frac{n}{\sigma_j^2} \sum_{a=1}^\infty \delta_{sa} \phi_a(\x_j)  \phi_b(\x_j)+ \delta_{sb} \right\}^2  \nonumber \\
	&= \sum_{b=1}^\infty \frac{\delta_{sb}^2}{\mu_b} + 2\sum_{b=1}^\infty \sum_{j=1}^m \frac{n}{\sigma_j^2} \sum_{a=1}^\infty \delta_{sa} \delta_{sb} \phi_a(\x_j) \phi_b(\x_j) 
	+ \sum_{b=1}^\infty \mu_b \left\{ \sum_{j=1}^m \frac{n}{\sigma_j^2} \sum_{a=1}^\infty \delta_{sa} \phi_a(\x_j)  \phi_b(\x_j)\right\}^2 \nonumber \\
	&= \|\eta_s\|_{\HH}^2 + 2\sum_{j=1}^m \frac{n}{\sigma_j^2} \left\{\sum_{a=1}^\infty \delta_{sa} \phi_a(\x_j) \right\}^2 
	+ \sum_{b=1}^\infty \mu_b \left\{\sum_{j=1}^m \frac{n}{\sigma_j^2} \sum_{a=1}^\infty \delta_{sa} \phi_a(\x_j)  \phi_b(\x_j)\right\}^2 \nonumber \\
	&\geq \|\eta_s\|_{\HH}^2, \nonumber \\
	\implies & \|\eta_s\|_{\HH} \leq \|\uf_s\|_{\HH}.
\end{align}

On the other hand, we take $\HH$-inner product of the left-hand-side of \eqref{krreqn2} with $\phi_l(\x)$ for any fixed $l$ with $\mu_l>0$, and obtain that
\begin{align}\label{krreqn3}
	0&= \sum_{j=1}^m \frac{n}{\sigma_j^2} \eta_s(\x_j) \langle \bSigma_M(\x_j,\x), \phi_l(\x) \rangle_{\HH} + \langle \eta_s(\x), \phi_l(\x) \rangle_{\HH} - \langle \uf_s(\x), \phi_l(\x) \rangle_{\HH}, \nonumber \\
	&= \sum_{j=1}^m \frac{n}{\sigma_j^2} \eta_s(\x_j) \phi_l(\x_j) + \frac{\delta_{sl}}{\mu_l} - \frac{\theta_{sl}}{\mu_l} , \nonumber \\
	&= \sum_{j=1}^m \frac{n}{\sigma_j^2} \sum_{a=1}^{\infty} \delta_{sa} \phi_a(\x_j) \phi_l(\x_j) + \frac{\delta_{sl}}{\mu_l} - \frac{\theta_{sl}}{\mu_l}  , \nonumber \\
	&= \sum_{j=1}^m  \sum_{a=1}^{\zeta} \frac{n}{\sigma_j^2}\delta_{sa} \phi_a(\x_j) \phi_l(\x_j) + \sum_{j=1}^m \frac{n}{\sigma_j^2} v_{sj}\phi_l(\x_j) + \frac{\delta_{sl}}{\mu_l} - \frac{\theta_{sl}}{\mu_l}
\end{align}
where we have used the reproducing property for the function $\phi_l \in \HH$. We can then stack \eqref{krreqn3} in a column for $l=1,\ldots,\zeta$ for some $\zeta\in \NN$ with $\mu_\zeta>0$, and obtain that
\begin{align}\label{krreqn4}
	& \bPhi^\top \bSigma_{\epsilon}(\x^m)^{-1} \bPhi \delta_s^\downarrow + \bPhi^\top \bSigma_{\epsilon}(\x^m)^{-1} \bbv_s + \M^{-1} \delta_s^\downarrow - \M^{-1} \theta_s^\downarrow = 0, \nonumber \\
	\implies  \delta_s^\downarrow & = \left(\bPhi^\top \bSigma_{\epsilon}(\x^m)^{-1} \bPhi + \M^{-1} \right)^{-1} \left( \M^{-1} \theta_s^\downarrow  - \bPhi^\top \bSigma_{\epsilon}(\x^m)^{-1} \bbv_s \right), \nonumber \\
	& = \left(\bPhi^\top \bSigma_{\epsilon}(\x^m)^{-1} \bPhi + \M^{-1} \right)^{-1}\Q \left( \Q^{-1}\M^{-1} \theta_s^\downarrow  - \Q^{-1}\bPhi^\top \bSigma_{\epsilon}(\x^m)^{-1} \bbv_s \right),
\end{align}
where $\Q=\left(\I_{\zeta}+\tfrac{\overline \sigma_0^2}{mn}\M^{-1}\right)^{1/2}$ as defined in Lemma \ref{zhang15d}.
Therefore,
\begin{align}\label{krrbound1}
	& \left\| \delta_s^\downarrow \right\| \leq \vertiii{\left(\bPhi^\top \bSigma_{\epsilon}(\x^m)^{-1} \bPhi + \M^{-1} \right)^{-1}\Q } \left(\left\| \Q^{-1}\M^{-1} \theta_s^\downarrow \right\| + \left\|\Q^{-1}\bPhi^\top \bSigma_{\epsilon}(\x^m)^{-1} \bbv_s \right\| \right).
\end{align}

By Assumption A.1, we have that $\bSigma_{\epsilon}(\x^m)\preceq \frac{\overline \sigma_0^2}{n}\I_m $. Therefore, $\bPhi^\top \bSigma_{\epsilon}(\x^m)^{-1} \bPhi + \M^{-1} \succeq \frac{n}{\overline \sigma_0^2} \bPhi^\top \bPhi + \M^{-1} \succ 0$. This implies that
\begin{align}\label{qmatinq1}
	& 0 \prec \Q^{1/2}\left(\bPhi^\top \bSigma_{\epsilon}(\x^m)^{-1} \bPhi + \M^{-1}\right)^{-1}\Q^{1/2} \preceq \Q^{1/2}\left(\frac{n}{\overline \sigma_0^2} \bPhi^\top \bPhi + \M^{-1} \right)^{-1}\Q^{1/2}.
\end{align}
Note that the matrices $\left(\bPhi^\top \bSigma_{\epsilon}(\x^m)^{-1} \bPhi + \M^{-1}\right)^{-1}$,  $\left(\frac{n}{\overline \sigma_0^2} \bPhi^\top \bPhi + \M^{-1} \right)^{-1}$, and $\Q$ are all symmetric and positive definite matrices. Furthermore, $\left(\bPhi^\top \bSigma_{\epsilon}(\x^m)^{-1} \bPhi + \M^{-1} \right)^{-1}\Q$ is similar to the symmetric positive definite matrix $\Q^{1/2}\left(\bPhi^\top \bSigma_{\epsilon}(\x^m)^{-1} \bPhi + \M^{-1} \right)^{-1}\Q^{1/2}$. Therefore,
\begin{align}\label{qmatinq2}
	&\lambda_{\max}\left\{\left(\bPhi^\top \bSigma_{\epsilon}(\x^m)^{-1} \bPhi + \M^{-1} \right)^{-1}\Q\right\}
	=\lambda_{\max}\left\{\Q^{1/2}\left(\bPhi^\top \bSigma_{\epsilon}(\x^m)^{-1} \bPhi + \M^{-1} \right)^{-1}\Q^{1/2}\right\},
\end{align}
and similarly
\begin{align}\label{qmatinq3}
	&\lambda_{\max}\left\{\left(\frac{n}{\overline \sigma_0^2} \bPhi^\top \bPhi + \M^{-1} \right)^{-1}\Q\right\}
	=\lambda_{\max}\left\{\Q^{1/2}\left(\frac{n}{\overline \sigma_0^2} \bPhi^\top \bPhi + \M^{-1} \right)^{-1}\Q^{1/2}\right\}.
\end{align}
\eqref{qmatinq1}, \eqref{qmatinq2}, and \eqref{qmatinq3} imply that
\begin{align}\label{krrterm11}
	& \vertiii{\left(\bPhi^\top \bSigma_{\epsilon}(\x^m)^{-1} \bPhi + \M^{-1} \right)^{-1}\Q} = \lambda_{\max}\left\{\left(\bPhi^\top \bSigma_{\epsilon}(\x^m)^{-1} \bPhi + \M^{-1} \right)^{-1}\Q\right\} \nonumber \\
	&\leq \lambda_{\max}\left\{\left(\frac{n}{\overline \sigma_0^2} \bPhi^\top \bPhi + \M^{-1} \right)^{-1}\Q\right\} = \vertiii{\left(\frac{n}{\overline \sigma_0^2}\bPhi^\top \bPhi + \M^{-1} \right)^{-1}\Q} \nonumber \\
	&= \frac{\overline \sigma_0^2}{mn} \vertiii{\left\{\left(\I_{\zeta} + \frac{\overline \sigma_0^2}{mn}\M^{-1}\right) + \left(\frac{1}{m}\bPhi^\top \bPhi-\I_{\zeta}\right)\right\}^{-1}\Q} \nonumber \\
	&= \frac{\overline \sigma_0^2}{mn} \vertiii{\Q^{-1}\left\{\I_{\zeta} + \Q^{-1}\left(\frac{1}{m}\bPhi^\top \bPhi-\I_{\zeta}\right)\Q^{-1}\right\}^{-1}}, \nonumber \\
	&\leq \frac{\overline \sigma_0^2}{mn} \vertiii{\Q^{-1}} \vertiii{\left\{\I_{\zeta} + \Q^{-1}\left(\frac{1}{m}\bPhi^\top \bPhi-\I_{\zeta}\right)\Q^{-1}\right\}^{-1}}
\end{align}
We consider the event defined in Lemma \ref{zhang15d} with $\delta=1/2$, i.e.
$$\Ecal_3=\left\{\vertiii{\Q^{-1}\left( \frac{1}{m} \bPhi^\top \bPhi - \I_{\zeta} \right) \Q^{-1}} \leq \frac{1}{2} \right\}.$$
Then on the event $\Ecal_3$, $\I_{\zeta} + \Q^{-1}\left(\frac{1}{m}\bPhi^\top \bPhi-\I_{\zeta}\right)\Q^{-1} \succeq (1-1/2)\I_{\zeta} = (1/2)\I_{\zeta}$. Moreover, $0\prec \Q^{-1} \prec \I_{\zeta}$. Therefore, \eqref{krrterm11} implies that
\begin{align}\label{krrterm12}
	& \vertiii{\left(\bPhi^\top \bSigma_{\epsilon}(\x^m)^{-1} \bPhi + \M^{-1} \right)^{-1}\Q } \leq \frac{2\overline \sigma_0^2}{mn} \vertiii{\Q^{-1}}\leq \frac{2\overline \sigma_0^2}{mn}.
\end{align}

In \eqref{krrbound1}, the term $\left\|\Q^{-1} \M^{-1} \theta_s^\downarrow \right\|$ can be bounded as
\begin{align}\label{krrterm21}
	& \left\|\Q^{-1} \M^{-1} \theta_s^\downarrow \right\| = \sqrt{\left(\theta_s^\downarrow\right)^\top \M^{-1}\Q^{-2}\M^{-1} \theta_s^\downarrow} = \sqrt{\left(\theta_s^\downarrow\right)^\top \left(\M^2+\frac{\overline \sigma_0^2}{mn}\M\right)^{-1} \theta_s^\downarrow} \nonumber \\
	& \leq  \sqrt{\left(\theta_s^\downarrow\right)^\top \left(\frac{\overline \sigma_0^2}{mn}\M\right)^{-1} \theta_s^\downarrow} = \sqrt{\frac{mn}{\overline \sigma_0^2}}\sqrt{\sum_{l=1}^{\zeta} \frac{\theta_{sl}^2}{\mu_l^2}} \leq \sqrt{\frac{mn}{\overline \sigma_0^2}} \left\|\uf_s\right\|_{\HH}.
\end{align}
For the term $\left\|\Q^{-1}\bPhi^\top \bSigma_{\epsilon}(\x^m)^{-1} \bbv_s \right\|$ in \eqref{krrbound1}, we first have that
\begin{align}\label{krrterm31}
	& \left\|\Q^{-1}\bPhi^\top \bSigma_{\epsilon}(\x^m)^{-1} \bbv_s \right\| = \left\|\left(\M+\frac{\overline \sigma_0^2}{mn}\I_{\zeta}\right)^{-1/2} \M^{1/2}\bPhi^\top \bSigma_{\epsilon}(\x^m)^{-1} \bbv_s \right\|\nonumber \\
	&\leq \vertiii{\left(\M+\frac{\overline \sigma_0^2}{mn}\I_{\zeta}\right)^{-1/2}} \cdot \left\|\M^{1/2}\bPhi^\top \bSigma_{\epsilon}(\x^m)^{-1} \bbv_s \right\| = \frac{1}{\sqrt{\mu_{\zeta} + \frac{\overline \sigma_0^2}{mn}}} \left\|\M^{1/2}\bPhi^\top \bSigma_{\epsilon}(\x^m)^{-1} \bbv_s \right\| \nonumber \\
	&\leq \sqrt{\frac{mn}{\overline \sigma_0^2}} \left\|\M^{1/2}\bPhi^\top \bSigma_{\epsilon}(\x^m)^{-1} \bbv_s \right\| = \sqrt{\frac{mn}{\overline \sigma_0^2}} \sqrt{\sum_{l=1}^{\zeta} \mu_l \left(\bphi_l^\top \bSigma_{\epsilon}(\x^m)^{-1} \bbv_s \right)^2 } \nonumber \\
	&\stackrel{(i)}{\leq} \sqrt{\frac{mn}{\overline \sigma_0^2}} \left\{\sum_{l=1}^{\zeta} \mu_l \left(\bphi_l^\top \bSigma_{\epsilon}(\x^m)^{-1}\bphi_l\right) \left(\bbv_s^\top \bSigma_{\epsilon}(\x^m)^{-1} \bbv_s \right) \right\}^{1/2} \stackrel{(ii)}{\leq} \frac{n}{\underline \sigma_0^2}\sqrt{\frac{mn}{\overline \sigma_0^2}} \sqrt{\sum_{l=1}^{\zeta} \mu_l \left\|\bphi_l\right\|^2 \left\|\bbv_s \right\|^2 },
\end{align}
where (i) follows from the Cauchy-Schwarz inequality, and (ii) follows from Assumption A.1 that $\sigma_j^2\geq \underline \sigma_0^2$ for all $j=1,\ldots,m$ and hence $\bSigma_{\epsilon}(\x^m)^{-1}\preceq \tfrac{n}{\underline \sigma_0^2}\I_{\zeta}$.

We can combine \eqref{krrbound1}, \eqref{krrterm12}, \eqref{krrterm21}, \eqref{krrterm31}, and apply the inequality $(a+b)^2\leq 2a^2+2b^2$ to obtain that
\begin{align}\label{krrbound2}
	& \left\| \delta_s^\downarrow \right\|^2 \leq 2\vertiii{\left(\bPhi^\top \bSigma_{\epsilon}(\x^m)^{-1} \bPhi + \M^{-1} \right)^{-1}\Q }^2 \left(\left\| \Q^{-1}\M^{-1} \theta_s^\downarrow \right\|^2 + \left\|\Q^{-1}\bPhi^\top \bSigma_{\epsilon}(\x^m)^{-1} \bbv_s \right\|^2 \right) \nonumber \\
	&\leq 2 \left(\frac{2\overline \sigma_0^2}{mn}\right)^2 \left\{ \frac{mn}{\overline \sigma_0^2} \left\|\uf_s\right\|_{\HH}^2 + \left(\frac{n}{\underline \sigma_0^2}\right)^2 \frac{mn}{\overline \sigma_0^2} \sum_{l=1}^d \mu_l \left\|\bphi_l\right\|^2 \left\|\bbv_s \right\|^2 \right\} \nonumber \\
	&= 8\left\{\frac{\overline \sigma_0^2}{mn} \left\|\uf_s\right\|_{\HH}^2 + \frac{n \overline \sigma_0^2}{m \underline \sigma_0^4} \sum_{l=1}^{\zeta} \mu_l \left\|\bphi_l\right\|^2 \left\|\bbv_s \right\|^2 \right\}.
\end{align}

Now we evaluate the expectation $\E_{\X^m} \| \delta_s^\downarrow \|^2$. From \eqref{krrbound2}, it suffices to control ${\E}_{\X^m}\left(\left\|\bphi_l\right\|^2 \left\|\bbv_s \right\|^2\right)$ for $l=1,\ldots,d$. By the Cauchy-Schwarz inquality,
\begin{align}\label{krrterm32}
	{\E}_{\X^m} \left(\left\|\bphi_l\right\|^2 \left\|\bbv_s \right\|^2\right) &\leq \sqrt{{\E}_{\X^m} \left(\left\|\bphi_l\right\|^4\right)} \sqrt{{\E}_{\X^m} \left(\left\|\bbv_s \right\|^4\right)}.
\end{align}
By Assumption A.3, ${\E}_{\P_{\X}}\left\{\phi_l^{2r_*}(\X)\right\}\leq \rho_*^{2r_*}$ for some $r_*\geq 2$. By Jensen's inequality, for all $l=1,2,\ldots$,
$${\E}_{\P_{\X}}\left\{\phi_l^4(\X)\right\}\leq \left[{\E}_{\P_{\X}}\left\{\phi_l^{2r_*}(\X)\right\}\right]^{2/r_*}\leq \rho_*^{2r_*\cdot 2/r_*}=\rho_*^4.$$
Since $\X_1,\ldots,\X_m$ are i.i.d. distributed as $\P_{\X}$ and ${\E}_{\P_{\X}}\left\{\phi_l^4(\X)\right\}\leq \rho_*^4$ for all $l$, we have that
\begin{align}\label{krrterm33}
	& {\E}_{\X^m} \left(\left\|\bphi_l\right\|^4\right) = {\E}_{\X^m} \left\{ \left(\sum_{j=1}^m \phi_l^2(\X_j)\right)^2\right\} \nonumber \\
	& \leq {\E}_{\X^m} \left( m \sum_{j=1}^m \phi_l^4(\X_j)\right) \leq m^2 {\E}_{\X^m} \left( \phi_l^4(\X_1)\right) \leq m^2 \rho_*^4.
\end{align}
On the other hand, by applying the Cauchy-Schwarz inequality, we have
\begin{align}\label{krrterm34}
	& {\E}_{\X^m} \left(\left\|\bbv_s \right\|^4\right) = {\E}_{\X^m} \left\{\left(\sum_{j=1}^m v_{sj}^2 \right)^2\right\}
	\leq m {\E}_{\X^m} \left(\sum_{j=1}^m v_{sj}^4 \right) = m^2 {\E}_{\X^m} \left(v_{s1}^4 \right) \nonumber \\
	& = m^2 {\E}_{\X^m} \left\{ \left(\sum_{l=\zeta+1}^{\infty} \delta_{sl} \phi_{l}(\X_1)\right)^4 \right\} \leq
	m^2 {\E}_{\X^m} \left[ \left\{\sum_{l=\zeta+1}^{\infty} \frac{\delta_{sl}^2}{\mu_l} \cdot \sum_{l=\zeta+1}^{\infty} \mu_l\phi_{l}^2(\X_1)\right\}^2 \right].
\end{align}
From \eqref{etanorm1}, we can get an upper bound $\sum_{l=\zeta+1}^{\infty} \tfrac{\delta_{sl}^2}{\mu_l} \leq \sum_{l=1}^{\infty} \frac{\delta_{sl}^2}{\mu_l} = \|\eta_s\|_{\HH}^2 \leq \|\uf_s\|_{\HH}^2$. Therefore, \eqref{krrterm34} further implies that
\begin{align}\label{krrterm35}
	& {\E}_{\X^m} \left(\left\|\bbv_s \right\|^4\right) \leq m^2 \| \uf_s \|_{\HH}^4 \cdot {\E}_{\X^m} \left[ \left\{\sum_{l=\zeta+1}^{\infty} \mu_l\phi_{l}^2(\X_1)\right\}^2 \right] \nonumber \\
	&= m^2 \| \uf_s \|_{\HH}^4 \cdot {\E}_{\X^m} \left\{\sum_{a=\zeta+1}^{\infty}\sum_{b=\zeta+1}^{\infty} \mu_a \mu_b \phi_{a}^2(\X_1) \phi_{b}^2(\X_1)\right\} \nonumber \\
	&\stackrel{(i)}{\leq} m^2 \| \uf_s \|_{\HH}^4 \cdot  \left\{\sum_{a=\zeta+1}^{\infty}\sum_{b=\zeta+1}^{\infty} \mu_a \mu_b \sqrt{{\E}_{\X^m} \phi_{a}^4(\X_1) \cdot {\E}_{\X^m} \phi_{b}^4(\X_1)}\right\} \nonumber \\
	&\stackrel{(ii)}{\leq} m^2 \rho_*^4 \| \uf_s \|_{\HH}^4 \sum_{a=\zeta+1}^{\infty}\sum_{b=\zeta+1}^{\infty} \mu_a \mu_b
	= m^2 \rho_*^4 \| \uf_s \|_{\HH}^4 \left\{\tr\left(\bSigma_M^{(\zeta)}\right)\right\}^2,
\end{align}
where (i) follows from the Cauchy-Schwarz inequality and the monotone convergence theorem, and (ii) follows from Assumption A.3.

We combine \eqref{krrbound2}, \eqref{krrterm32}, \eqref{krrterm33}, and \eqref{krrterm35}, and to obtain that
\begin{align}\label{krrbound3}
	{\E}_{\X^m} \left(\left\| \delta_s^\downarrow \right\|^2 ~\Big |~ \Ecal_3\right)
	&\leq 8 \left\{\frac{\overline \sigma_0^2}{mn} \left\|\uf_s\right\|_{\HH}^2 + \frac{n \overline \sigma_0^2}{m \underline \sigma_0^4} \sum_{l=1}^{\zeta} \mu_l \cdot m^2 \rho_*^4 \| \uf_s \|_{\HH}^2 \tr\left(\bSigma_M^{(\zeta)}\right)\right\} \nonumber \\
	& \leq 8\| \uf_s \|_{\HH}^2 \left\{\frac{\overline \sigma_0^2}{mn} + \frac{mn \overline \sigma_0^2}{ \underline \sigma_0^4}   \rho_*^4 \tr\left(\bSigma_M \right) \tr\left(\bSigma_M^{(\zeta)}\right)\right\}.
\end{align}
We also have the coarse upper bound for ${\E}_{\X^m}  \|\delta_s^\downarrow \|^2$ using \eqref{etanorm1}:
\begin{align}\label{krrbound4}
	{\E}_{\X^m} \left\| \delta_s^\downarrow \right\|^2 &= \sum_{l=1}^{\zeta} \delta_{sl}^2 \leq  \sum_{l=1}^{\infty} \delta_{sl}^2 
	\leq \mu_1 \sum_{l=1}^{\infty} \frac{\delta_{sl}^2}{\mu_l} = \mu_1 \|\eta_s\|_{\HH}^2 \leq \mu_1 \|\uf_s\|_{\HH}^2 \leq \|\uf_s\|_{\HH}^2 \tr\left(\bSigma_M\right).
\end{align}
This together with the upper bound for $\PP_{\X^m}(\Ecal_3^c)$ in Lemma \ref{zhang15d} (with $\delta=1/2$) implies that
\begin{align}\label{krrbound5}
	{\E}_{\X^m} \left\| \delta_s^\downarrow \right\|^2 &= {\E}_{\X^m} \left\{\left\| \delta_s^\downarrow \right\|^2 \1(\Ecal_3)\right\} + {\E}_{\X^m} \left\{\left\| \delta_s^\downarrow \right\|^2 \1(\Ecal_3^c)\right\} \nonumber \\
	&\leq {\E}_{\X^m} \left(\left\| \delta_s^\downarrow \right\|^2 ~\Big |~ \Ecal_3\right) \cdot \PP_{\X^m}(\Ecal_3) + {\E}_{\X^m} \left\| \delta_s^\downarrow \right\|^2 \cdot\PP_{\X^m}(\Ecal_3^c)\nonumber \\
	&\leq {\E}_{\X^m} \left(\left\| \delta_s^\downarrow \right\|^2 ~\Big |~ \Ecal_3\right) + {\E}_{\X^m} \left\| \delta_s^\downarrow \right\|^2 \cdot \PP_{\X^m}(\Ecal_3^c)\nonumber \\
	&\leq 8\| \uf_s \|_{\HH}^2 \left\{\frac{\overline \sigma_0^2}{mn} + \frac{mn \overline \sigma_0^2}{ \underline \sigma_0^4}   \rho_*^4 \tr\left(\bSigma_M \right) \tr\left(\bSigma_M^{(\zeta)}\right)\right\} \nonumber \\
	&~~ + \|\uf_s\|_{\HH}^2 \tr\left(\bSigma_M\right) \left\{ 200\rho_*^2 \frac{ b(m,\zeta,r_*)  \gamma(\tfrac{\overline \sigma_0^2}{mn})}{\sqrt{m}} \right\}^{r_*}.
\end{align}

On the other hand, from \eqref{etanorm1}, we have that
\begin{align}\label{krrbound6}
	\left\| \delta_s^\uparrow \right\|^2 & = \sum_{l=\zeta+1}^{\infty} \delta_{sl}^2 \leq \mu_{\zeta+1} \sum_{l=\zeta+1}^{\infty} \frac{\delta_{sl}^2}{\mu_l} \leq \mu_{\zeta+1} \sum_{l=1}^{\infty} \frac{\delta_{sl}^2}{\mu_l} \nonumber \\
	&= \mu_{\zeta+1} \|\eta_s\|_{\HH}^2 \leq \mu_{\zeta+1} \|\uf_s\|_{\HH}^2 \leq \tr\left(\bSigma_M^{(\zeta)}\right) \|\uf_s\|_{\HH}^2.
\end{align}
Therefore, \eqref{krrbound5} and \eqref{krrbound6} together imply that
\begin{align}\label{krrbound7}
	{\E}_{\X^m} \left\| \eta_s \right\|_2^2 & ={\E}_{\X^m} \left\| \delta_s \right\|^2 = {\E}_{\X^m} \left\| \delta_s^\downarrow \right\|^2 + {\E}_{\X^m} \left\| \delta_s^\uparrow \right\|^2 \nonumber \\
	&\leq 8\| \uf_s \|_{\HH}^2 \frac{\overline \sigma_0^2}{mn} + 8\| \uf_s \|_{\HH}^2 \frac{mn \overline \sigma_0^2}{ \underline \sigma_0^4}   \rho_*^4 \tr\left(\bSigma_M \right) \tr\left(\bSigma_M^{(\zeta)}\right) + \|\uf_s\|_{\HH}^2 \tr\left(\bSigma_M^{(\zeta)}\right)\nonumber \\
	&~~ + \|\uf_s\|_{\HH}^2 \tr\left(\bSigma_M\right)\left\{ 200\rho_*^2 \frac{ b(m,\zeta,r_*)  \gamma(\tfrac{\overline \sigma_0^2}{mn})}{\sqrt{m}} \right\}^{r_*}.
\end{align}
Taking the infimum with respect to $\zeta$ leads to the result. \hfill $\Box$

\vspace{10mm}

\begin{lemma}\label{numeratorbound}
	Under Assumptions A.1-A.4, for any $\xi\in (0,1)$, there exists a large integer $m_0\in \NN$ that depends on $\xi$, $\bSigma_M$, $\bbf$, $n$, $\overline\sigma_0^2$ in Assumption A.1, and $\rho_*$ in Assumption A.3, such that for all $m>m_0$,
	\begin{align*}
		& \PP_{\X^m}\left(\lambda_{\min}\left\{\F^\top \left(\bSigma_{M}(\X^m,\X^m)+\bSigma_{\epsilon}(\X^m)\right)^{-1}\F\right\} \geq \frac{\lambda_{\min}\left({\E}_{\X}[\bbf(\X)\bbf(\X)^\top]\right)}{8\tr(\bSigma_M)} \right) \geq  1-\xi.
	\end{align*}
\end{lemma}

\noindent \textbf{Proof of Lemma \ref{numeratorbound}}:
\begin{align*}
	& \lambda_{\min}\left\{\F^\top \left(\bSigma_{M}(\X^m,\X^m)+\bSigma_{\epsilon}(\X^m)\right)^{-1}\F\right\} = \min_{\|a\|=1} a^\top \F^\top \left(\bSigma_{M}(\X^m,\X^m)+\bSigma_{\epsilon}(\X^m)\right)^{-1}\F a \\
	&\geq \lambda_{\min}\left\{m\left(\bSigma_{M}(\X^m,\X^m)+\bSigma_{\epsilon}(\X^m)\right)^{-1}\right\} \cdot \min_{\|a\|=1} a^\top \left(\frac{1}{m}\F^\top \F \right) a \\
	&= \lambda_{\min}\left\{m\left(\bSigma_{M}(\X^m,\X^m)+\bSigma_{\epsilon}(\X^m)\right)^{-1}\right\} \cdot \lambda_{\min}\left(\frac{1}{m}\F^\top \F \right).
\end{align*}
Therefore, for any constants $c_1,c_2>0$,
\begin{align}\label{numlow1}
	& \PP_{\X^m}\left(\lambda_{\min}\left\{\F^\top \left(\bSigma_{M}(\X^m,\X^m)+\bSigma_{\epsilon}(\X^m)\right)^{-1}\F\right\} < c_1c_2 \right) \nonumber \\
	\leq {}& \PP_{\X^m}\left(\lambda_{\min}\left(\frac{1}{m}\F^\top \F\right) < c_1 \right) + \PP_{\X^m}\left(\lambda_{\min}\left\{m\left(\bSigma_{M}(\X^m,\X^m)+\bSigma_{\epsilon}(\X^m)\right)^{-1} \right\} < c_2 \right) \nonumber \\
	={}& \PP_{\X^m}\left(\lambda_{\min}\left(\frac{1}{m}\F^\top \F\right) < c_1 \right) + \PP_{\X^m}\left(\lambda_{\max} \left(\bSigma_{M}(\X^m,\X^m)+\bSigma_{\epsilon}(\X^m)\right) > m/c_2 \right).
\end{align}

We choose the values of $c_1$ and $c_2$ and bound the two terms separately. Since $\tfrac{1}{m}\F^\top \F = \tfrac{1}{m}\sum_{j=1}^m \bbf(\X_j)\bbf(\X_j)^\top$, by the strong law of large numbers, $\tfrac{1}{m}\F^\top \F \xrightarrow{a.s.} {\E}_{\X}[\bbf(\X)\bbf(\X)^\top]$ as $m\to\infty$, where $\xrightarrow{a.s.}$ means the almost sure convergence. Since $\lambda_{\min}(\cdot)$ is a continuous function, by the continuous mapping theorem, $\lambda_{\min}\left(\tfrac{1}{m}\F^\top \F\right) \xrightarrow{a.s.} \lambda_{\min}\left({\E}_{\X}[\bbf(\X)\bbf(\X)^\top]\right)$ as $m\to\infty$. Therefore, we can set $c_1=\tfrac{1}{2}\lambda_{\min}\left({\E}_{\X}[\bbf(\X)\bbf(\X)^\top]\right)$, and for any given constant $\xi\in (0,1)$, there exists a large integer $m_1=m_1(\xi)\in \NN$, such that for all $m\geq m_1$,
\begin{align}\label{numlow21}
	&\PP_{\X^m}\left(\left|\lambda_{\min}\left(\frac{1}{m}\F^\top \F\right) - \lambda_{\min}\left({\E}_{\X}[\bbf(\X)\bbf(\X)^\top]\right) \right| > \frac{1}{2}\lambda_{\min}\left({\E}_{\X}[\bbf(\X)\bbf(\X)^\top]\right)  \right) < \frac{\xi}{2} \nonumber\\
	\implies & \PP_{\X^m}\left( \lambda_{\min}\left(\frac{1}{m}\F^\top \F\right) < \frac{1}{2}\lambda_{\min}\left({\E}_{\X}[\bbf(\X)\bbf(\X)^\top]\right)  \right) < \frac{\xi}{2}.
\end{align}

On the other hand, we know that by Assumption A.1, $\bSigma_{\epsilon}(\X^m)\preceq \tfrac{\overline \sigma_0^2}{n}\I_m$. Moreover, using the monotone convergence theorem, the expectation and the variance of $\tr(\bSigma_M(\X^m,\X^m))$ can be controlled as follows:
\begin{align}
	& {\E}_{\X^m} \tr \left(\bSigma_{M}(\X^m,\X^m)\right) = {\E}_{\X^m} \left\{\sum_{j=1}^m \sum_{l=1}^{\infty} \mu_l \phi_l^2(\X_j)\right\} = \sum_{j=1}^m  \sum_{l=1}^{\infty} \mu_l {\E}_{\X^m}\left\{\phi_l^2(\X_j)\right\} \nonumber \\
	& = m \sum_{l=1}^{\infty} \mu_l = m\tr(\bSigma_M), \label{trbsigma_mean} \\
	& {\var}_{\X^m} \left\{ \tr \left(\bSigma_{M}(\X^m,\X^m)\right)\right\} =  {\var}_{\X^m} \left\{\sum_{j=1}^m \sum_{l=1}^{\infty} \mu_l \phi_l^2(\X_j)\right\} \nonumber \\
	&\stackrel{(i)}{=} \sum_{j=1}^m {\var}_{\X^m} \left\{ \sum_{l=1}^{\infty} \mu_l \phi_l^2(\X_j)\right\} \stackrel{(ii)}{\leq}  \sum_{j=1}^m {\E}_{\X^m} \left\{ \sum_{l=1}^{\infty} \mu_l \phi_l^2(\X_j)\right\}^2 \nonumber \\
	&\stackrel{(iii)}{=} \sum_{j=1}^m \sum_{a=1}^{\infty}\sum_{a=1}^{\infty} \mu_a\mu_b {\E}_{\X^m} \left\{ \phi_a^2(\X_j)\phi_b^2(\X_j) \right\} \nonumber \\
	&\stackrel{(iv)}{\leq}  \sum_{j=1}^m \sum_{a=1}^{\infty}\sum_{b=1}^{\infty} \mu_a\mu_b  \sqrt{{\E}_{\X^m} \phi_a^4(\X_j) \cdot {\E}_{\X^m} \phi_b^4(\X_j) } \nonumber\\
	&\stackrel{(v)}{\leq} \sum_{j=1}^m \sum_{a=1}^{\infty}\sum_{b=1}^{\infty} \mu_a\mu_b \rho_*^4 = m \rho_*^4 \left\{\tr(\bSigma_M)\right\}^2, \label{trbsigma_var}
\end{align}
where (i) follows from the independence between $\X_1,\ldots,\X_m$, (ii) follows from the inequality $\var(Z)\leq \E(Z^2)$ for any random variable $Z$, (iii) follows from the monotone convergence theorem, (iv) follows from the Cauchy-Schwarz inequality, and (v) follows from Assumption A.3. Now we set $c_2=1/\left[4\tr\left(\bSigma_M\right)\right]$, and $m_2=m_2(\xi)\equiv \max\left\{2c_2\tfrac{\overline\sigma_0^2}{n}, 2\rho_*^4/\xi\right\} = \max\left\{\tfrac{\overline\sigma_0^2}{2n\tr\left(\bSigma_M\right)}, 2\rho_*^4/\xi\right\}$. Then for all $m>m_2$, we have that
\begin{align}\label{numlow22}
	&\PP_{\X^m}\left(\lambda_{\max} \left(\bSigma_{M}(\X^m,\X^m)+\bSigma_{\epsilon}(\X^m)\right) > m/c_2 \right)  \leq \PP_{\X^m}\left(\lambda_{\max} \left(\bSigma_{M}(\X^m,\X^m)\right) + \frac{\overline\sigma_0^2}{n} > m/c_2 \right) \nonumber \\
	&\stackrel{(i)}{\leq} \PP_{\X^m}\left(\lambda_{\max} \left(\bSigma_{M}(\X^m,\X^m)\right) > \frac{m}{2c_2} \right)  \leq \PP_{\X^m}\left(\tr \left(\bSigma_{M}(\X^m,\X^m)\right) > \frac{m}{2c_2} \right) \nonumber \\
	&\leq \PP_{\X^m}\left(\big| \tr\left(\bSigma_{M}(\X^m,\X^m)\right) - \E \tr\left(\bSigma_{M}(\X^m,\X^m)\right) \big|  > \frac{m}{2c_2} - \E \tr\left(\bSigma_{M}(\X^m,\X^m)\right)\right)  \nonumber \\
	&\stackrel{(ii)}{=} \PP_{\X^m}\left(\big| \tr\left(\bSigma_{M}(\X^m,\X^m)\right) - \E \tr\left(\bSigma_{M}(\X^m,\X^m)\right) \big|  > m \tr\left(\bSigma_{M}\right)\right)  \nonumber \\
	&\stackrel{(iii)}{\leq} \frac{\var\left\{\tr\left(\bSigma_{M}(\X^m,\X^m)\right)\right\}}{m^2 \left\{\tr\left(\bSigma_{M}\right)\right\}^2 } \stackrel{(iv)}{\leq} \frac{\rho_*^4}{m} \stackrel{(v)}{<} \xi,
\end{align}
where (i) follows from the choice of $m_2$, (ii) follows from \eqref{trbsigma_mean} and the choice of $c_2$, (iii) follows from the Chebyshev's inequality, (iv) follows from \eqref{trbsigma_var}, and (v) follows from the choice of $m_2$ again.

We combine \eqref{numlow1}, \eqref{numlow21}, and \eqref{numlow22} to obtain that for any given $\xi\in (0,1)$, for all $m>m_0=m_0(\xi)\equiv \max\left\{m_1(\xi), m_2(\xi)\right\}$,
\begin{align} 
	& \PP_{\X^m}\left(\lambda_{\min}\left\{\F^\top \left(\bSigma_{M}(\X^m,\X^m)+\bSigma_{\epsilon}(\X^m)\right)^{-1}\F\right\} < \frac{\lambda_{\min}\left({\E}_{\X}[\bbf(\X)\bbf(\X)^\top]\right)}{8\tr(\bSigma_M)} \right) < \frac{\xi}{2} + \frac{\xi}{2} = \xi. \nonumber
\end{align}
Taking the probability of the complement leads to the conclusion. \hfill $\Box$

\vspace{3mm}

\begin{lemma}(\citet{Zhaetal15} Lemma 10)\label{zhang15d}
	Let $\bphi_l = \left[\phi_l(\X_1),\ldots,\phi_l(\X_m)\right]^\top$, for $l=1,2,\ldots$. For a given $\zeta\in \NN$, let $\bPhi = \left[\bphi_1, \ldots, \bphi_{\zeta} \right]$. Let $\M = \text{diag}\left(\mu_1,\ldots,\mu_{\zeta}\right)$ and let $\Q = \left(\I_{\zeta} + \frac{\overline \sigma_0^2}{mn}\M^{-1}\right)^{1/2}$ be the symmetric positive definite square root of $\I_{\zeta} + \frac{\overline \sigma_0^2}{mn}\M^{-1}$. For any given $\delta>0$, define the event
	$$\Ecal=\left\{\vertiii{\Q^{-1}\left( \frac{1}{m} \bPhi^\top \bPhi - \I_{\zeta} \right) \Q^{-1}} \leq \delta \right\}.$$
	Then under Assumptions A.1-A.3,

	\begin{align*}
		& \PP_{\X^m}(\Ecal^c) \leq \left\{ 100\rho_*^2\frac{ b(m,\zeta,r_*)  \gamma(\tfrac{\overline \sigma_0^2}{mn})}{\delta\sqrt{m}} \right\}^{r_*},
	\end{align*}
	where $b(m,\zeta,r_*)$ and $\gamma(\cdot)$ are defined in Theorem 1.
\end{lemma}

\vspace{8mm}

\noindent \textbf{Proof of Theorem 3}:

We use Theorems 1 and 2 to prove the results for three different types of kernels. The results of Theorems 1 and 2 will be applied to each of $k$ individual design first and then combined.

First note that by the Markov's inequality, Theorem 1 implies that for the $i$th design ($i=1,\ldots,k$),
\begin{align}\label{varbound11}
	& {\E}_{\X_0} \left[\mse_{i, \opt}^{(M)} (\X_0)\right] ~{\lesssim}_{\PP_{\X^m}} ~  \frac{2\overline \sigma_0^2}{mn} \gamma_i \left( \frac{\overline \sigma_0^2}{mn} \right) \nonumber \\
	& ~~ + \underset{\zeta \in \NN}{\inf} \,\left[ \left\{\frac{3mn}{\overline \sigma_0^2}\tr(\bSigma_{M,i})+1\right\}\tr\left(\bSigma_{M,i}^{(\zeta)}\right) + \tr(\bSigma_{M,i}) \left\{ 300 \rho_*^2 \frac{b(m,\zeta,r_*) \gamma_i(\tfrac{\overline \sigma_0^2}{mn})}{\sqrt{m}} \right\}^{r_*}  \right],
\end{align}
where $\gamma_i(a)=\sum_{l=1}^{\infty} \mu_{i,l}/(\mu_{i,l}+a)$ for any $a>0$. And similarly,
Theorem 2 implies that for the $i$th design ($i=1,\ldots,k$),
\begin{align}\label{varbound12}
	{\E}_{\X_0}\left[\mse_{i,\opt}^{(\bbeta)} (\X_0)\right] & {\lesssim}_{\PP_{\X^m}}  \frac{8q\tr(\bSigma_{M,i})}{\lambda_{\min}\left({\E}_{\X}[\bbf(\X)\bbf(\X)^\top]\right)}
	\Bigg\{8C_{\uf}^2 \frac{\overline \sigma_0^2}{mn} \nonumber \\
	&\quad + \inf_{\zeta \in \NN} \Bigg[8C_{\uf}^2 \frac{mn \overline \sigma_0^2}{ \underline \sigma_0^4} \rho_*^4 \tr\left(\bSigma_{M,i} \right) \tr\left(\bSigma_{M,i}^{(\zeta)}\right) \nonumber \\
	&\quad + C_{\uf}^2 \tr\left(\bSigma_{M,i}^{(\zeta)}\right)+ C_{\uf}^2  \tr\left(\bSigma_{M,i}\right)\left\{200\rho_*^2 \frac{ b(m,\zeta,r_*)  \gamma_i(\tfrac{\overline \sigma_0^2}{mn})}{\sqrt{m}} \right\}^{r_*} \Bigg]\Bigg\},
\end{align}

The subsequent proofs are based on evaluating the right-hand-sides of \eqref{varbound11} and \eqref{varbound12}. To obtain the upper bound for the maximum IMSE over $i=1,\ldots,k$, we notice that if for every $i=1,\ldots,k$, ${\E}_{\X_0} \left[\mse_{i,\opt}(\X_0)\right] {\lesssim}_{\PP_{\X^m}} a(m,n) $ for some sequence of $a(m,n)$ that does not depend on $i$, then $\max_{i\in\{1,\ldots,k\}}{\E}_{\X_0} \left[\mse_{i,\opt}(\X_0)\right] {\lesssim}_{\PP_{\X^m}} a(m,n)$.

Since we only care about the asymptotic orders of ${\E}_{\X_0} \left[\mse_{i,\opt}(\X_0)\right]$ in terms of $m$ and $n$, in the analysis below, we will use $C_1,C_2,\ldots$ to denote the constants whose values may vary from case to case but do not depend on $m$ and $n$.
\vspace{2mm}

\noindent (i) If the $i$th covariance kernel $\Sigma_{M,i}$ has finite rank $l_{*i}$, then for the $\inf_{\zeta \in \NN}$ terms in both \eqref{varbound11} and \eqref{varbound12}, we let $l_*=\max_{i\in\{1,\ldots,k\}} l_{*i}$ and choose $\zeta=l_*$, which leads to $\tr\left(\bSigma_{M,i}^{(\zeta)}\right)=0$ for all $i=1,\ldots,k$. Furthermore, since $r_*\geq 2$ in Assumption A.3, with $\zeta=l_*$,
\begin{align*}
	b(m,\zeta,r_*) &= \max \left( \sqrt{\max(r_*, \log \zeta)}, \frac{\max(r_*,\log \zeta)}{m^{1/2 - 1/r_*}} \right)\leq \max(r_*,\log l_*), \\
	\gamma_i\left(\frac{\overline \sigma_0^2}{mn}\right) &= \sum_{l=1}^{l_*} \frac{\mu_{i,l}}{\mu_{i,l}+\frac{\overline \sigma_0^2}{mn}} \leq l_*.
\end{align*}
\eqref{varbound11} and \eqref{varbound12} imply that for every $i=1,\ldots,k$,
\begin{align*}
	{\E}_{\X_0} \left[\mse_{i,\opt}^{(M)} (\X_0)\right] ~{\lesssim}_{\PP_{\X^m}} ~ & \frac{2l_*\overline \sigma_0^2}{mn} + \tr(\bSigma_{M,i}) \left\{ 300  \rho_*^2 \frac{l_* \max(r_*,\log l_*)}{\sqrt{m}} \right\}^{r_*} \\
	{\lesssim}_{\PP_{\X^m}} ~ & \frac{C_1}{mn} + \frac{C_2}{m^{r_*/2}}, \\
	{\E}_{\X_0}\left[\mse_{i,\opt}^{(\bbeta)} (\X_0)\right] ~{\lesssim}_{\PP_{\X^m}}~ & \frac{8q\tr(\bSigma_{M,i})}{\lambda_{\min}\left({\E}_{\X}[\bbf(\X)\bbf(\X)^\top]\right)}  \Bigg[8C_{\uf}^2 \frac{\overline \sigma_0^2}{mn} \\
	& + C_{\uf}^2  \tr\left(\bSigma_{M,i}\right)\left\{ 200\rho_*^2 \frac{ l_* \max(r_*,\log l_*) }{ \sqrt{m}} \right\}^{r_*} \Bigg] \\
	~{\lesssim}_{\PP_{\X^m}}~ & \frac{C_3}{mn} + \frac{C_4}{m^{r_*/2}},
\end{align*}
where $C_1,C_2,C_3,C_4$ are constants (note that $\max_{i\in\{1,\ldots,k\}} \tr\left(\bSigma_{M,i}\right)$ is also a finite constant by Assumption A.2).
Therefore,
\begin{align*}
	\max_{i\in\{1,\ldots,k\}} {\E}_{\X_0} \left[\mse_{i,\opt} (\X_0)\right] & \leq \max_{i\in\{1,\ldots,k\}} {\E}_{\X_0} \left[\mse_{i,\opt}^{(M)} (\X_0)\right] + \max_{i\in\{1,\ldots,k\}} {\E}_{\X_0}\left[\mse_{i,\opt}^{(\bbeta)} (\X_0)\right] \\
	& ~{\lesssim}_{\PP_{\X^m}}~ \frac{C_5}{mn} + \frac{C_6}{m^{r_*/2}} ~{\lesssim}_{\PP_{\X^m}} \max\left(\frac{1}{mn}, \frac{1}{m^{r_*/2}}\right).
\end{align*}
\vspace{4mm}

\noindent (ii) If the $i$th covariance kernel $\Sigma_{M,i}$ satisfies $\mu_{i,l} \leq c_{1i} \exp\left(-c_{2i} l^{\kappa_i/d} \right)$ for all $l \in \NN$, then for the $\inf_{\zeta \in \NN}$ terms in both \eqref{varbound11} and \eqref{varbound12}, we can choose $\zeta=(mn)^2$. Let $c_{1*}=\max_{i\in\{1,\ldots,k\}}c_{1i}$, $c_{2*}=\min_{i\in\{1,\ldots,k\}}c_{2i}$, and $\kappa_*=\min_{i\in\{1,\ldots,k\}}\kappa_{i}$. This definition implies that for any $z\geq 1$, $c_{1i} \exp\left(-c_{2i} z^{\kappa_i/d} \right)\leq c_{1*} \exp\left(-c_{2*} z^{\kappa_*/d} \right)$. Then for sufficiently large $m$,
\begin{align*}
	b(m,\zeta,r_*) &= \max \left\{ \sqrt{\max(r_*, \log \zeta)}, \frac{\max(r_*,\log \zeta)}{m^{1/2 - 1/r_*}} \right\} \\
	&=\max \left\{ \sqrt{\max(r_*, 2\log (mn))}, \frac{\max(r_*,2\log (mn))}{m^{1/2 - 1/r_*}} \right\}
	\leq 2\log (mn), \\
	\tr\left(\bSigma_{M,i}^{(\zeta)}\right) &=\sum_{l=(mn)^2+1}^{\infty} \mu_{i,l} \leq \sum_{l=(mn)^2+1}^{\infty} c_{1i} \exp\left(-c_{2i} l^{\kappa_i/d} \right)\\
	&\leq \int_{(mn)^2}^{\infty} c_{1i} \exp\left(-c_{2i} z^{\kappa_i/d} \right) dz \leq \int_{(mn)^2}^{\infty} c_{1*} \exp\left(-c_{2*} z^{\kappa_*/d} \right) dz\\
	&\stackrel{(i)}{\leq} \frac{c_{1*} d}{\kappa_*} \int_{(mn)^{2\kappa_*}}^{\infty} t^{\frac{d}{\kappa_*}-1} \exp\left(-c_{2*} t \right) dt,
\end{align*}
where in (i), we use the change of variable $t=z^{\kappa_*/d}$. If $\kappa_*/d \geq 1$, then since $t\geq (mn)^{2\kappa_*/d}\geq 1$, we have $t^{\frac{d}{\kappa_*}-1}\leq 1$. If $0<\kappa_*/d<1$, then there exists a large $m_0\in \NN$ that depends on only $c_{2*},\kappa_*,d$, such that for all $m \geq m_0$ and $t\geq (mn)^{2\kappa_*/d} \geq m^{2\kappa_*/d}$, we have $t^{\frac{d}{\kappa_*}-1}\leq \exp(c_{2*}t/2)$. Therefore, in all cases,
\begin{align} \label{trCd2}
	\tr\left(\bSigma_{M,i}^{(\zeta)}\right) &\leq  \frac{c_{1*}d}{\kappa} \int_{(mn)^{2\kappa_*/d}}^{\infty}  \exp\left(-c_{2*} t /2 \right) dt = \frac{2c_{1*}d}{c_{2*}\kappa_*} \exp\left\{ -c_{2*} (mn)^{2\kappa_*/d} /2 \right\}.
\end{align}
Let $l_1=\left\{\tfrac{2}{c_{2*}} \log (mn)\right\}^{d/\kappa_*}$. For sufficiently large $m$ and every $i=1,\ldots,k$, $\gamma_i \left(\frac{\overline \sigma_0^2}{mn}\right)$ can be bounded by
\begin{align*}
	\gamma_i\left(\frac{\overline \sigma_0^2}{mn}\right) &= \sum_{l=1}^{\infty} \frac{\mu_{i,l}}{\mu_{i,l}+\frac{\overline \sigma_0^2}{mn}} = \sum_{l=1}^{\lfloor l_1 \rfloor + 1} \frac{\mu_{i,l}}{\mu_{i,l} + \frac{\overline \sigma_0^2}{mn}} + \sum_{l=\lfloor l_1 \rfloor +2}^{\infty} \frac{\mu_{i,l}}{\mu_{i,l} + \frac{\overline \sigma_0^2}{mn}}  \\
	& \leq l_1  + 1 + \frac{mn}{\overline \sigma_0^2} \sum_{l=\lfloor l_1 \rfloor +1}^{\infty} c_{1i} \exp\left(-c_{2i} l^{\kappa_i/d} \right)   \\
	& \leq l_1  + 1  + \frac{mn}{\overline \sigma_0^2} \int_{l_1}^{\infty} c_{1*} \exp\left(-c_{2*} z^{\kappa_*/d} \right) dz    \\
	& = l_1 + 1  +  \frac{mn c_{1*}d}{\kappa_* \overline \sigma_0^2 } \int_{l_1^{\kappa_*/d}}^{\infty} t^{\frac{d}{\kappa_*}-1} \exp\left(-c_{2*} t  \right) dt  \\
	& \leq l_1 + 1  +  \frac{mn c_{1*}d}{\kappa_* \overline \sigma_0^2 } \int_{l_1^{\kappa_*/d}}^{\infty} \exp\left(-c_{2*} t/2  \right) dt  \\
	& = l_1 + 1  +  \frac{mn c_{1*}d}{2 c_{2*}\kappa_* \overline \sigma_0^2 } \exp \left( -c_{2*} l_1^{\kappa_*/d} /2 \right)  \\
	& = l_1 + 1  + \frac{c_{1*}d}{2 c_{2*}\kappa \overline \sigma_0^2 } \leq C_1 \log^{\frac{d}{\kappa_*}}(mn),
\end{align*}
for some constant $C_1>0$ that does not depend on $i$. Therefore, \eqref{varbound11} and \eqref{varbound12} imply that for every $i=1,\ldots,k$,
\begin{align*}
	{\E}_{\X_0} \left[\mse_{i,\opt}^{(M)} (\X_0)\right] ~{\lesssim}_{\PP_{\X^m}} ~ & \frac{2C_1\overline \sigma_0^2\log^{\frac{d}{\kappa_*}}(mn)}{mn} \\
	& + \left\{\frac{3mn}{\overline \sigma_0^2}\tr(\bSigma_{M,i})+1\right\}\frac{2c_{1*}d}{c_{2*}\kappa_*} \exp\left\{-\frac{c_{2*}}{2} (mn)^{2\kappa_*/d} \right\} \\
	&+ \tr(\bSigma_{M,i}) \left\{  300 \rho_*^2  \frac{2\log (mn)\cdot C_1\log^{\frac{d}{{\kappa_*}}}(mn)}{ \sqrt{m}} \right\}^{r_*} \\
	{\lesssim}_{\PP_{\X^m}} ~ & \frac{C_2\log^{\frac{d}{{\kappa_*}}}(mn)}{mn} + C_3 mn\exp\left\{-c_{2*} (mn)^{2\kappa_*/d} /2  \right\} + C_4 \frac{\log^{\frac{r_*({\kappa_*}+d)}{{\kappa_*}}}(mn)}{m^{r_*/2}}, \\
	{\E}_{\X_0}\left[\mse_{i,\opt}^{(\bbeta)} (\X_0)\right] ~{\lesssim}_{\PP_{\X^m}}~ & \frac{8q\tr(\bSigma_{M,i})}{\lambda_{\min}\left({\E}_{\X}[\bbf(\X)\bbf(\X)^\top]\right)}  \Bigg[8C_{\uf}^2 \frac{\overline \sigma_0^2}{mn} \\
	& + 8C_{\uf}^2 \frac{mn \overline \sigma_0^2}{ \underline \sigma_0^4} \rho_*^4 \tr\left(\bSigma_{M,i} \right)\frac{c_{1*}}{c_{2*}} \exp\left\{-c_{2*} (mn)^2 \right\}  \\
	& + C_{\uf}^2 \frac{2c_{1*}d}{c_{2*}\kappa_*} \exp\left\{-c_{2*} (mn)^{2\kappa_*/d} /2 \right\} \\
	& + C_{\uf}^2  \tr\left(\bSigma_{M,i}\right)\left\{ 200\rho_*^2 \frac{2\log (mn)\cdot C_1\log^{\frac{d}{{\kappa_*}}}(mn)}{ \sqrt{m}} \right\}^{r_*} \Bigg] \\
	~{\lesssim}_{\PP_{\X^m}}~ & \frac{C_5}{mn} + C_6 mn\exp\left\{-c_{2*} (mn)^{2\kappa_*/d} /2  \right\} +C_7\frac{\log^{\frac{r_*({\kappa_*}+d)}{{\kappa_*}}}(mn)}{m^{r_*/2}},
\end{align*}
for some positive constants $C_2,C_3,C_4,C_5,C_6,C_7$. Therefore,
\begin{align*}
	&\quad~	\max_{i\in \{1,\ldots,k\}} {\E}_{\X_0} \left[\mse_{i,\opt} (\X_0)\right] \\
	& \leq \max_{i\in \{1,\ldots,k\}}  {\E}_{\X_0} \left[\mse_{i,\opt}^{(M)} (\X_0)\right] + \max_{i\in \{1,\ldots,k\}} {\E}_{\X_0}\left[\mse_{i,\opt}^{(\bbeta)} (\X_0)\right] \\
	& ~{\lesssim}_{\PP_{\X^m}}~ \frac{C_2\log^{\frac{d}{{\kappa_*}}}(mn)}{mn} + C_3 mn\exp\left\{-c_{2*}(mn)^{2\kappa_*/d} /2  \right\} + C_4 \frac{\log^{\frac{r_*({\kappa_*}+d)}{{\kappa_*}}}(mn)}{m^{r_*/2}} \\
	& \qquad + \frac{C_5}{mn} + C_6 mn\exp\left\{-c_{2*} (mn)^{2\kappa_*/d} /2  \right\} +C_7\frac{\log^{\frac{r_*({\kappa_*}+d)}{{\kappa_*}}}(mn)}{m^{r_*/2}} \\
	& ~{\lesssim}_{\PP_{\X^m}} \max\left\{\frac{\log^{\frac{d}{{\kappa_*}}}(mn)}{mn}, \frac{\log^{\frac{r_*({\kappa_*}+d)}{{\kappa_*}}}(mn)}{m^{r_*/2}}\right\},
\end{align*}
where the last inequality follows because $\tfrac{mn\exp\left\{-c_{2*} (mn)^{2\kappa_*/d} /2 \right\}}{\log^{\frac{d}{{\kappa_*}}}(mn)/(mn)}\to 0$ and $\tfrac{1/(mn)}{\log^{\frac{d}{{\kappa_*}}}(mn)/(mn)}\to 0$ as $mn\to \infty$.
\vspace{6mm}

\noindent (iii) If the $i$th covariance kernel $\Sigma_{M,i}$ satisfies $\mu_{i,l} \leq c_{i} l^{-2\nu_i/d-1}$ for all $l\in\NN$, then $\mu_{i,l}\leq c_* l^{-2\nu_*/d-1}$ for all $l\in \NN$ and all $i=1,\ldots,k$, where $c_*=\max_{i\in\{1,\ldots,k\}} c_i$ and $\nu_*=\min_{i\in\{1,\ldots,k\}} \nu_i$. For the $\inf_{\zeta \in \NN}$ terms in both \eqref{varbound11} and \eqref{varbound12}, we choose $\zeta = \lfloor (mn)^{3d/(2\nu_*)} \rfloor$. Then for sufficiently large $m$,
\begin{align*}
	b(m,\zeta,r_*) &= \max \left\{ \sqrt{\max(r_*, \log \zeta)}, \frac{\max(r_*,\log \zeta)}{m^{1/2 - 1/r_*}} \right\} \\
	&=\max \left\{ \sqrt{\max\left(r_*, \frac{3d}{2\nu_*}\log (mn)\right)}, \frac{\max\left(r_*, \frac{3d}{2\nu_*}\log (mn)\right)}{m^{1/2 - 1/r_*}} \right\}
	\leq \frac{3d}{2\nu_*}\log (mn), \\
	\tr\left(\bSigma_{M,i}^{(\zeta)}\right) &= \sum_{l=\zeta+1}^{\infty} \mu_{i,l} \leq \sum_{l=\zeta+1}^{\infty} c_* l^{-2\nu_*/d-1} \leq \int_{\zeta}^{\infty} c_* z^{-2\nu_*/d-1} dz = \frac{c_*d}{2\nu_*} \zeta^{-2\nu_*/d} \leq \frac{c_*d}{2\nu_*} (mn)^{-3} , \\
	\gamma_i\left(\frac{\overline \sigma_0^2}{mn}\right) &= \sum_{l=1}^{\infty} \frac{1}{1+\frac{\overline \sigma_0^2}{mn\mu_{i,l}}} = \sum_{l=1}^{\infty} \frac{1}{1 + \frac{\overline \sigma_0^2 l^{2\nu_*/d+1}}{c_* mn}}
	\leq  (mn)^{d/(2\nu_*+d)} +1 + \sum_{l=\lfloor (mn)^{d/(2\nu_*+d)}\rfloor +2} \frac{c_*mn}{\overline \sigma_0^2 l^{2\nu_*/d+1}} \\
	&\leq (mn)^{d/(2\nu_*+d)} +1 +  \frac{c_*mn}{\overline \sigma_0^2} \int_{(mn)^{d/(2\nu_*+d)}}^{\infty} \frac{1}{z^{2\nu_*/d+1}} dz  \\
	&= (mn)^{d/(2\nu_*+d)}+1 +  \frac{c_*dmn}{ 2\nu_*\overline \sigma_0^2} (mn)^{-\frac{2\nu_*}{2\nu_*+d}}
	\leq C_1 (mn)^{d/(2\nu_*+d)},
\end{align*}
for some large constant $C_1>0$ that does not depend on $i$. Therefore, \eqref{varbound11} and \eqref{varbound12} imply that for every $i=1,\ldots,k$,
\begin{align*}
	{\E}_{\X_0} \left[\mse_{i,\opt}^{(M)} (\X_0)\right] ~{\lesssim}_{\PP_{\X^m}} ~ & \frac{2C_1\overline \sigma_0^2(mn)^{d/(2\nu_*+d)}}{mn} + \left\{\frac{3mn}{\overline \sigma_0^2}\tr(\bSigma_{M,i})+1\right\} \frac{c_*d}{2\nu_*} (mn)^{-3}  \\
	&+ \tr(\bSigma_{M,i}) \left\{ 300\rho_*^2  \frac{\frac{3d}{2\nu_*}\log (mn) \cdot C_1 (mn)^{d/(2\nu_*+d)}}{\sqrt{m}} \right\}^{r_*} \\
	{\lesssim}_{\PP_{\X^m}} ~ & C_2 (mn)^{-\frac{2\nu_*}{2\nu_*+d}} + C_3 (mn)^{-2} + C_4 \frac{n^{\frac{dr_*}{2\nu_*+d}}\log^{r_*}(mn)}{m^{\frac{r_*(2\nu_*-d)}{2(2\nu_*+d)}}}, \\
	{\E}_{\X_0}\left[\mse_{i,\opt}^{(\bbeta)} (\X_0)\right] ~{\lesssim}_{\PP_{\X^m}}~ & \frac{8q\tr(\bSigma_{M,i})}{\lambda_{\min}\left({\E}_{\X}[\bbf(\X)\bbf(\X)^\top]\right)}  \Bigg[8C_{\uf}^2 \frac{\overline \sigma_0^2}{mn} \\
	& + 8C_{\uf}^2 \frac{mn \overline \sigma_0^2}{ \underline \sigma_0^4} \rho_*^4 \tr\left(\bSigma_{M,i} \right) \frac{c_*d}{2\nu_*} (mn)^{-3}  + C_{\uf}^2 \frac{c_*d}{2\nu_*} (mn)^{-3} \\
	& + C_{\uf}^2  \tr\left(\bSigma_{M,i}\right)\left\{ 200\rho_*^2 \frac{\frac{3d}{2\nu_*}\log (mn) \cdot C_1 (mn)^{d/(2\nu_*+d)}}{ \sqrt{m}} \right\}^{r_*} \Bigg] \\
	~{\lesssim}_{\PP_{\X^m}}~ & \frac{C_5}{mn} + C_6 (mn)^{-2} + C_7 \frac{n^{\frac{dr_*}{2\nu_*+d}} \log^{r_*}(mn)}{m^{\frac{r_*(2\nu_*-d)}{2\nu_*+d}}},
\end{align*}
for some positive constants $C_2,C_3,C_4,C_5,C_6,C_7$. Therefore,
\begin{align*}
	\max_{i\in \{1,\ldots,k\}} {\E}_{\X_0} \left[\mse_{i,\opt} (\X_0)\right] & \leq  \max_{i\in \{1,\ldots,k\}} {\E}_{\X_0} \left[\mse_{i,\opt}^{(M)} (\X_0)\right] + \max_{i\in \{1,\ldots,k\}} {\E}_{\X_0}\left[\mse_{i,\opt}^{(\bbeta)} (\X_0)\right] \\
	& ~{\lesssim}_{\PP_{\X^m}}~ C_2 (mn)^{-\frac{2\nu_*}{2\nu_*+d}} + C_3 (mn)^{-2} + C_4 \frac{n^{\frac{dr_*}{2\nu_*+d}}\log^{r_*}(mn)}{m^{\frac{r_*(2\nu_*-d)}{2\nu_*+d}}} \\
	& \qquad + \frac{C_5}{mn} + C_6 (mn)^{-2} + C_7 \frac{n^{\frac{dr_*}{2\nu+d}}\log^{r_*}(mn)}{m^{\frac{r_*(2\nu_*-d)}{2\nu_*+d}}} \\
	& ~{\lesssim}_{\PP_{\X^m}} \max\left\{\frac{1}{(mn)^{\frac{2\nu_*}{2\nu_*+d}}}, \frac{n^{\frac{dr_*}{2\nu_*+d}}\log^{r_*}(mn)}{m^{\frac{r_*(2\nu_*-d)}{2\nu_*+d}}} \right\},
\end{align*}
where the last inequality follows because $\tfrac{1/(mn)}{(mn)^{-\frac{2\nu_*}{2\nu_*+d}}} = (mn)^{-d/(2\nu_*+d)}\to 0$ and $\tfrac{1/(mn)^2}{(mn)^{-\frac{2\nu_*}{2\nu_*+d}}} = (mn)^{-d/(2\nu_*+d)-1}\to 0$ as $mn\to \infty$.  \hfill $\Box$

\noindent \textbf{Proof of Theorem 4}:

For $i=1,\ldots,k$, let $\widetilde \X_{i,0}=\big(\sqrt{b_i}, \X_0^\top\big)^\top$ be the $\RR^{d+1}$ random vector version of $\widetilde \x_{i,0}$ with $\X_0$ following the distribution $\PP_{\X}$. For the covariance kernel $\bSigma_{M,i}(\x,\x')=a_i\left(\x^\top \x'+b_i\right)$, using the definition of $\widetilde \x_{i,0}$ and $\Z_i$ in Theorem 4, we have that
\begin{align*}
	\bSigma_{M,i}(\X^m,\X_0) & = \left(\bSigma_{M,i}(\X_1,\X_0), \ldots, \bSigma_{M,i}(\X_m,\X_0)\right)^\top \\
	&= \left(a_i(\X_1^\top \X_0 + b_i), \ldots, a_i(\X_m^\top \X_0 + b_i) \right)^\top = a_i \Z_i \widetilde \X_0, \\
	\bSigma_{M,i}(\X^m,\X^m) & = \left(
	\begin{array}{ccc}
		\bSigma_{M,i}(\X_1,\X_1) & \ldots & \bSigma_{M,i}(\X_1,\X_m) \\
		& \ldots & \\
		\bSigma_{M,i}(\X_m,\X_1) & \ldots & \bSigma_{M,i}(\X_m,\X_m)
	\end{array}
	\right) \\
	& =
	\left(
	\begin{array}{ccc}
		a_i\big(\X_1^\top \X_1 + b_i\big) & \ldots & a_i\big(\X_1^\top  \X_m + b_i\big) \\
		& \ldots & \\
		a_i\big(\X_m^\top \X_1 + b_i\big) & \ldots & a_i\big(\X_m^\top \X_m + b_i\big)
	\end{array}
	\right) \\
	& = a_i \Z_i \Z_i^\top.
\end{align*}
Therefore, we plug in $\bbf_i(\X)\equiv 0$ to (2) of the manuscript and obtain that
\begin{align*}
	\widehat y_i(\X_0) & = \bSigma_{M,i} (\X^m,\X_0)^\top \left[\bSigma_{M,i}(\X^m,\X^m)+ \frac{\sigma^2}{n}\I_m \right]^{-1} \overline Y_i \\
	& = a_i \widetilde \X_{i,0}^{\top} \Z_i^\top \left(a_i\Z_i \Z_i^\top + \frac{\sigma^2}{n}\I_m \right)^{-1} \overline Y_i.
\end{align*}
Similarly we obtain from (3) of the manuscript that
\begin{align*}
	\mse_{i,\opt}(\X_0) & = \bSigma_{M,i}(\X_0,\X_0) - \bSigma_{M,i}^\top(\X^m,\X_0) \left[\bSigma_{M,i}(\X^m,\X^m) + \frac{\sigma^2}{n}\I_m \right]^{-1} \bSigma_{M,i}(\X^m,\X_0) \\
	& = a_i \widetilde \X_{i,0}^{\top} \widetilde \X_{i,0} - a_i\widetilde \X_{i,0}^{\top} \Z_i^\top \left(a_i\Z_i \Z_i^\top + \frac{\sigma^2}{n}\I_m \right)^{-1} a_i\Z_i \widetilde \X_{i,0} \\
	& = a_i\widetilde \X_{i,0}^{\top} \left[\I_{d+1} -\Z_i^\top \left(\Z_i \Z_i^\top + \frac{\sigma^2}{a_i n}\I_m \right)^{-1} \Z_i \right]\widetilde \X_{i,0} \\
	& \stackrel{(i)}{=} a_i\widetilde \X_{i,0}^{\top} \left(\I_{d+1} + \frac{a_i n}{\sigma^2}\Z_i^\top \Z_i\right)^{-1} \widetilde \X_{i,0},
\end{align*}
where we have applied the Woodbury matrix inversion formula (Rasmussen and Williams \citeyear{RasWil06}, Appendix A.3) in the step (i). This has proved (12) of the main text.

Now we turn to (13) of the manuscript. Note that
\begin{align} \label{mserank11}
	{\E}_{\X_0} \left[\mse_{i,\opt}(\X_0)\right] &= {\E}_{\X_0} \left[ a_i\widetilde \X_{i,0}^{\top} \left(\I_{d+1} + \frac{a_in}{\sigma^2}\Z_i^\top \Z_i\right)^{-1} \widetilde \X_{i,0}  \right] \nonumber \\
	&= \tr\left\{ \left(\I_{d+1} + \frac{a_i n}{\sigma^2}\Z_i^\top \Z_i\right)^{-1} \cdot a_i {\E}_{\X_0} \left(\widetilde \X_{i,0} \widetilde \X_{i,0}^{\top}\right) \right\}.
\end{align}
According to the definition of $\Z_i$ and the fact that $\X^m,\X_0$ are i.i.d. draws from $\PP_{\X}$, by the strong law of large numbers, as $m\to\infty$, almost surely in $\PP_{\X^m}$,
\begin{align}\label{zzlimit}
	\frac{1}{m} \Z_i^\top \Z_i & = \left(
	\begin{array}{cc}
		b_i & \frac{\sqrt{b_i}}{m}\sum_{j=1}^m \X_j^\top \nonumber \\
		\frac{\sqrt{b_i}}{m}\sum_{j=1}^m \X_j & \frac{1}{m}\sum_{j=1}^m \X_j \X_j^\top
	\end{array}
	\right) \\
	&\rightarrow
	\left(
	\begin{array}{cc}
		b_i & \sqrt{b_i} {\E}_{\X_1} (\X_1^\top) \\
		\sqrt{b_i}{\E}_{\X_1} (\X_1) & {\E}_{\X_1} (\X_1 \X_1^\top)
	\end{array}
	\right) = {\E}_{\X_0} \left(\widetilde \X_{i,0} \widetilde \X_{i,0}^{\top}\right) .
\end{align}

Therefore, \eqref{mserank11} and \eqref{zzlimit} together imply that for each $i=1,\ldots,k$, as $m\to\infty$, almost surely in $\PP_{\X^m}$,
\begin{align} \label{eq:i.conv}
	mn \cdot {\E}_{\X_0} \left[\mse_{i,\opt}(\X_0)\right] & = \tr\left\{ \left(\frac{1}{mn}\I_{d+1} + \frac{a_i}{\sigma^2}\cdot \frac{1}{m}\Z_i^\top \Z_i\right)^{-1} \cdot a_i{\E}_{\X_0} \left(\widetilde \X_{i,0} \widetilde \X_{i,0}^{\top}\right) \right\} \nonumber \\
	&\rightarrow \tr\left\{ \left[\frac{a_i}{\sigma^2}\cdot {\E}_{\X_0} \left(\widetilde \X_{i,0} \widetilde \X_{i,0}^{\top}\right) \right]^{-1} \cdot a_i {\E}_{\X_0} \left(\widetilde \X_{i,0} \widetilde \X_{i,0}^{\top}\right) \right\} \nonumber  \\
	& = \tr\left(\sigma^2 \I_{d+1}\right) = (d+1)\sigma^2 .
\end{align}
Define the event $\Acal_i=\left\{\text{The convergence in \eqref{eq:i.conv} happens as }n\to\infty\right\}$ for $i=1,\ldots,k$. Then the almost sure convergence in \eqref{eq:i.conv} implies $\PP_{\X^m}(\Acal_i)=1$ for every $i=1,\ldots,k$. This further implies that
$$\PP_{\X^m}\left(\cap_{i=1}^k \Acal_i\right) = 1 - \PP_{\X^m}\left(\cup_{i=1}^k \Acal_i^c\right) \geq 1 - \sum_{i=1}^k \PP_{\X^m}\left(\Acal_i^c\right) = 1 - \sum_{i=1}^k 0 = 1,$$
which implies that $\PP_{\X^m}\left(\cap_{i=1}^k \Acal_i\right)=1$, i.e. the convergence in \eqref{eq:i.conv} happens jointly over $i=1,\ldots,k$ as $m\to\infty$ almost surely in $\PP_{\X^m}$. Therefore, on the event $\cap_{i=1}^k \Acal_i$, \eqref{eq:i.conv} implies that $mn \cdot \max_{i\in \{1,\ldots,k\}}{\E}_{\X_0} \left[\mse_{i,\opt}(\X_0)\right]\rightarrow (d+1)\sigma^2$ as $m\to\infty$. This has proved (13) of the main text. \hfill $\Box$

\vspace{5mm}

\noindent \textbf{Proof of Theorem 5}:

We first derive a natural bound for $\pfs(\x_0)$. For any $\x_0\in \Xcal$, any $i,i'\in \{1,\ldots,k\}$, define the random variable $W_{i,i'}(\x_0)=[\widehat y_i(\x_0)- y_i(\x_0)]-[\widehat y_{i'}(\x_0)- y_{i'}(\x_0)]$ (so $W_{i,i}(\x_0)=0$). According to our definition in (4) of the manuscript, $y^\circ(\x_0) = y_{i^\circ(\x_0)}(\x_0)$. Therefore,
\begin{align}\label{pfsbound1}
	&\pfs(\x_0) = \PP_{\epsilon}\left(y_{\widehat i^\circ(\x_0)} (\x_0)-y^\circ(\x_0) \geq \delta_0\right) \nonumber \\
	&= \PP_{\epsilon}\left\{\big[\widehat y_{i^\circ(\x_0)}(\x_0) - y_{i^\circ(\x_0)}(\x_0)\big] - \big[\widehat y^\circ(\x_0) - y_{\widehat i^\circ (\x_0)}(\x_0)\big]  \geq \delta_0 + \big[\widehat y_{ i^\circ(\x_0)}(\x_0)-\widehat y^\circ(\x_0)\big]\right\} \nonumber \\
	&\stackrel{(i)}{\leq} \PP_{\epsilon}\left(W_{i^\circ(\x_0), \widehat i^\circ(\x_0)} (\x_0) \geq \delta_0\right)
	\leq \PP_{\epsilon} \left(\max_{1\leq i,i'\leq k} W_{i, i'} (\x_0) \geq \delta_0\right) \nonumber \\
	&\leq \sum_{1\leq i,i'\leq k} \PP_{\epsilon} \left( W_{i, i'} (\x_0) \geq \delta_0\right) =  \sum_{1\leq i< i'\leq k} \PP_{\epsilon} \left( \left| W_{i, i'} (\x_0)\right| \geq \delta_0\right),
\end{align}
where $\widehat y^\circ(\x_0)=\min_{i\in\{1,2,...,k\}} \widehat y_i (\x_0)$. Inequality (i) holds because $\widehat i^\circ(\x_0)=\arg\min_{i\in \{1,\ldots,k\}} \widehat y_i(\x_0)$ and hence $\widehat y_{ i^\circ(\x_0)}(\x_0)\geq \widehat y^\circ(\x_0)$. Now since $y_i(\x)=\bbf_i(\x)^\top \bbeta_i + M_i(\x)$ in (1) of the manuscript includes $M_i(\x)$, it is clear that $W_{i,i'}$ depends on $M_i(\cdot)$, $M_{i'}(\cdot)$, $\X^m$ and $\x_0$, which are all random. We first remove the randomness from $M_i(\x)$'s ($i=1,\ldots,k$) by taking the expectation of $\pfs(\x_0)$ with respect to the joint Gaussian measure $\PP_M$ induced by the $k$ independent Gaussian processes with mean zero and covariance function $\bSigma_{M,i}(\cdot,\cdot)$ for $i=1,\ldots,k$. Then from \eqref{pfsbound1} we can obtain that
\begin{align}\label{pfsbound2}
	&{\E}_M \left[\pfs(\x_0)\right] \leq {\E}_M \Big[\sum_{1\leq i< i'\leq k} \PP_{\epsilon} \left( \left| W_{i, i'} (\x_0)\right| \geq \delta_0\right) \Big] \nonumber \\
	&= \sum_{1\leq i< i'\leq k} {\E}_M {\E}_{\epsilon} \left[\1\left\{\left| W_{i, i'} (\x_0)\right| \geq \delta_0 \right\}\right] = \sum_{1\leq i< i'\leq k} \PP_{M,\epsilon} \left(\left| W_{i, i'} (\x_0)\right| \geq \delta_0 \right),
\end{align}
where $\PP_{M,\epsilon}$ denotes the joint (independent) probability measure of all $M_i(\cdot)$'s from Gaussian processes and the error terms. The inequality of \eqref{pfsbound2} allows us to directly consider all randomness in $W_{i,i'}$'s given fixed $\X^m$ and $\X_0$.

Let $M_i(\X^m)=(M_i(\X_1),\ldots,M_i(\X_m))^\top$ and $\overline \epsilon(\X^m) = (\overline \epsilon_i(\X_1),\ldots,\overline \epsilon_i(\X_m))^\top$, for $i=1,\ldots,k$. Under the joint measure $\PP_{M,\epsilon}$ (with expectation ${\E}_{M,\epsilon}$), based on (2) of the manuscript, we have that for any given $\X^m$ and $\x_0\in \Xcal$,
\begin{align}
	&{\E}_{M,\epsilon} (\overline \Y_i) = {\E}_{M,\epsilon} \left[\F_i \bbeta_i + M_i(\X^m) + \overline \epsilon(\X^m)\right] =\F_i \bbeta_i, \nonumber \\
	&{\E}_{M,\epsilon} \left[\widehat y_i(\x_0) - y_i(\x_0)\right] \nonumber \\
	&= {\E}_{M,\epsilon} \left[\bbf_i(\x_0)^\top \widehat\bbeta_{i} + \bSigma_{M,i}(\X^m,\x_0) ^\top \bSigma_{y,i}^{-1} \left(\overline \Y_i - \F_i \widehat\bbeta_{i}\right) - \bbf_i(\x_0)^\top \bbeta_{i} - M_i(\x_0) \right] \nonumber \\
	&= \bbf_i(\x_0)^\top \left(\F_i^\top \bSigma_{y,i}^{-1} \F_i\right)^{-1}\F_i^\top \bSigma_{y,i}^{-1} \F_i \bbeta_i +   \bSigma_{M,i}(\X^m,\x_0) ^\top \bSigma_{y,i}^{-1} \F_i \bbeta_i   \nonumber \\
	&~ - \bSigma_{M,i}(\X^m,\x_0) ^\top \bSigma_{y,i}^{-1} \F_i \left(\F_i^\top \bSigma_{y,i}^{-1} \F_i\right)^{-1}\F_i^\top \bSigma_{y,i}^{-1} \F_i \bbeta_i - \bbf_i(\x_0)^\top \bbeta_{i} \nonumber \\
	&=0. \nonumber
\end{align}
Hence ${\E}_{M,\epsilon} (W_{i,i'})=0$ for all $1\leq i<i'\leq k$. Furthermore, the variance of $\widehat y_i(\x_0) - y_i(\x_0)$ is ${\var}_{M,\epsilon}[\widehat y_i(\x_0) - y_i(\x_0)] ={\E}_{M,\epsilon}[\widehat y_i(\x_0) - y_i(\x_0)]^2$, which is the MSE of $\widehat y_i(\x_0)$ and hence is equal to $\mse_{i,\opt}(\x_0)$ given in (3) of the manuscript. For $W_{i,i'}$ ($1\leq i<i'\leq k$), the independence between different $M_i(\cdot)$'s and errors implies that
\begin{align*}
	{\var}_{M,\epsilon}(W_{i,i'})&= {\var}_{M,\epsilon}[\widehat y_i(\x_0) - y_i(\x_0)] + {\var}_{M,\epsilon}[\widehat y_{i'}(\x_0) - y_{i'}(\x_0)] \\
	& = \mse_{i,\opt}(\x_0) + \mse_{i',\opt}(\x_0).
\end{align*}

From \eqref{pfsbound2}, we apply the Markov's inequality and obtain that
\begin{align} \label{Wbound1}
	& {\E}_{\X_0} {\E}_M \left[\pfs(\X_0)\right]  \leq \sum_{1\leq i< i'\leq k} {\E}_{\X_0} \left[ \PP_{M,\epsilon} \left(\left| W_{i, i'} (\X_0)\right| \geq \delta_0 \right)\right] \nonumber \\
	&\leq \sum_{1\leq i< i'\leq k} {\E}_{\X_0} \left[ \frac{{\E}_{M,\epsilon} \left| W_{i, i'} (\X_0)\right|^2 }{\delta_0^2} \right] = \sum_{1\leq i< i'\leq k} {\E}_{\X_0} \left[ \frac{ \mse_{i,\opt}(\X_0) + \mse_{i',\opt}(\X_0)}{\delta_0^2} \right] \nonumber \\
	&\leq \frac{k(k-1)}{\delta_0^2} \max_{i\in \{1,\ldots,k\}} {\E}_{\X_0} \left[\mse_{i,\opt}(\X_0)\right]
\end{align}

We now prove Part (i) of Theorem 5. Under Assumptions A.1-A.4, Part (i) of Theorem 3 says that $\max_{i\in\{1,2,...,k\}}{\E}_{\X_0}\left[\mse_{i,\opt}(\X_0)\right] {\lesssim}_{\PP_{\X^m}} R(m,n)$ as $m\to \infty$. This is to say that for any $\xi\in (0,1/2)$, there exist $m_0\geq 1$ and $c_1>0$ that depends on $\xi$, such that for all $m\geq m_0$,
\begin{align}\label{Rbound1}
	& \PP_{\X^m}\left(\max_{i\in\{1,...,k\}}{\E}_{\X_0}\left[\mse_{i,\opt}(\X_0)\right] \leq c_1 R(m,n) \right) \geq 1-\xi.
\end{align}
\eqref{Wbound1} and \eqref{Rbound1} together implies that
\begin{align}\label{Rbound2}
	& \PP_{\X^m}\left({\E}_{\X_0} {\E}_M \left[\pfs(\X_0)\right] \leq \frac{c_1 k(k-1)}{\delta_0^2}  R(m,n) \right) \nonumber \\
	\geq{} & \PP_{\X^m}\left(\max_{i\in\{1,...,k\}}{\E}_{\X_0}\left[\mse_{i,\opt}(\X_0)\right] \leq c_1 R(m,n) \right) \geq 1-\xi.
\end{align}
This is to say that for any $\xi\in (0,1/2)$, there exist $m_0\geq 1$ and $c_1>0$ that depends on $\xi$, such that for all $m\geq m_0$, the relation \eqref{Rbound2} holds. In other words, we have proved that ${\E}_{\X_0} {\E}_M \left[\pfs(\X_0)\right] {\lesssim}_{\PP_{\X^m}}  R(m,n)$.
\vspace{2mm}

Next we prove Part (ii) of Theorem 5 with the additional Assumptions A.5 and A.6. The simulation errors $\epsilon_{il}(\x)$'s are all normally distributed by Assumption A.5. Also $M_i(\x)$'s are normally distributed due to the Gaussian process model. Hence we know that for given $\X^m$ and $\x_0$, $\widehat y_i(\x_0) - y_i(\x_0)$ as a linear function of $\overline \Y_i$ and $y_i(\x_0)$, is normally distributed as $N(0,\mse_{i,\opt}(\x_0))$. The independence of $\widehat y_i(\x_0) - y_i(\x_0)$ and $\widehat y_{i'}(\x_0) - y_{i'}(\x_0)$ for $1\leq i<i'\leq k$ further implies that for given $\X^m$ and $\x_0$, ${\var}_{M,\epsilon}(W_{i,i'}) = \mse_{i,\opt}(\x_0) + \mse_{i',\opt}(\x_0)$ and thus $W_{i,i'}\sim N(0, \mse_{i,\opt}(\x_0) + \mse_{i',\opt}(\x_0))$. We can apply the tail probability bound of normal distributions ($\PP(|Z|>z)\leq \exp(-z^2/2)$ if $Z\sim N(0,1)$ and $z>0$) and obtain that
\begin{align}\label{normaltail1}
	\PP_{M,\epsilon} \left(\left| W_{i, i'} (\x_0)\right| \geq \delta_0\right) & \leq \exp\left(-\frac{\delta_0^2}{2\left[\mse_{i,\opt}(\x_0) + \mse_{i',\opt}(\x_0)\right]}\right).
\end{align}
\eqref{pfsbound2} and \eqref{normaltail1} together imply that
\begin{align}\label{normaltail2}
	& {\E}_M \left[\pfs(\x_0)\right] \leq \sum_{1\leq i< i'\leq k} \exp\left(-\frac{\delta_0^2}{2\left[\mse_{i,\opt}(\x_0) + \mse_{i',\opt}(\x_0)\right]}\right) \nonumber \\
	&\leq \frac{k(k-1)}{2} \exp\left(-\frac{\delta_0^2}{4 \max_{i\in \{1,\ldots,k\}}\mse_{i,\opt}(\x_0) }\right).
\end{align}
For abbreviation, we let $V=\max_{i\in \{1,\ldots,k\}}\mse_{i,\opt}(\x_0)$. Assumption A.6 says that for any given $\xi\in(0,1/2)$, there exist constants $w_1>0,w_2>0,m_0\geq 1$ that depend on $\xi$, such that for $m\geq m_0$, for any $t>0$, we have $\PP_{\X^m}(\Ecal_4)\geq 1-\xi$, where $\Ecal_4$ is defined as
$$\Ecal_4=\Big\{\PP_{\X_0}\left(V\geq tR(m,n)\right) \leq w_1 \exp\left(-w_2 t \right)\Big\}.$$
Conditional on the event $\Ecal_4$, from \eqref{normaltail2}, we can derive that
\begin{align} \label{pfsbound3}
	& {\E}_{\X_0} {\E}_M \left[\pfs(\X_0)\right] \leq {\E}_{\X_0} \left[\frac{k(k-1)}{2} \exp\left(-\frac{\delta_0^2}{4V}\right)\right] \nonumber \\
	& \stackrel{(i)}{=} \frac{k(k-1)}{2} \int_0^{+\infty} \PP_{\X_0}\left\{\exp\left(-\frac{\delta_0^2}{4V}\right) > u\right\} \ud u \nonumber \\
	& = \frac{k(k-1)}{2} \int_0^{+\infty} \PP_{\X_0}\left\{ V > \frac{\delta_0^2}{-4\log u}\right\} \ud u \nonumber \\
	& \stackrel{(ii)}{\leq} \frac{k(k-1)}{2} \int_0^{+\infty} w_1 \exp\left\{-w_2 \left(\frac{\delta_0^2}{-4R(m,n)\log u}\right)\right\} \ud u \nonumber \\
	& \stackrel{(iii)}{\leq} \frac{w_1k(k-1)}{2} \int_0^{+\infty}  \exp\left\{- v - \left(\frac{w_2\delta_0^2}{4R(m,n)}\right) \frac{1}{v}\right\} \ud v, \nonumber \\
	& \stackrel{(iv)}{=} \frac{w_1k(k-1)}{2} \cdot \sqrt{\frac{w_2\delta_0^2}{R(m,n)}} \cdot K_1\left(\sqrt{\frac{w_2\delta_0^2}{R(m,n)}}\right),
\end{align}
where (i) uses the relation $\E(Z)=\int_0^\infty P(Z>t)\ud t$ for any nonnegative random variable $Z$, (ii) follows from Assumption A.6 and the relation on the event $\Ecal_4$, and (iii) uses a change of variable $v=-\log u$ in the integral. (iv) follows because the integral in \eqref{pfsbound3} can be recognized as the density of a generalized inverse Gaussian distribution without normalizing constant, and here $K_1(\cdot)$ is the modified Bessel function of the second kind with parameter $1$.

Theorem 2.13 of \citet{Kre12} has shown that
$$\lim_{x\to+\infty} \frac{K_1(x)}{\sqrt{\frac{\pi}{2x}}e^{-x}} =1,$$
which implies that there exists a constant $x_0>0$, such that for all $x>x_0$, $K_1(x) < 2\sqrt{\frac{\pi}{2x}}e^{-x} = \sqrt{\frac{2\pi}{x}}e^{-x}$. Since $R(m,n)\to 0$ for fixed $n$ as $m\to\infty$, we can take $m\geq m_1$ for some large integer $m_1\geq m_0$ such that $\sqrt{\frac{w_2\delta_0^2}{R(m,n)}} > x_0$ and meanwhile
$$\left[R(m,n)\right]^{-1/4} \leq \exp\left\{\frac{1}{2}w_2^{1/2}\delta_0\left[R(m,n)\right]^{-1/2}\right\}.$$
As a result, we can derive from \eqref{pfsbound3} that on the event $\Ecal_4$, for all $m>m_1$,
\begin{align}\label{pfsbound4}
	& {\E}_{\X_0} {\E}_M \left[\pfs(\X_0)\right] \leq \frac{w_1k(k-1)}{2} \cdot \sqrt{\frac{w_2\delta_0^2}{R(m,n)}} \cdot \sqrt{\frac{2\pi}{ \sqrt{\frac{w_2\delta_0^2}{R(m,n)}}}}\exp\left\{- \sqrt{\frac{w_2\delta_0^2}{R(m,n)}}\right\} \nonumber \\
	&\leq \sqrt{\frac{\pi}{2}} w_1w_2^{1/4}k(k-1)\delta_0^{1/2} \left[R(m,n)\right]^{-1/4} \exp\left\{- w_2^{1/2}\delta_0\left[R(m,n)\right]^{-1/2}\right\} \nonumber \\
	&\leq \sqrt{\frac{\pi}{2}} w_1w_2^{1/4}k(k-1)\delta_0^{1/2} \exp\left\{-\frac{1}{2}w_2^{1/2}\delta_0\left[R(m,n)\right]^{-1/2}\right\}.
\end{align}
Thus, ${\E}_{\X_0} {\E}_M \left[\pfs(\X_0)\right] {\lesssim}_{\PP_{\X^m}} \exp\left\{-\frac{1}{2}w_2^{1/2}\delta_0\left[R(m,n)\right]^{-1/2}\right\}$ with probability at least $1-\xi$ for all $m\geq m_1$, which has proved Part (ii) of Theorem 5.
\vspace{2mm}

Finally, we prove Part (iii) of Theorem 5 with the additional Assumptions A.5 and A.7. Similar to the derivation of Part (ii), we define the quantity $\widetilde V =\max_{i\in \{1,\ldots,k\}}\sup_{\x_0\in \Xcal}\mse_{i,\opt}(\x_0) $ for abbreviation. Then Assumption A.7 says that
for any given $\xi\in(0,1/2)$, there exist constants $w_3>0,m_0\geq 1$ that depend on $\xi$, such that for $m\geq m_0$, for any $t>0$, we have $\PP_{\X^m}(\Ecal_5)\geq 1-\xi$, where $\Ecal_5$ is defined as
$\Ecal_5 = \big\{\widetilde V \leq w_3 R(m,n)\big\}$. Therefore, from \eqref{normaltail2}, we can derive that on the event $\Ecal_5$, for all $m\geq m_0$,
\begin{align*}
	& {\E}_{\X_0}{\E}_M \left[\pfs(\X_0)\right] \leq \frac{k(k-1)}{2} {\E}_{\X_0}\exp\left(-\frac{\delta_0^2}{4 \max_{i\in \{1,\ldots,k\}}\mse_{i,\opt}(\X_0) }\right) \\
	&\leq \frac{k(k-1)}{2} \sup_{\x_0\in \Xcal}\exp\left(-\frac{\delta_0^2}{4 \max_{i\in \{1,\ldots,k\}}\mse_{i,\opt}(\x_0) }\right) \\
	&= \frac{k(k-1)}{2} \exp\left(-\frac{\delta_0^2}{4 \max_{i\in \{1,\ldots,k\}} \sup_{\x_0\in \Xcal} \mse_{i,\opt}(\x_0) }\right) \\
	&= \frac{k(k-1)}{2} \exp\left(-\frac{\delta_0^2}{4 \widetilde V}\right) \leq \frac{k(k-1)}{2} \exp\left(-\frac{\delta_0^2}{4w_3R(m,n)}\right).
\end{align*}
Thus, ${\E}_{\X_0}{\E}_M \left[\pfs(\X_0)\right] {\lesssim}_{\PP_{\X^m}} \exp\left\{-\frac{\delta_0^2}{4w_3}\left[R(m,n)\right]^{-1}\right\}$ with probability at least $1-\xi$ for all $m\geq m_0$, which has proved Part (iii) of Theorem 5.  \hfill $\Box$

\vspace{10mm}

Now we discuss the restrictiveness of Assumptions A.6 and A.7 in the main text. We present Theorem 6 below to illustrate that A.6 and A.7 can hold, by using the finite-rank kernel example as described in Remark 2 and Theorem 4 of the main text.

\setcounter{theorem}{5}
\begin{theorem}
	(Exponentially decaying IPFS for finite-rank kernels) For a fixed positive integer $k$, consider the same model setup in Remark 2 of the main text with $k$ finite-rank kernels $\bSigma_{M,i}=a_i \left(\x^\top \x' + b_i\right)$ for any $\x,\x'\in \Xcal\subseteq \RR^d$, where $a_i>0$ and $b_i>0$ are known constants for $i=1,\ldots,k$.  Let $\PP_{\X}$ be any non-degenerate sampling distribution on $\Xcal$ for $\X^m$ and $\X_0$.
	\begin{itemize}
		\item[(i)] Suppose that there exist constants $c_1>0,c_2>0,t_0>0$, such that $\PP_{\X}$ has the tail bound $\PP_{\X}(\|\X\|>t)\leq c_1\exp(-c_2 t^2)$ for all $t>t_0$. Then for the optimal MSE given in (12) of the main text, for any given $\xi\in(0,1/2)$, there exist constants $w_1>0,w_2>0,m_0\geq 1$ that depend on $\xi$, such that for all $m\geq m_0$, for any $t>0$,
		\begin{align}
			&\PP_{\X^m} \left\{\PP_{\X_0}\left(mn\cdot \max_{i\in \{1,\ldots,k\}}\mse_{i,\opt}(\X_0)\geq t\right) \leq w_1 \exp\left(-w_2 t\right)\right\}\geq 1-\xi.
		\end{align}
		\item[(ii)] Suppose that $\Xcal$ is a compact set in $\RR^d$. Then for the optimal MSE given in (12) of the main text, for any given $\xi\in(0,1/2)$, there exist constants $w_3>0,m_0\geq 1$ that depend on $\xi$, such that for all $m\geq m_0$,
		\begin{align}
			&\PP_{\X^m} \left\{ mn\cdot \max_{i\in \{1,\ldots,k\}} \sup_{\x_0\in \Xcal} \mse_{i,\opt}(\x_0)\leq w_3 \right\}\geq 1-\xi.
		\end{align}
	\end{itemize}
\end{theorem}

Note that the rate $1/(mn)$ here is a tight convergence rate given Theorem 4, in the sense that it cannot be improved to any faster rate. The tail condition in Part (i) of Theorem 6 is satisfied by any $d$-dimensional multivariate normal distribution by the Hanson-Wright inequality \citep{Hsuetal12}. Theorem 6 shows that Assumption A.6 holds for the finite-rank kernel if the sampling distribution of $\X^m$ and $\X_0$ has tail decaying like the Gaussian distribution. Similarly, Assumption A.7 holds when the covariance kernel and the $\uf$-functions are continuous with a compact domain.

\vspace{10mm}
\newpage
\noindent \textbf{Proof of Theorem 6}:

First we show Part (i). We note that the tail condition $\PP_{\X}(\|\X\|>t)\leq c_1\exp(-c_2 t^2)$ implies the finite second moment for $\PP_{\X}$, because,
\begin{align*}
	{\E}_{\X} \left[\|\X\|^{2}\right] & = \int_0^{+\infty} \PP_{\X}\left(\|\X\|^2>u\right) \ud u \leq \int_0^{+\infty} c_1\exp(-c_2 u ) \ud u  = \frac{c_1}{c_2}<+\infty.
\end{align*}
Furthermore, since $\PP_{\X}$ is a non-degenerate sampling distribution on $\RR$, the covariance matrix $\V_{\X} \equiv {\E}_{\X_0} \left\{ [\X_0- {\E}_{\X_0}(\X_0)][\X_0- {\E}_{\X_0}(\X_0)]^{\top}\right\}$ must be positive definite. This is because otherwise, there exists a vector $\bba\in\RR^d$, such that
$$0 = \bba^\top {\E}_{\X_0} \left\{ [\X_0- {\E}_{\X_0}(\X_0)][\X_0- {\E}_{\X_0}(\X_0)]^{\top}\right\}\bba = {\E}_{\X_0} \left\{ \bba^\top[\X_0- {\E}_{\X_0}(\X_0)]\right\}^2 ,$$
which implies that $\bba^\top \X_0$ is almost surely a constant, contradicting the assumption that $\PP_{\X}$ is not degenerate.

For every $i=1,\ldots,k$, we define
\begin{align*}
	\widetilde \x_{i,0} = \left( \sqrt{b_i},~ \x_0^\top \right)^\top \in \RR^{d+1}, \quad  \Z_i = \left(
	\begin{array}{ccc}
		\sqrt{b_i} & \ldots & \sqrt{b_i} \\
		\x_1 & \ldots & \x_m
	\end{array}
	\right)^\top \in \RR^{(d+1)\times m} ,
\end{align*}
and $\widetilde \X_{i,0}$ is the $\RR^{d+1}$ random vector version of $\widetilde \x_{i,0}$ with $\X_0$ following the distribution $\PP_{\X}$. Define $\widetilde \V_i = {\E}_{\X_0} \left(\widetilde \X_{i,0} \widetilde \X_{i,0}^{\top}\right)$ for $i=1,\ldots,k$. Then we can write that
\begin{align*}
	\widetilde \V_i & = {\E}_{\X_{i,0}} \left(\widetilde \X_{i,0} \widetilde \X_{i,0}^{\top}\right) = \left(
	\begin{array}{cc}
		b_i & \sqrt{b_i} {\E}_{\X_{i,0}}(\X_{i,0}^\top) \\
		\sqrt{b_i} {\E}_{\X_{i,0}}(\X_{i,0}) & {\E}_{\X_{i,0}}(\X_{i,0}\X_{i,0}^\top)
	\end{array}
	\right) \\
	&=
	\left(
	\begin{array}{cc}
		\sqrt{b_i} & 0 \\
		{\E}_{\X_{i,0}}(\X_{i,0}) & \I_{d}
	\end{array}
	\right)
	\left(
	\begin{array}{cc}
		1 & 0 \\
		0 & \V_{\X}
	\end{array}
	\right)
	\left(
	\begin{array}{cc}
		\sqrt{b_i} & {\E}_{\X_{i,0}}(\X_{i,0}^\top) \\
		0 & \I_{d}
	\end{array}
	\right).
\end{align*}
From the last expression, we can see that the matrix $\widetilde \V_i = {\E}_{\X_{i,0}} \left(\widetilde \X_{i,0} \widetilde \X_{i,0}^{\top}\right)$ must be positive definite since it is congruent to a block diagonal matrix which is positive definite. Let $\underline\lambda_{\min} \equiv \min_{i\in \{1,\ldots,k\}} \lambda_{\min} \big(\widetilde \V_i\big) $ which is strictly positive.

Similar to the convergence in \eqref{zzlimit} in the proof of Theorem 4, by the strong law of large numbers, $\frac{1}{m} \Z_i^\top \Z_i$ converges to $\widetilde \V_i = {\E}_{\X_{i,0}} \left(\widetilde \X_{i,0} \widetilde \X_{i,0}^{\top}\right)$ entry-wise as $m\to\infty$ for each $i=1,\ldots,k$. Furthermore, for a fixed $k$, we have that for each $i=1,\ldots,k$, for any given $\xi\in (0,1/2)$, there exists a large integer $m_{i,0}>0$ that depends on $\xi$, such that for all $m\geq m_{i,0}$,
\begin{align*}
	\PP_{\X^m} \left(\vertiii{\frac{1}{m} \Z_i^\top \Z_i - \widetilde \V_i} > \frac{1}{2}\lambda_{\min}\left(\widetilde\V_i\right) \right) < \frac{\xi}{k}.
\end{align*}
Taking a union bound over all $k$ designs implies that for all $m\geq m_0\equiv \max_{i\in \{1,\ldots,k\}} m_{i,0}$ implies that
\begin{align*}
	\PP_{\X^m} \left(\vertiii{\frac{1}{m} \Z_i^\top \Z_i - \widetilde \V_i} > \frac{1}{2}\lambda_{\min}\left(\widetilde\V_i\right),\text{ for all } i=1,\ldots,k \right) < \sum_{i=1}^k \frac{\xi}{k} = \xi.
\end{align*}
This further implies that with $\PP_{\X^m}$-probability at least $1-\xi$, for all $m\geq m_0$,
\begin{align*}
	\lambda_{\min}\left(\I_{d+1} + \frac{a_i n}{\sigma^2}\Z_i^\top \Z_i\right) & = \lambda_{\min}\left[\I_{d+1} + \frac{a_i mn}{\sigma^2}\left\{\frac{1}{m}\Z_i^\top \Z_i - \widetilde\V_i\right\} + \frac{a_i mn}{\sigma^2}\widetilde\V_i\right] \\
	&\geq \lambda_{\min}\left(\I_{d+1} + \frac{a_i mn}{\sigma^2}\left[ \widetilde\V_i - \frac{1}{2}\lambda_{\min}\left(\widetilde\V_i\right) \I_{d+1} \right] \right) \\
	&\geq \lambda_{\min}\left[\I_{d+1} + \frac{a_i mn}{2\sigma^2} \lambda_{\min}\left(\widetilde\V_i\right) \I_{d+1}\right] \\
	&> \frac{a_i mn}{2\sigma^2} \underline\lambda_{\min}.
\end{align*}
Therefore, using the expression of $\mse_{i,\opt}(\X_0)$ derived in Theorem 4, we have that with $\PP_{\X^m}$-probability at least $1-\xi$, for any $t>0$, for all $m\geq m_0$,
\begin{align}\label{ppt1}
	&\quad ~ \PP_{\X_0}\left(mn\cdot \max_{i\in \{1,\ldots,k\}}\mse_{i,\opt}(\X_0)\geq t\right) \nonumber \\
	&= \PP_{\X_0}\left(mn\cdot a_i \widetilde \X_{i,0}^{\top} \left(\I_{d+1} + \frac{a_i n}{\sigma^2}\Z_i^\top \Z_i\right)^{-1} \widetilde \X_{i,0} \geq t\right) \nonumber \\
	& \leq \PP_{\X_0}\left(mna_i \cdot \left( \frac{a_i mn}{2\sigma^2}\underline\lambda_{\min} \right)^{-1} \widetilde \X_{i,0}^{\top} \widetilde \X_{i,0} \geq t\right) \nonumber \\
	& = \PP_{\X_0}\left(b_i + \|\X_0\|^{2} \geq \frac{\underline\lambda_{\min}}{2\sigma^2} t\right)
\end{align}
Let $\overline b=\max_{i\in \{1,\ldots,k\}} b_i$. If $t> \max\left(4 \sigma^2 \underline \lambda_{\min}^{-1} \overline b,t_0\right)$, then for all $i=1,\ldots,k$,
$$\frac{\underline\lambda_{\min}}{2\sigma^2} t - b_i > \frac{\underline\lambda_{\min}}{4\sigma^2} t,$$
and from the tail assumption $\PP_{\X}(\|\X\|>t)\leq c_1\exp(-c_2 t^2)$ in Theorem 6, we have that
\begin{align}\label{ppt2}
	& \PP_{\X_0}\left(b_i + \|\X_0\|^{2} \geq \frac{\underline\lambda_{\min}}{2\sigma^2} t\right) \leq \PP_{\X_0}\left(\|\X_0\|^{2} \geq \frac{\underline\lambda_{\min}}{4\sigma^2} t \right) \nonumber \\
	&=\PP_{\X_0}\left(\|\X_0\| \geq \frac{\sqrt{\underline\lambda_{\min}}}{2\sigma}  \sqrt{t} \right) \leq c_1 \exp\left( -\frac{c_2\underline\lambda_{\min}}{4\sigma^2} t\right).
\end{align}
If $0<t\leq  \max\left(4 \sigma^2 \underline \lambda_{\min}^{-1} \overline b,t_0\right)$, then we use the simple bound
\begin{align}\label{ppt3}
	& \PP_{\X_0}\left(b_i + \|\X_0\|^{2} \geq \frac{\underline\lambda_{\min}}{2\sigma^2} t\right) \leq 1\leq e^{c_2+1}\cdot \exp\left\{- t / \max\left(4 \sigma^2 \underline \lambda_{\min}^{-1} \overline b,t_0\right) \right\}.
\end{align}
Now let $w_1= \max(e^{c_2+1},c_1)$, $w_2=\min\left\{c_2\underline \lambda_{\min}/(4\sigma^2), \underline \lambda_{\min}/(4\sigma^2\overline b), 1/t_0\right\}$, then \eqref{ppt1}, \eqref{ppt2}, and \eqref{ppt3} together imply that with $\PP_{\X^m}$-probability at least $1-\xi$, for any $t>0$, for all $m\geq m_0$,
\begin{align*}
	&\quad ~\PP_{\X_0}\left(mn\cdot \max_{i\in \{1,\ldots,k\}}\mse_{i,\opt}(\X_0) \geq t\right) \\
	&\leq \1\left(0<t\leq  4 \sigma^2 \underline \lambda_{\min}^{-1} \overline b \right)\cdot e^{c_2+1 }\cdot \exp\left\{- t / \max\left(4 \sigma^2 \underline \lambda_{\min}^{-1} \overline b,t_0\right) \right\}  \\
	& \quad ~ + \1\left(t> 4 \sigma^2 \underline \lambda_{\min}^{-1} \overline b  \right) \cdot c_1 \exp\left( -\frac{c_2\underline\lambda_{\min}}{4\sigma^2} t\right) \\
	&\leq w_1 \exp(-w_2t) .
\end{align*}
This has proved Part (i) of Theorem 6.

\vspace{2mm}

Next we show Part (ii). Let $\underline a = \min_{i\in \{1,\ldots,k\}} a_i$ which is strictly positive given a fixed $k$. From the proof above, with $\PP_{\X^m}$-probability at least $1-\xi$, there exists a large integer $m_0$ such that uniformly for all $i=1,\ldots,k$ and all $m\geq m_0$,
\begin{align*}
	\lambda_{\min}\left(\I_{d+1} + \frac{a_i n}{\sigma^2}\Z_i^\top \Z_i\right) &> \frac{a_i mn}{2\sigma^2}\underline \lambda_{\min}.
\end{align*}
Since $\Xcal$ is a compact set, there exists a constant $c_3>0$ such that $|\x|\leq c_3$ for all $\x\in \Xcal$. Recall that $\widetilde \x_{i,0}=(\sqrt{b_i},\x_0^\top)^\top$ for any $\x_0\in \Xcal$. Therefore, with $\PP_{\X^m}$-probability at least $1-\xi$, for all $m\geq m_0$,
\begin{align*}
	& \quad ~ mn\cdot \max_{i\in \{1,\ldots,k\}} \sup_{\x_0\in\Xcal} \mse_{i,\opt}(\x_0) \\
	& = mn\cdot \max_{i\in \{1,\ldots,k\}} \sup_{\x_{0}\in \Xcal} \widetilde \x_{i,0}^{\top} \left(\I_{d+1} + \frac{a_i n}{\sigma^2}\Z_i^\top \Z_i \right)^{-1} \widetilde \x_{i,0} \\
	& \leq mn\cdot \max_{i\in \{1,\ldots,k\}} \left\{\lambda_{\min}^{-1}\left(\I_{d+1} + \frac{a_i n}{\sigma^2}\Z_i^\top \Z_i\right) \sup_{\x_0\in\Xcal} \widetilde \x_{i,0}^{\top} \widetilde \x_{i,0}\right\} \\
	& \leq mn\cdot \max_{i\in \{1,\ldots,k\}}  \left\{ \frac{2\sigma^2}{a_i mn} \underline \lambda_{\min}^{-1} \cdot \sup_{\x_0\in\Xcal}\left(b_i+\|\x_0\|^{2}\right) \right\} \\
	& \leq \frac{2\sigma^2(\overline b +c_3^2)\underline\lambda_{\min}^{-1}}{\underline a}.
\end{align*}
Set $w_3=2\sigma^2(\overline b +c_3^2)\underline\lambda_{\min}^{-1}/\underline a$ and then Part (ii) of Theorem 6 is proved. \hfill $\Box$

\vspace{8mm}

\section{Estimators of IMSE and IPFS} \label{sec:Estimators}

In this section, we propose simple estimators of IMSE and IPFS based on Monte Carlo draws from the sampling distribution $\PP_{\X}$. Suppose that we already have the covariate sample $\X^m=\{\X_1,\ldots,\X_m\}$. To estimate MSE, we draw another random sample $\widetilde \X^{m'} = \{\widetilde \X_1,\ldots,\widetilde \X_{m'}\}$ from the distribution $\PP_{\X}$. The two samples $\X^m$ and $\tilde \X^m$ are independent. The sample size $m'$ can be different from $m$. Then, according the definition of MSE in Equation (3) of the main text, we estimate the IMSE under the $i$th design ($i=1,\ldots,k$) as
\begin{align} \label{eq:IMSE_est}
	& \widehat{\text{IMSE}}_i  = \frac{1}{m'} \sum_{j=1}^{m'} \mse_{i,\opt}(\widetilde \X_j), \qquad \text{where for }~ j=1,\ldots,m', \nonumber \\
	&\mse_{i,\opt}(\widetilde \X_j) =  \bSigma_{M,i}(\widetilde \X_j,\widetilde \X_j) - \bSigma_{M,i}^\top(\X^m,\widetilde \X_j) \left[\bSigma_{M,i}(\X^m,\X^m) +  \bSigma_{\epsilon,i}(\X^m)\right]^{-1} \bSigma_{M,i}(\X^m,\widetilde \X_j) \nonumber \\
	& ~ + \eta_i(\widetilde \X_j)^\top \left[\F_i^\top \left(\bSigma_{M,i}(\X^m,\X^m)+\bSigma_{\epsilon,i}(\X^m)\right)^{-1}\F_i\right]^{-1} \eta_i(\widetilde \X_j), \nonumber \\
	&\text{and }~ \eta_i(\tilde \X_j) = \bbf_i(\widetilde \X_j) - \F_i^\top \left[\bSigma_{M,i}(\X^m,\X^m)+\bSigma_{\epsilon,i}(\X^m)\right]^{-1} \bSigma_{M,i}(\X^m,\widetilde \X_j) .
\end{align}
It is straightforward to see that since $\widetilde \X^{m'}$ is an i.i.d. sample from $\PP_{\X}$ and is independent of the sample $\X^m$, the proposed estimator $\widehat{\text{IMSE}}_i$ in \eqref{eq:IMSE_est} is unbiased for the IMSE defined as ${\E}_{\X^m}{\E}_{\X_0} \left[\mse_{i,\opt} (\X_0)\right]$. The maximal IMSE among the $k$ designs can be then estimated by $\max_{i\in\{1,\ldots,k\}} \widehat{\text{IMSE}}_i $.

For IPFS with an IZ parameter $\delta_0>0$, we first need to estimate the PFS at a given covariate point $\x_0$, which can be approximated by the following quantity:
\begin{align} \label{eq:apfs1}
	\text{APFS}(\x_0) = \sum_{i \ne \widehat i^\circ (\x_0)} \PP \left( N(0,1) < -\frac{ \widehat{y}_i(\x_0) - \widehat{y}_{\widehat i^\circ (\x_0)}(\x_0) + \delta_0 }{ \sqrt{\mse_{i,\opt}(\x_0) + \mse_{\widehat i^\circ (\x_0),\opt}(\x_0)} } \right),
\end{align}
where $\widehat i^\circ (\x_0)$ and $\widehat{y}_i(\x_0)$ are defined in Equation (4) of the main text, $\widehat{y}_i(\x_0)$ is defined in Equation (2) of the main text, and $\mse_{i,\opt}(\x_0)$ is defined in Equation (3) of the main text. Then, based on the random sample $\widetilde \X^{m'} = \{\widetilde \X_1,\ldots,\widetilde \X_{m'}\}$ from the distribution $\PP_{\X}$ independent of $\X^m$, we can estimate the IPFS as
\begin{align} \label{eq:IPFS_est}
	& \widehat {\text{IPFS}} = \frac{1}{m'} \sum_{j=1}^{m'} \text{APFS}(\widetilde \X_j),
\end{align}
where $\text{APFS}(\cdot)$ is defined in \eqref{eq:apfs1}. The $\widehat {\text{IPFS}}$ in \eqref{eq:IPFS_est} is a consistent estimator of $\text{IPFS}={\E}_M{\E}_{\X_0}  \left[\pfs(\X_0)\right]$.

\vspace{8mm}

\section{Analysis for the Case of Unequal $n_i$'s}

Let $n_{i}$ be the number of simulation replications allocated to each of the $m$ covariate points with design $i$, $i=1,\ldots,k$. In this section, we fix the number of covariate points $m$, allow $n_i$ to be unequal among different designs $i$, and develop a ranking and selection (R\&S) framework for optimizing the simulation budget allocation $n_i$'s in simulation with covariates introduced in the main text.

Suppose that the $m$ covariate points collected are $\x_1,\ldots,\x_m$, and the total simulation budget to be allocated among pairs of covariate points and designs is $n_{tot}$, i.e., $m\sum_{i=1}^k n_i=n_{tot}$. With the target measures of the maximal IMSE and IPFS, the corresponding R\&S problems can be formulated as
\begin{align} \label{eq:rs1}
	& \min \max_{i\in\{1,2,...,k\}}{\E}_{\X_0}\left[\mse_{i,\opt}(\X_0)\right]  \\
	\text{s.t.} \ \ & m\sum_{i=1}^k n_i=n_{tot}, \text{ and } n_i\geq 0, \text{ for } i=1,\ldots,k,  \nonumber
\end{align}
and
\begin{align} \label{eq:rs2}
	& \min {\E}_M{\E}_{\X_0}  \left[\pfs(\X_0)\right]  \\
	\text{s.t.} \ \ & m\sum_{i=1}^k n_i=n_{tot}, \text{ and } n_i\geq 0, \text{ for } i=1,\ldots,k,  \nonumber
\end{align}

However, both optimization problems (\ref{eq:rs1}) and (\ref{eq:rs2}) cannot be directly solved due to the lack of analytical expressions of the objective functions $\max_{i\in\{1,2,...,k\}}{\E}_{\X_0}\left[\mse_{i,\opt}(\X_0)\right]$ and ${\E}_M{\E}_{\X_0}  \left[\pfs(\X_0)\right]$. Here we propose two methods to approximate them.

Our first proposal is to replace the maximal IMSE and IPFS in \eqref{eq:rs1} and \eqref{eq:rs2} with their Monte Carlo estimators proposed in Section \ref{sec:Estimators}. Both $\max_{i\in\{1,\ldots,k\}} \widehat{\text{IMSE}}_i$ defined in \eqref{eq:IMSE_est} and $\widehat {\text{IPFS}}$ defined in \eqref{eq:IPFS_est} have already taken into account the unequal $n_i$'s in the matrix $\bSigma_{\epsilon,i}(\X^m)$. We can choose the Monte Carlo sample size $m'$ according to the optimization budget. Then \eqref{eq:rs1} and \eqref{eq:rs2} can be solved using numerical optimization methods.

Our second proposal is to approximate them by the analytical upper bounds in our Theorems 1 and 2. Note that analytical approximations are common in solving R\&S problems, especially in the OCBA method \citep{chen2000,chen2008}. They make the optimization problem tractable, and can often lead to efficient budget allocation rules.

Let $\{\mu_{i,l}:l=1,2,\ldots\}$ be the eigenvalues of the linear operator $T_{\bSigma_{M,i}}$ defined in Section 2.1 of the main text. We recall from the second paragraph after Assumptions A.1-A.4 that the constants $r_*$ and $\rho_*$ in Assumption A.3 can be made common for all the $k$ designs. Using the results in Theorems 1, 2 and 5, we can prove the following proposition.

\begin{proposition} \label{prop:budget.alloc}
	Suppose that Assumptions A.1 - A.4 in the main text hold for all the $k$ designs. Let $\varrho_i=mn_i/n_{tot}$. For any $0\leq \varrho\leq 1$, define the following quantities for $i=1,\ldots,k$:
	\begin{align} \label{eq:R.design.i}
		& R_i(\varrho)  = \frac{2\overline \sigma_0^2}{n_{tot}\varrho} \gamma_i \left( \frac{\overline \sigma_0^2}{n_{tot}\varrho} \right) + \frac{64C^{\dagger}_i q\overline \sigma_0^2 \tr(\bSigma_{M,i})}{n_{tot}\varrho} \nonumber \\
		& ~~ + \underset{\zeta \in \NN}{\inf} \Bigg[ \left\{\frac{64C^{\dagger}_i q \rho_*^4 \overline \sigma_0^2}{ \underline \sigma_0^4} \tr\left(\bSigma_{M,i} \right)^2 + 8C^{\dagger}_i q\tr\left(\bSigma_{M,i} \right) +  \frac{3}{\overline \sigma_0^2}\tr(\bSigma_{M,i}) + 1  \right\} \tr\left(\bSigma_{M,i}^{(\zeta)}\right) n_{tot}\varrho \nonumber \\
		&~~ + \left[ 8C^{\dagger}_i q \tr\left(\bSigma_{M,i}\right)^2 + \tr\left(\bSigma_{M,i}\right) \right] \left\{300  \rho_*^2 \frac{b(m,\zeta,r_*)}{\sqrt{m}}  \gamma_i\left( \frac{\overline \sigma_0^2}{n_{tot}\varrho}\right)\right\}^{r_*}  \Bigg],
	\end{align}
	where $A$ is the universal constant in Theorem 1 and
	\begin{align*}
		& C^{\dagger}_i= C_{\uf,i}^2 / \lambda_{\min}\left({\E}_{\X}[\bbf(\X)\bbf(\X)^\top]\right), \quad C_{\uf,i}=\max_{1\leq s\leq q}\| \uf_s \|_{\HH_i}  \\
		& \gamma_i(a) = \sum_{l=1}^{\infty} \frac{\mu_{i,l}}{\mu_{i,l}+a} \text{ for any } a>0 , \\
		& \tr\left(\bSigma_{M,i}\right) =\sum_{l=1}^{\infty} \mu_{i,l} ,  \quad \tr\left(\bSigma_{M,i}^{(\zeta)}\right) = \sum_{l=\zeta+1}^{\infty} \mu_{i,l}  \text{ for any } \zeta \in \NN , \\
		& b(m,\zeta,r_*) = \max \left( \sqrt{\max(r_*, \log \zeta)},~ \frac{\max(r_*,\log \zeta)}{m^{1/2 - 1/r_*}} \right) .
	\end{align*}
	Then, for the measures of the maximal IMSE and IPFS, we have
	\begin{align}
		&\max_{i\in \{1,\ldots,k\}} {\E}_{\X_0} \left[\mse_{i,\opt} (\X_0)\right]  {\lesssim}_{\PP_{\X^m}} \max_{i\in \{1,\ldots,k\}} R_i(\varrho_i),\label{eq:max.ineq1}\\
		& {\E}_M{\E}_{\X_0}  \left[\pfs(\X_0)\right] {\lesssim}_{\PP_{\X^m}} \max_{i\in \{1,\ldots,k\}} R_i(\varrho_i),\label{eq:max.ineq2}.
	\end{align}
\end{proposition}

\noindent \textbf{Proof of Proposition \ref{prop:budget.alloc}:}
\vspace{1mm}

By directly combining the upper bounds in Theorems 1 and 2 together with the MSE decomposition in Equation (6) of the main text, we have that with $\PP_{\X^m}-$ probability approaching 1, for each $i=1,\ldots,k$,
\begin{align*}
	{\E}_{\X_0} \left[\mse_{i,\opt} (\X_0)\right]  = {\E}_{\X_0} \left[\mse_{i,\opt}^{(M)} (\X_0)\right] + {\E}_{\X_0}\left[\mse_{i,\opt}^{(\bbeta)} (\X_0)\right] \leq R_i(mn_i/n_{tot})=R_i(\varrho_i),
\end{align*}
where $R_i(\cdot)$ is defined in \eqref{eq:R.design.i} above and is slightly larger than the combined upper bounds from Theorems 1 and 2 by adjusting some constants. This implies the following upper bound
\begin{align*}
	\max_{i\in \{1,\ldots,k\}} {\E}_{\X_0} \left[\mse_{i,\opt} (\X_0)\right]  {\lesssim}_{\PP_{\X^m}} \max_{i\in \{1,\ldots,k\}} R_i(\varrho_i),
\end{align*}
which proves (\ref{eq:max.ineq1}).

For the IPFS measure, we notice that for each of the three cases in Theorem 5, the upper bound is a monotone increasing function of $R(m,n)$, which is defined as a probablistic upper bound for $\max_{i\in \{1,\ldots,k\}} {\E}_{\X_0} \left[\mse_{i,\opt} (\X_0)\right]$ in Theorem 3 of the main text. Therefore, the inequality in (\ref{eq:max.ineq2}) holds as well.  \hfill $\Box$

With Proposition (\ref{prop:budget.alloc}), we can build an analytical R\&S model for both (\ref{eq:rs1}) and (\ref{eq:rs2}),
\begin{align} \label{eq:optimize.imse}
	& \min \max_{i\in \{1,\ldots,k\}} R_i(\varrho_i)  \\
	\text{s.t.} & \sum_{i=1}^k \varrho_i = 1, \text{ and } \varrho_i\geq 0, \text{ for } i=1,\ldots,k.  \nonumber
\end{align}
This is a typical nonlinear optimization problem. Its optimal solution $\varrho_i^*$ gives us an approximately optimal allocation of the simulation budget among pairs of covariate points and designs with $n_i^*=\frac{\varrho_i^* n_{tot}}{m}$, $i=1,2,...,k$.

Problem (\ref{eq:optimize.imse}) involves a number of constants that depend on the properties of the covariance kernels used in the $k$ designs. These constants can be made concrete when the covariate space $\Xcal$, the covariance kernels $\bSigma_{M,i}$, the sampling distribution $\PP_{\X}$, and the regression functions $\bbf_{i1},\ldots,\bbf_{iq}$ are fully specified in practice. Problem (\ref{eq:optimize.imse}) is not necessarily a convex optimization problem. Since it is built based on a different setting (fixed $m$ and unequal $n_i$'s) from that of the main questions in this research, we do not pursue further development of it in this paper. We emphasize that our proposed theoretical analysis and results can be used to formulate and solve R\&S type of problems that arise in simulation with covariates.

\vspace{1mm}

\section{Additional Numerical Results}

This section provides additional numerical results to the main text. Section \ref{sec:plot} plots the two test functions in Section 5.1 of the main text. Section \ref{sec:comp} compares our static sampling with an adaptive design procedure, under the target measures of the maximal IMSE and IPFS. Section \ref{sec:proc_targ} provides a procedure that can help the analyst make the design decision for achieving a target precision of the maximal IMSE.

\subsection{Plots of the Test Functions used in Section 5.1 of the Main Text}\label{sec:plot}

\begin{figure}[htbp]
	\begin{center}
		\includegraphics[width=1.05\textwidth]{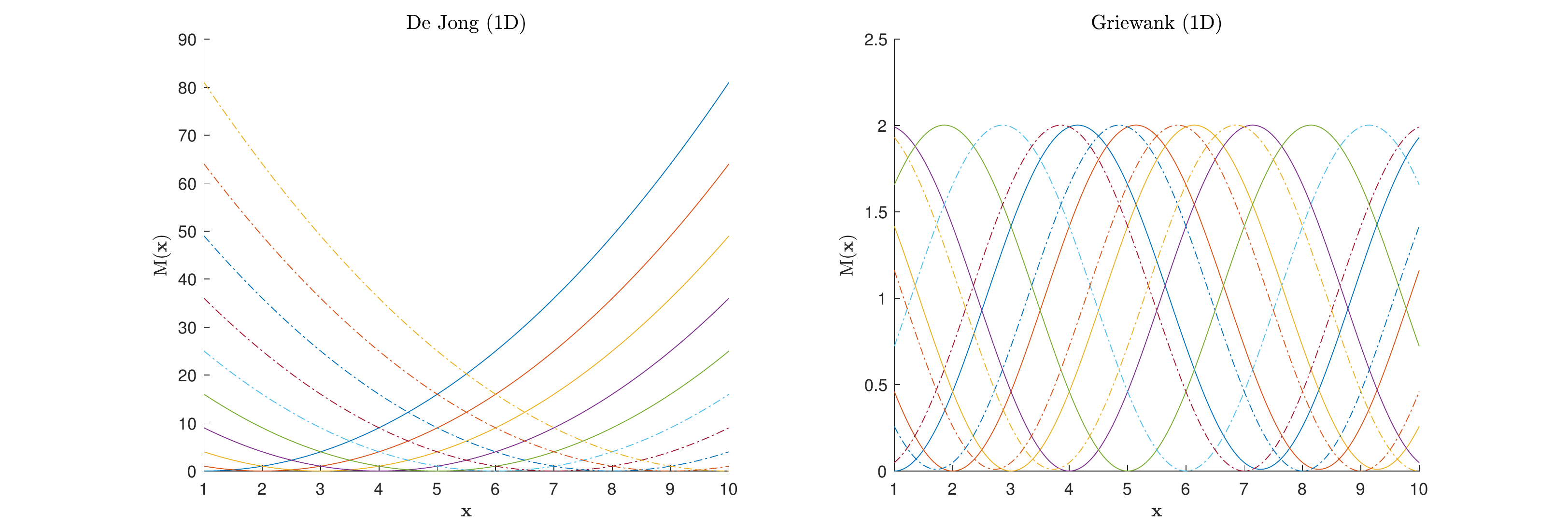}
	\end{center}
	\caption{Plots of the two test functions for 1-dimensional $\x$. Ten different curves stand for the ten designs.}
	\label{fig:4fun1d:truth}
\end{figure}

\begin{figure}[htbp]
	\begin{center}
		\includegraphics[width=1.05\textwidth]{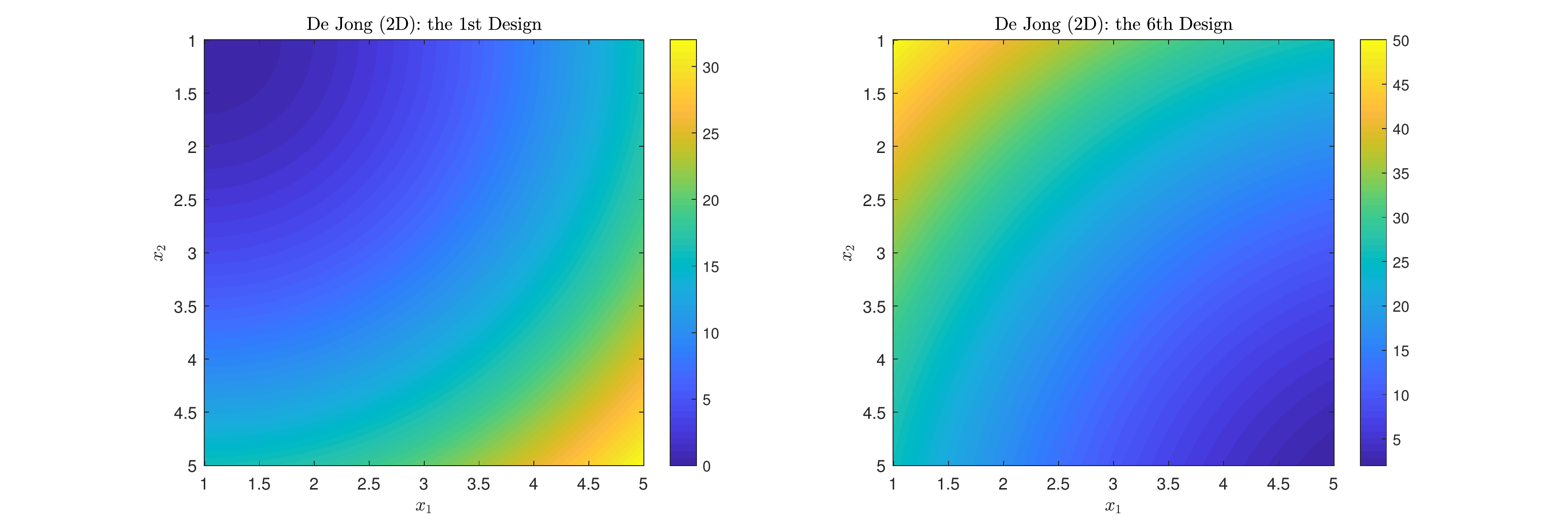}
	\end{center}
	\caption{Heatmaps of De Jong's function for 2-dimensional $\x$ under the 1st and 6th design ($i=1$ and $i=6$).}
	\label{fig:4fun2d:truth}
\end{figure}

The 1-d De Jong's function and Griewank's function without noise are shown in Figure \ref{fig:4fun1d:truth}.  The 2-d De Jong's function under selected designs without noise is shown in Figure \ref{fig:4fun2d:truth}. 

In the 1-d case (Figure \ref{fig:4fun1d:truth}), we present ten curves for the two test functions, each corresponding to $M(\x)$ at one of the ten designs. It can be observed that no design can dominate the others in the tested functions, and the best design might not be unique for some $\x$. The De Jong's functions are smooth while the Griewank's functions are highly nonlinear with many oscillations, which brings difficulty to SK modeling when the number of covariate points $m$ is small. In the 2-d case (Figure \ref{fig:4fun2d:truth}), we present the heatmap of the 1st and 6th design ($i=1$ and $i=6$) for the De Jong's function. We can see that $M(\x)$ varies a lot with $\x$.

\subsection{Comparison between Static Sampling and an Adaptive Procedure}\label{sec:comp}

In this section, we compare our static sampling (i.e., fixed-distribution sampling; this is the sampling method studied in this research) with an intuitive adaptive design procedure. The adaptive procedure works in a greedy manner and iteratively collects the covariate point that maximizes the largest MSE of the fitted SK models. In this way, it sequentially explores the whole covariate space and reduces the overall MSE of the SK prediction. We call it Adaptive MSE Procedure.

Suppose that $m_1$ covariate points $\x^{m_1} = \{\x_1,\ldots,\x_{m_1}\}$ have been sampled already. For the illustration purpose, we will use the superscript $[m_1]$ to indicate that the SK estimators are derived from the current simulation samples $\x^{m_1}$. From Equation (3) in the main text, the mean squared error of the current-stage SK predictor of design $i$ at $\x_0$ is
\begin{align}
	\mse_{i,\opt}^{[m_1]}(\x_0) &= \bSigma_{M,i}(\x_0,\x_0) - \bSigma_{M,i}^\top(\x^{m_1},\x_0) \left[\bSigma_{M,i}(\x^{m_1},\x^{m_1}) +  \bSigma_{\epsilon,i}(\x^{m_1})\right]^{-1} \bSigma_{M,i}(\x^{m_1},\x_0) \nonumber \\
	& ~ + \eta_i^{[m_1]}(\x_0)^\top \left[(\F_i^{[m_1]})^\top \left(\bSigma_{M,i}(\x^{m_1},\x^{m_1})+\bSigma_{\epsilon,i}(\x^{m_1})\right)^{-1}\F_i^{[m_1]}\right]^{-1} \eta_i^{[m_1]}(\x_0), \label{eq:msem}
\end{align}
where $\eta_i^{[m_1]}(\x_0)=\bbf_i(\x_0) - (\F_i^{[m_1]})^\top \left(\bSigma_{M,i}(\x^{m_1},\x^{m_1})+\bSigma_{\epsilon,i}(\x^{m_1})\right)^{-1} \bSigma_{M,i}(\x^{m_1},\x_0)$, $\F_i^{[m_1]}=(\bbf_i(\x_1), $ $\ldots,\bbf_i(\x_{m_1}))^\top$, and $\bSigma_{\epsilon,i}(\x^{m_1})$ is the $m_1\times m_1$ covariance matrix of the averaged simulation errors across $m_1$ covariate points under design $i$.

The Adaptive MSE Procedure samples the next covariate point $\x_{m_1+1}$ with the largest maximal $\mse_{i,\opt}^{[m_1]}(\x_0)$, where ``largest" is over the covariate space $\Xcal$ and ``maximal'' is over the $k$ SK models. That is,
\begin{align}\label{eq:imseproc}
	\x_{m_1+1} = \arg\max_{\x_0 \in \Xcal} \max_{i\in\{1,\ldots,k\}}  \mse_{i,\opt}^{[m_1]}(\x_0).
\end{align}
The formal description of the Adaptive MSE Procedure is given as follows.

\textbf{Adaptive MSE Procedure}
\begin{enumerate}
	\item[1.] Specify the covariate space $\Xcal$ and the total number of covariate points $m$. Perform $n_0$ replications for the pair of the center point of the covariate space and design $i$, $i=1,\ldots,k$. $m_1 \leftarrow 0$.
	\item[2.] If $m_1>m$, stop. Otherwise,
	\begin{enumerate}
		\item[a.] Obtain $\x_{m_1+1}$ by \eqref{eq:imseproc}.
		\item[b.] Perform $n_0$ replications for the pair of covariate point $\x_{m_1+1}$ and design $i$, $i=1,\ldots,k$.
		\item[c.] Update the SK model for each design $i=1,\ldots,k$. $m_1\leftarrow m_1+1$.
	\end{enumerate}
\end{enumerate}

We use the De Jong's and Griewank's functions under the same parameter settings as in Section 5.1 of the main text for testing, i.e., the covariate space is $\Xcal = [1,10]^d$ and there are $k=10$ designs. Meanwhile, we vary the domain dimension $d$ and the sampling distribution $\PP_{\X}$. Specifically, we test the following examples on our static sampling from $\PP_{\X}$ and the Adaptive MSE Procedure:
\begin{itemize}
	\item[(i)] De Jong's functions, for dimension $d=1$, $\PP_{\X}$ being the truncated $N(5.5,1^2)$;
	\item[(ii)] De Jong's functions, for dimension $d=1$, $\PP_{\X}$ being the truncated $N(5.5,0.25^2)$;
	\item[(iii)] De Jong's functions, for dimension $d=2$, $\PP_{\X}$ being the truncated $N(5.5,0.3^2)$ in each dimension;
	\item[(iv)] De Jong's functions, for dimension $d=3$, $\PP_{\X}$ being the truncated $N(2.5,0.3^2)$ in each dimension;
	\item[(v)] Griewank's functions, for dimension $d=1$, $\PP_{\X}$ being the uniform distribution on $[1,10]$;
	\item[(vi)] Griewank's functions, for dimension $d=1$, $\PP_{\X}$ being the truncated $N(5.5,1^2)$;
	\item[(vii)] Griewank's functions, for dimension $d=10$, $\PP_{\X}$ being the uniform distribution on $[1,10]^{10}$;
	\item[(viii)] Griewank's functions, for dimension $d=10$, $\PP_{\X}$ being the truncated $N(2.5,0.75^2)$ in each dimension.
\end{itemize}

\begin{figure}[p]
	\begin{center}
		\includegraphics[width=0.6\textwidth]{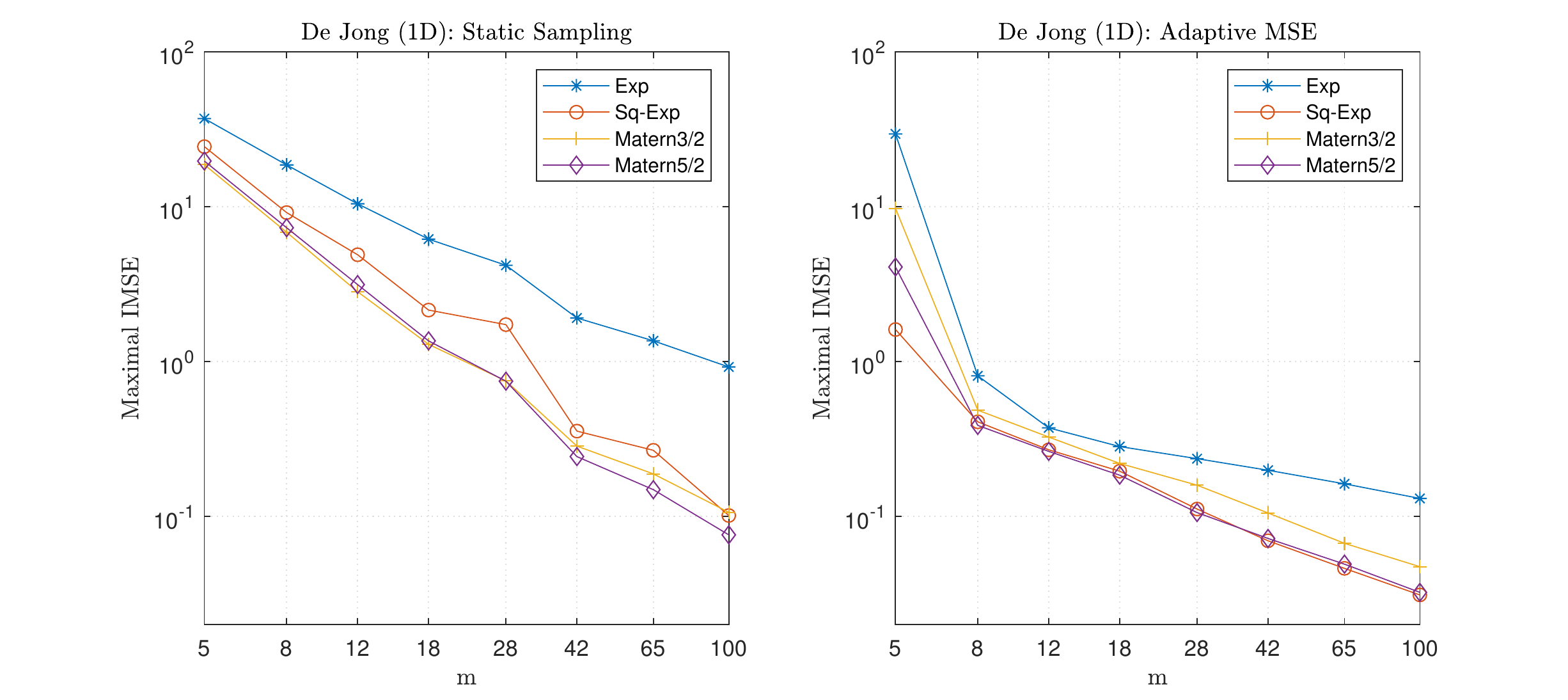}
		\includegraphics[width=0.6\textwidth]{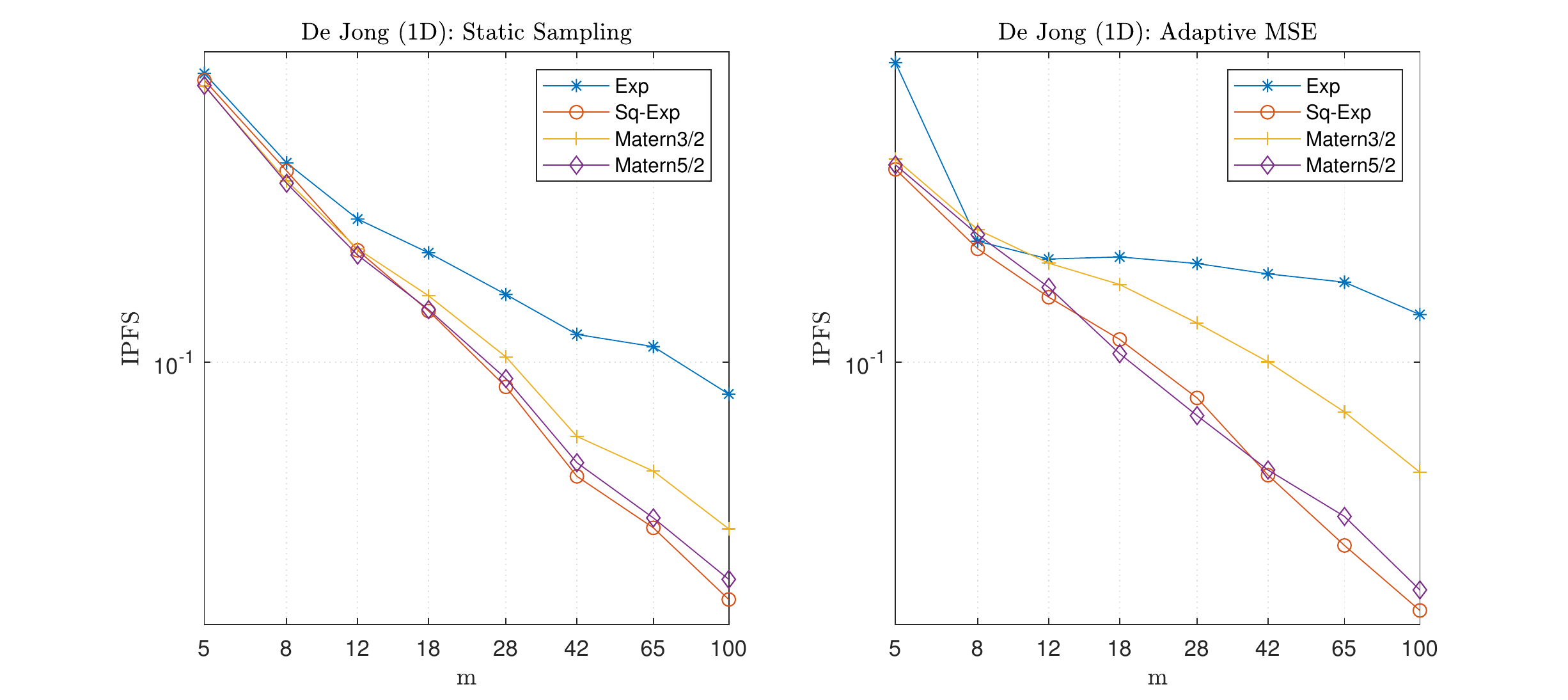}
	\end{center}
	\caption{Truncated $N(5.5, 1^2)$ of $d=1$: The maximal IMSE and IPFS for the 1-dimensional De Jong's functions and four covariance kernels. }
	\label{fig:test1comp1}
\end{figure}

\begin{figure}[p]
	\begin{center}
		\includegraphics[width=0.6\textwidth]{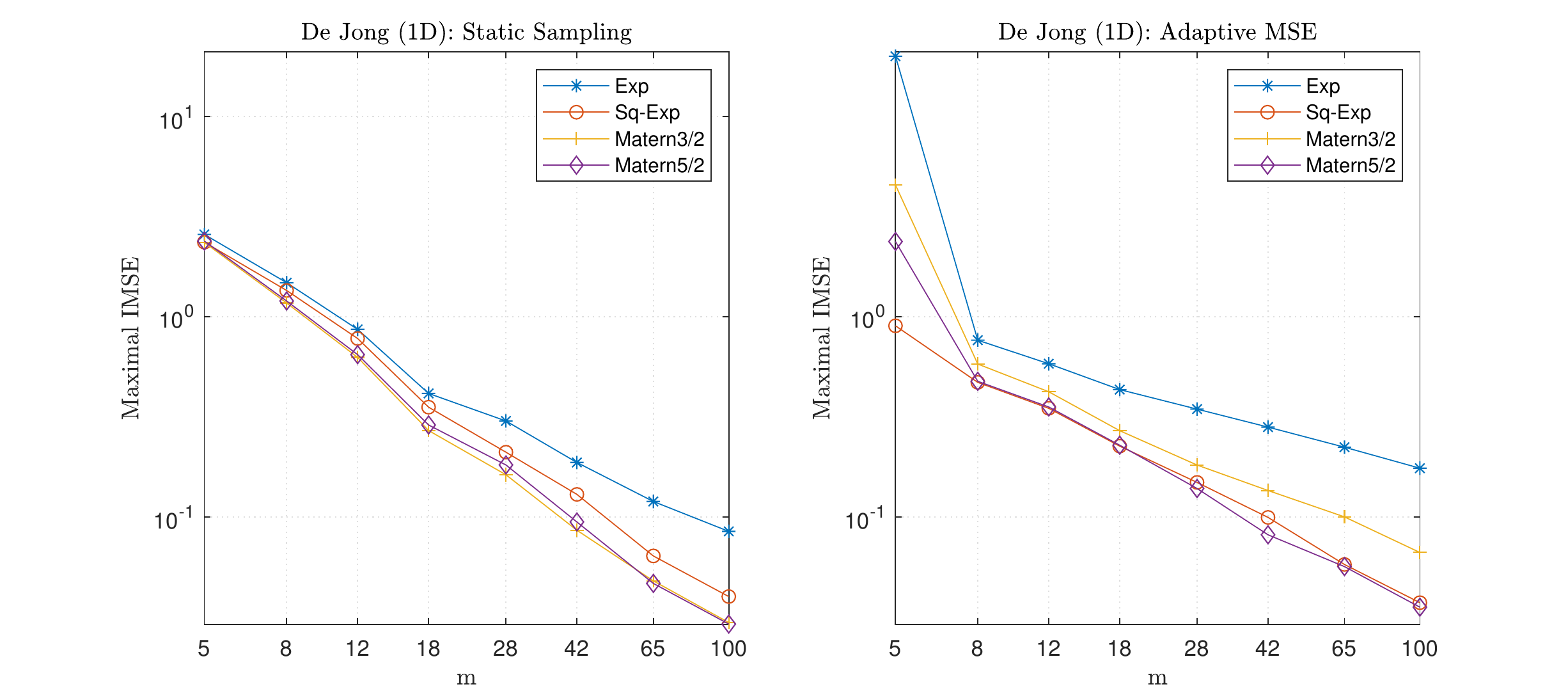}
		\includegraphics[width=0.6\textwidth]{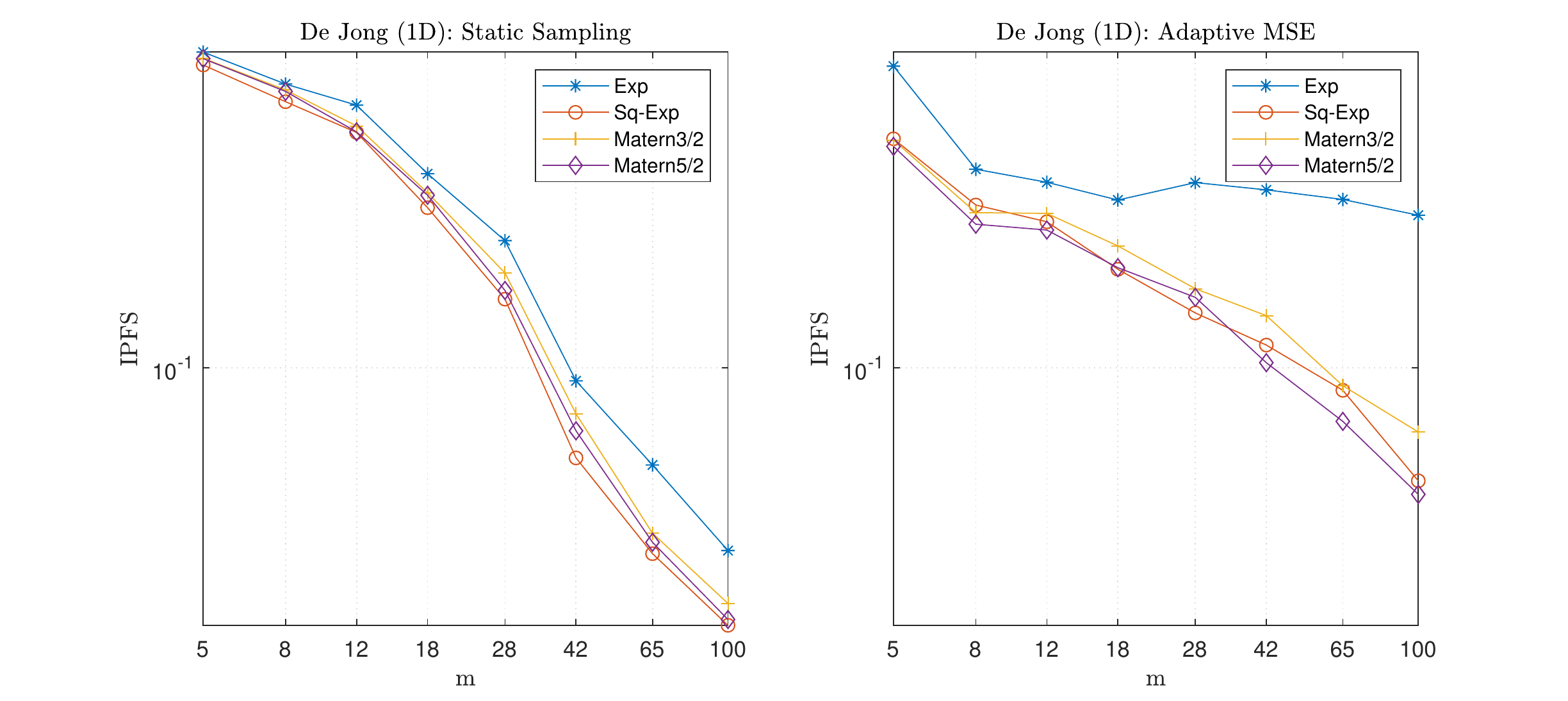}
	\end{center}
	\caption{Truncated $N(5.5, 0.25^2)$ of $d=1$: The maximal IMSE and IPFS for the 1-dimensional De Jong's functions and four covariance kernels. }
	\label{fig:test1comp2}
\end{figure}

\begin{figure}[htbp]
	\begin{center}
		\includegraphics[width=0.6\textwidth]{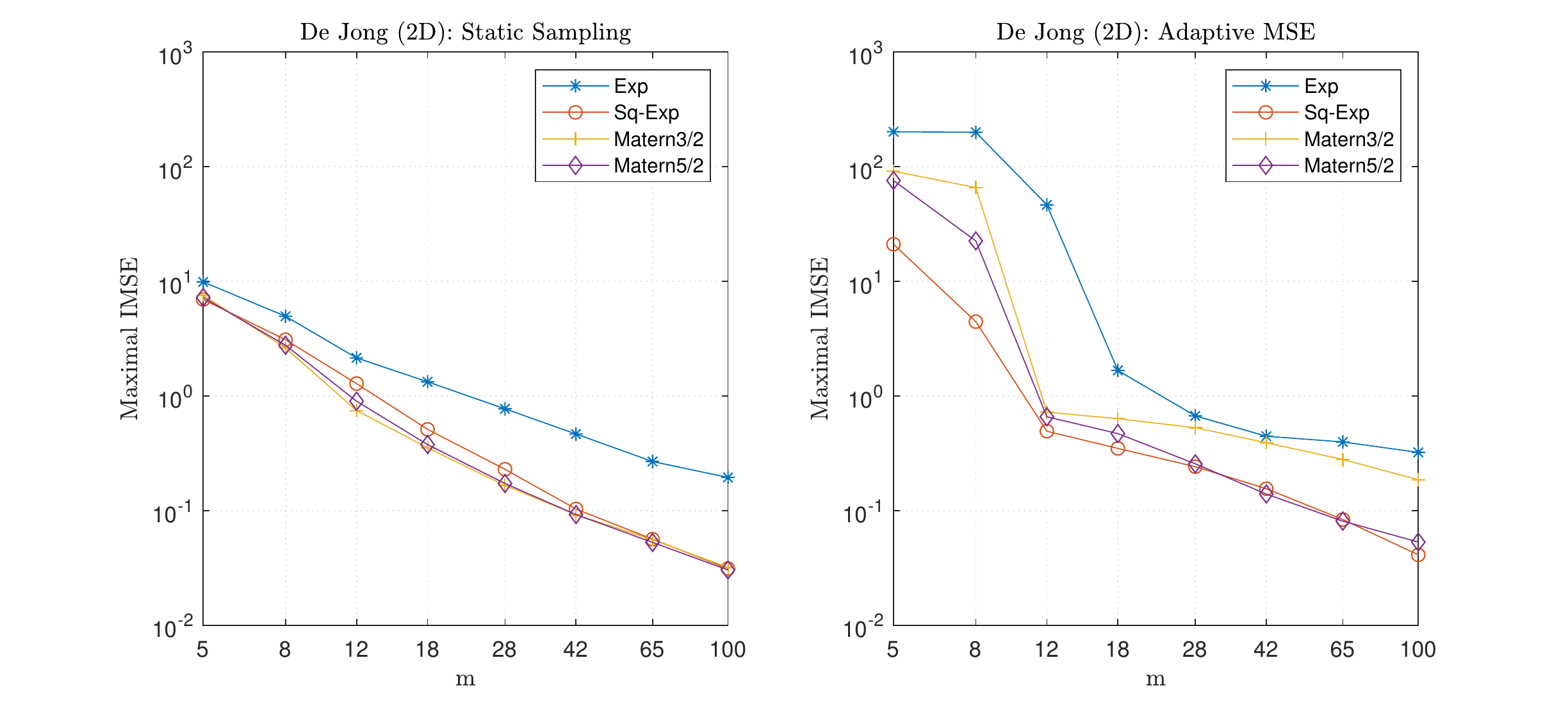}
		\includegraphics[width=0.6\textwidth]{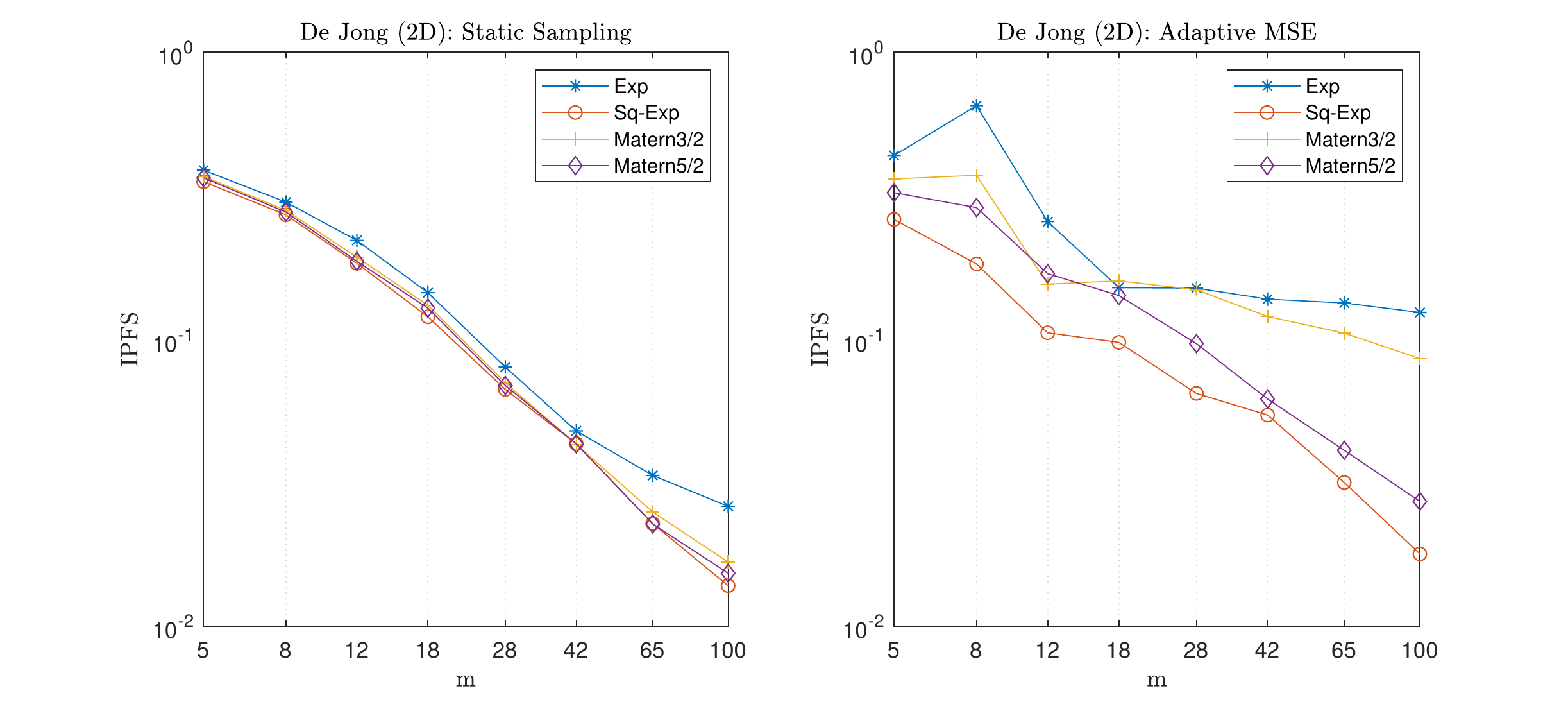}
	\end{center}
	\caption{Truncated $N(5.5, 0.3^2)$ on each dimension of $d=2$: The maximal IMSE and IPFS for the 2-dimensional De Jong's functions and four covariance kernels.}
	\label{fig:test1d2comp}
\end{figure}

\begin{figure}[htbp]
	\begin{center}
		\includegraphics[width=0.6\textwidth]{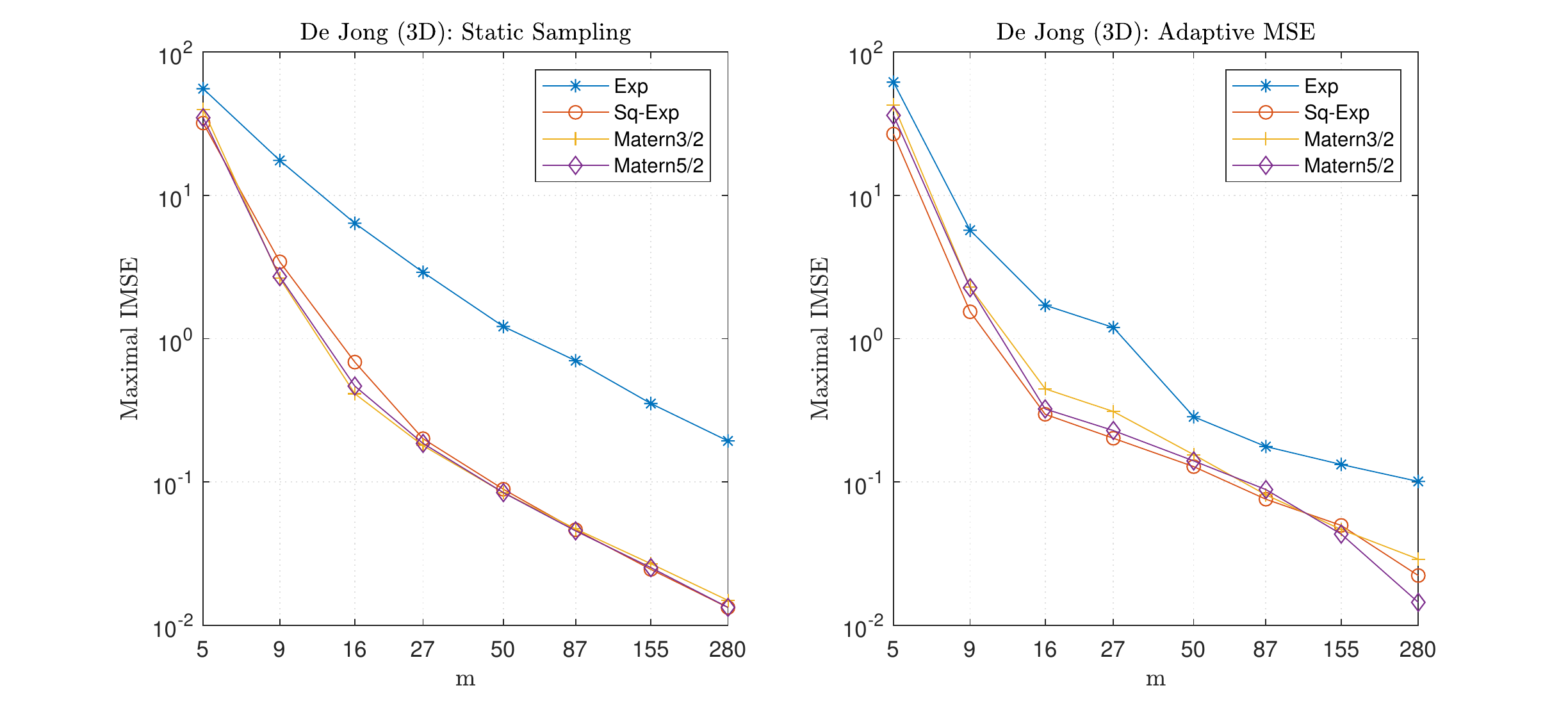}
		\includegraphics[width=0.6\textwidth]{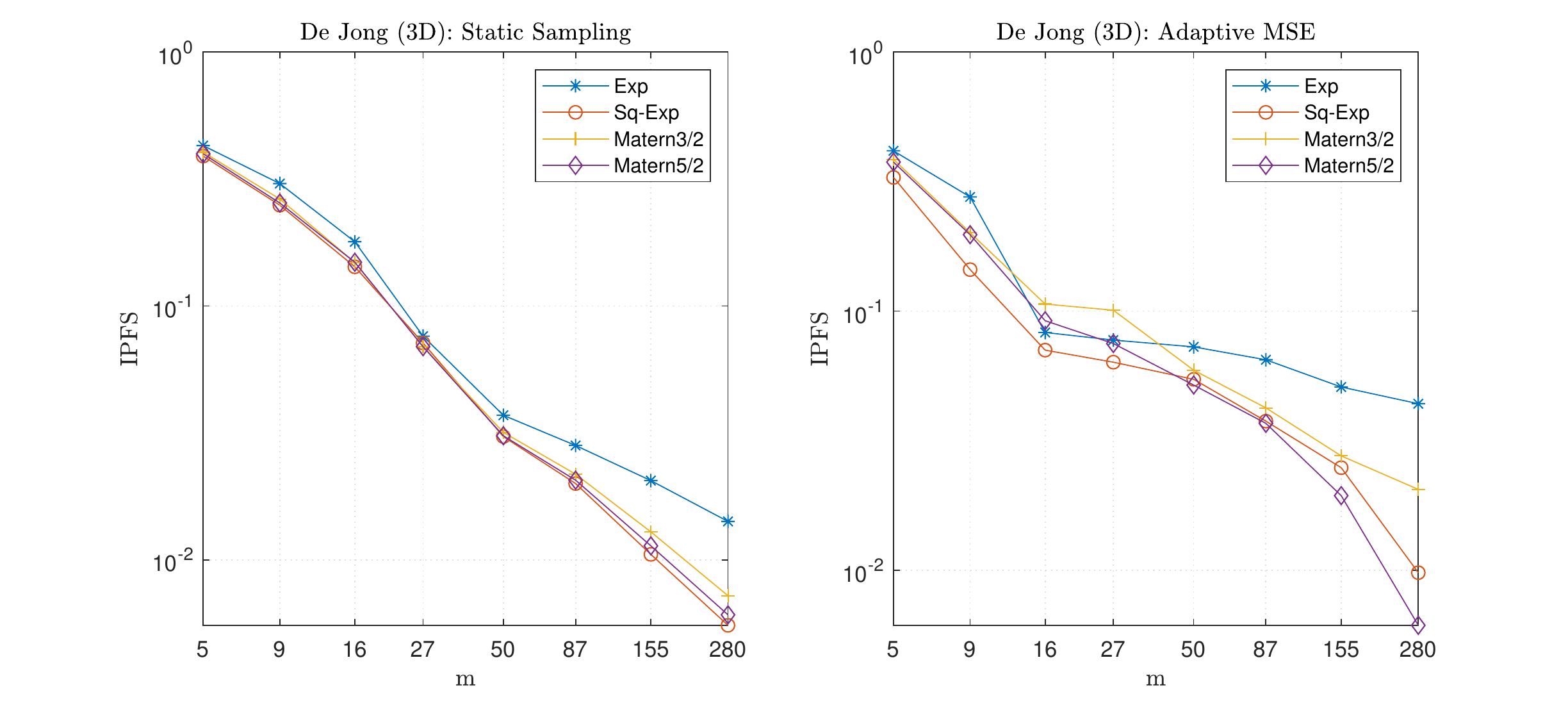}
	\end{center}
	\caption{Truncated $N(2.5, 0.3^2)$ on each dimension of $d=3$: The maximal IMSE and IPFS for the 3-dimensional De Jong's functions and four covariance kernels.}
	\label{fig:test1d3comp}
\end{figure}

\begin{figure}[htbp]
	\begin{center}
		\includegraphics[width=0.6\textwidth]{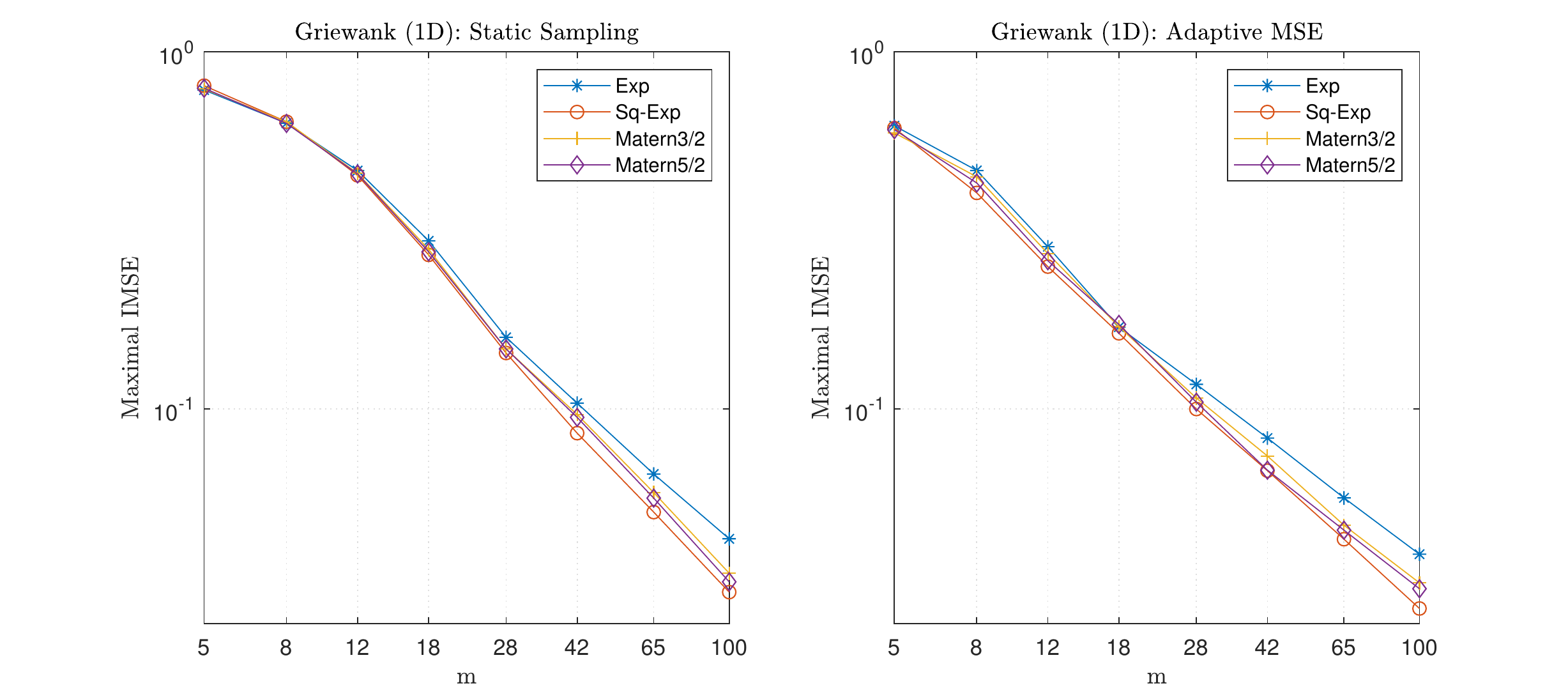}
		\includegraphics[width=0.6\textwidth]{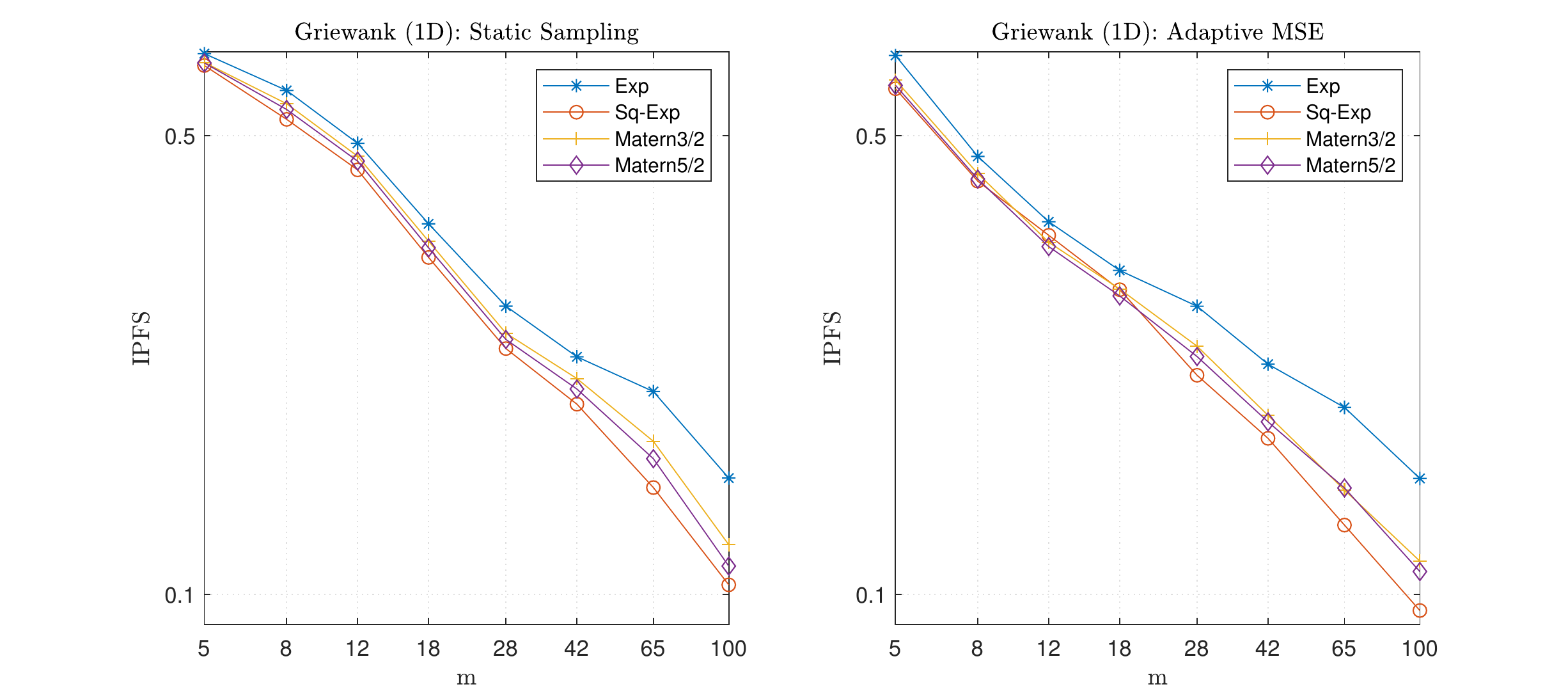}
	\end{center}
	\caption{Uniform distribution of $d=1$: The maximal IMSE and IPFS for the 1-dimensional Griewank's functions and four covariance kernels.}
	\label{fig:test2d1unif}
\end{figure}

\begin{figure}[htbp]
	\begin{center}
		\includegraphics[width=0.6\textwidth]{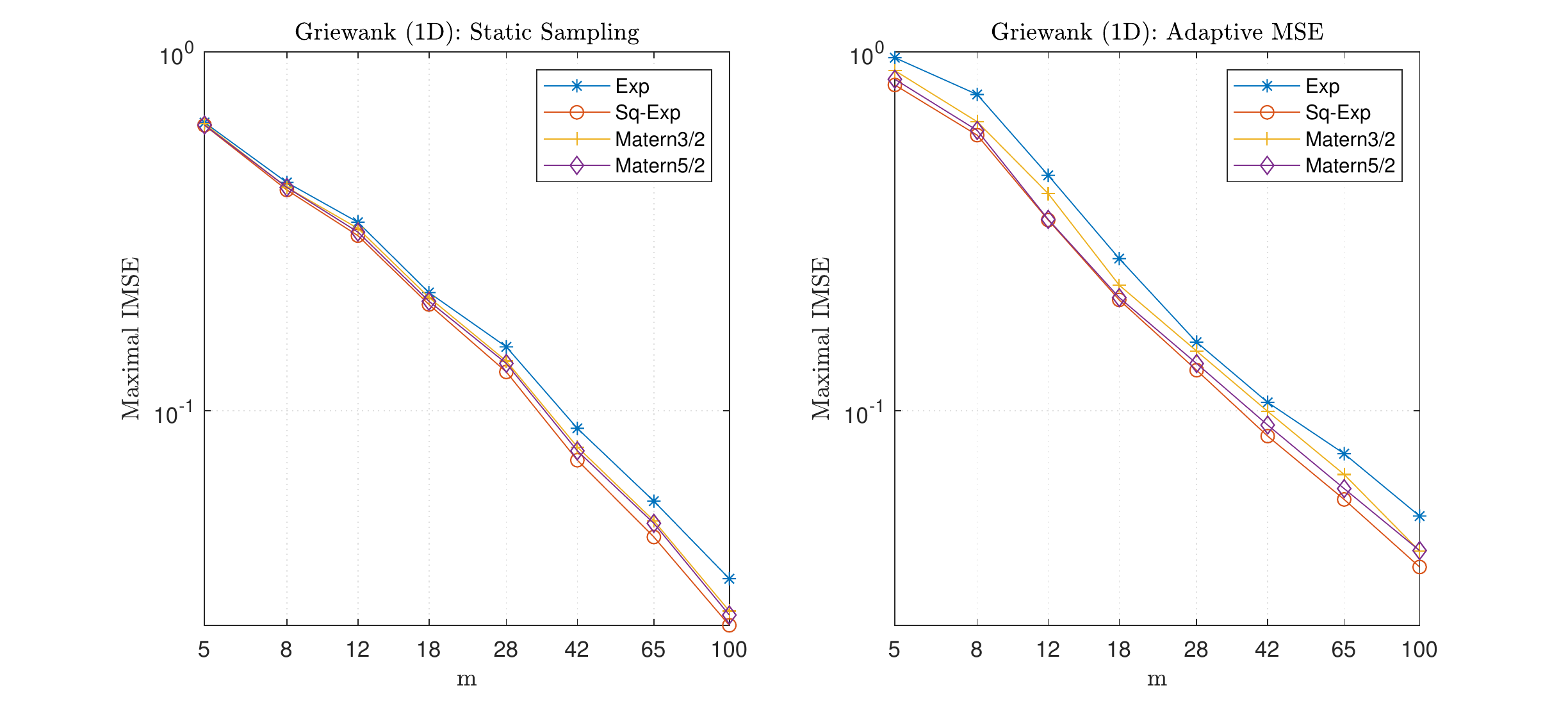}
		\includegraphics[width=0.6\textwidth]{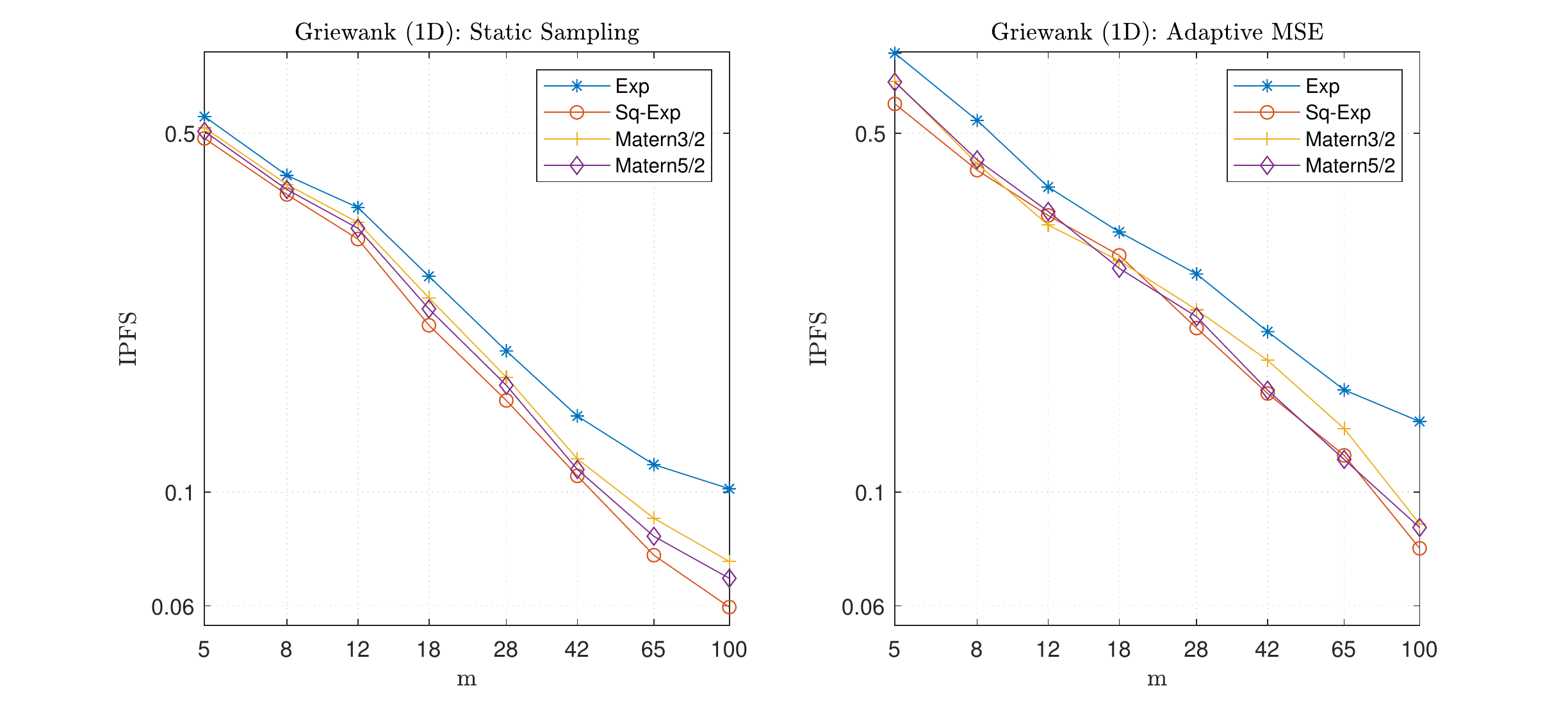}
	\end{center}
	\caption{Truncated $N(5.5, 1^2)$ of $d=1$: The maximal IMSE and IPFS for the 1-dimensional Griewank's functions and four covariance kernels. }
	\label{fig:test2comp1}
\end{figure}

\begin{figure}[htbp]
	\begin{center}
		\includegraphics[width=0.6\textwidth]{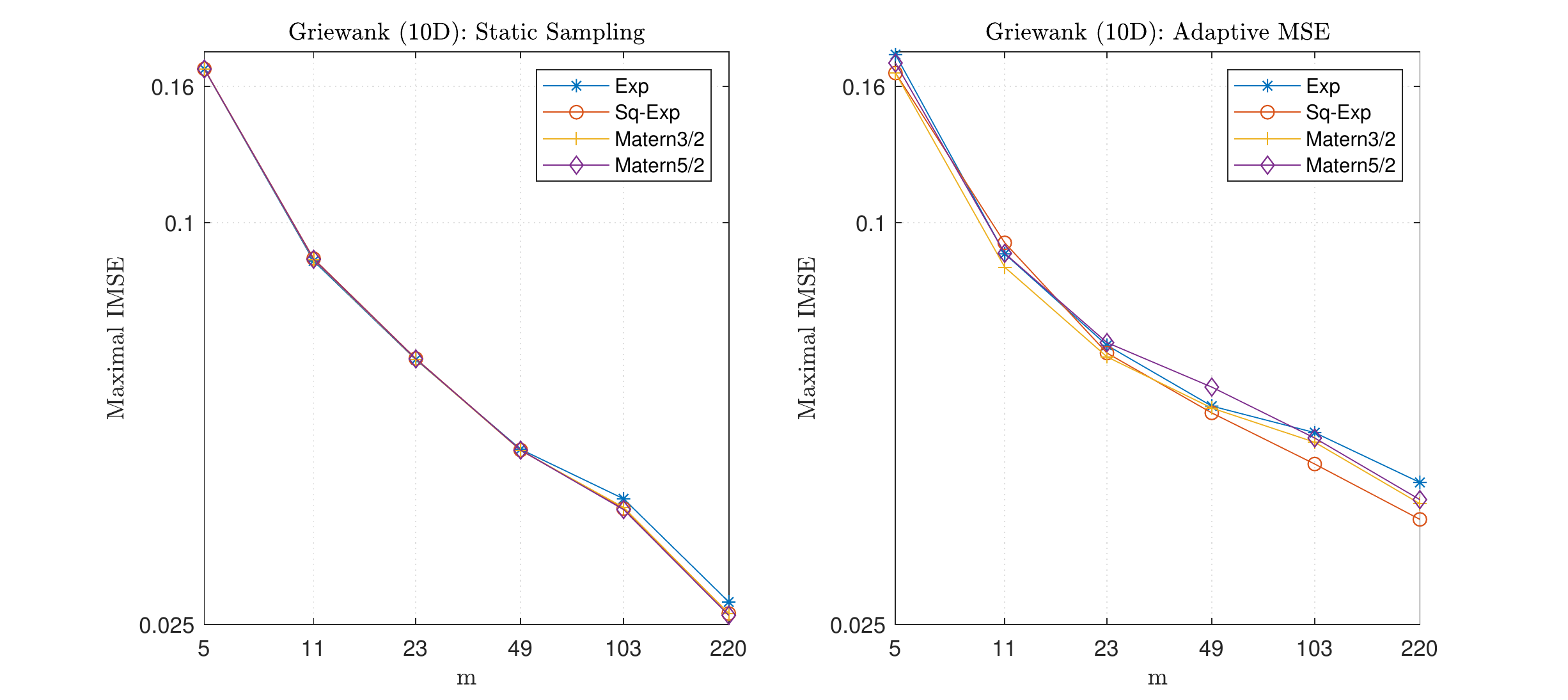}
		\includegraphics[width=0.6\textwidth]{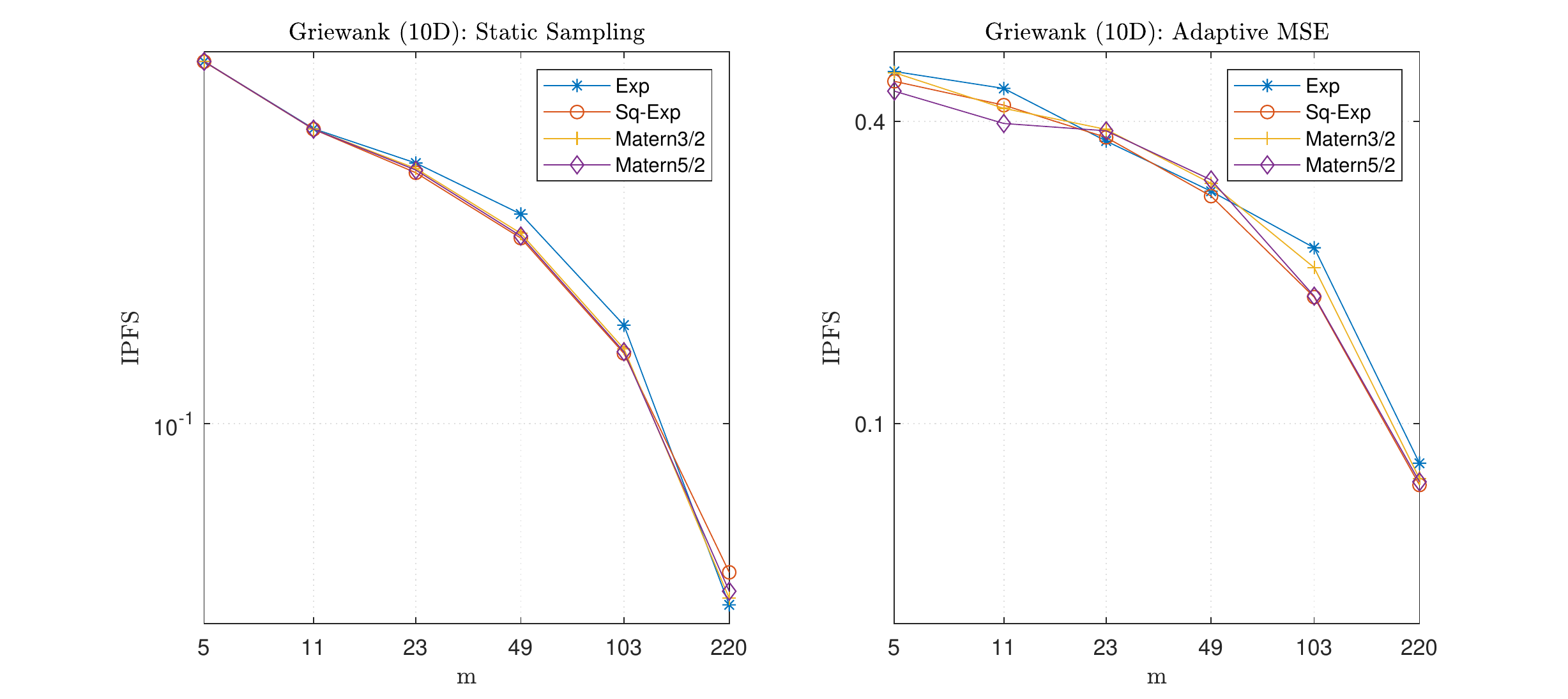}
	\end{center}
	\caption{Uniform distribution of $d=10$: The maximal IMSE and IPFS for the 10-dimensional Griewank's functions and four covariance kernels.}
	\label{fig:test2d10unif}
\end{figure}

\begin{figure}[htbp]
	\begin{center}
		\includegraphics[width=0.6\textwidth]{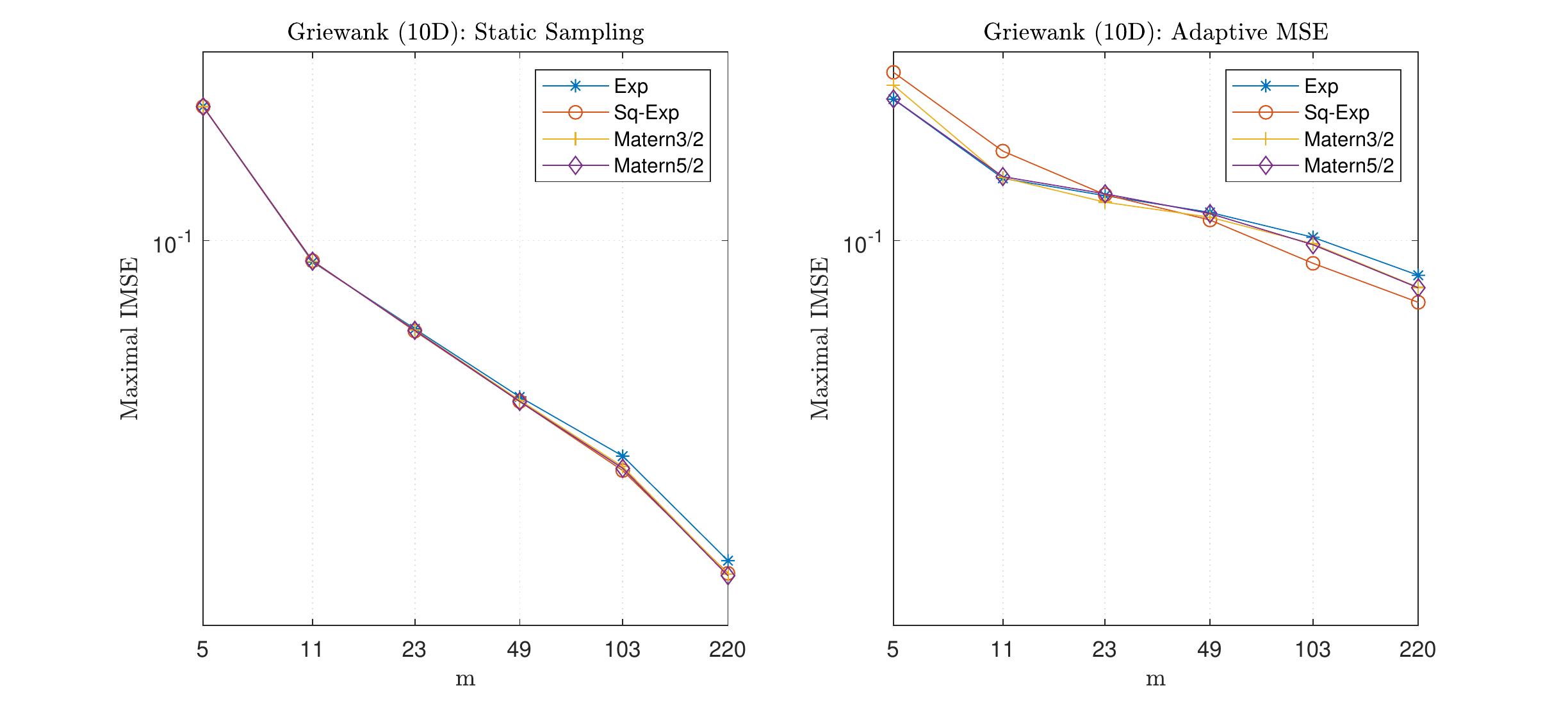}
		\includegraphics[width=0.6\textwidth]{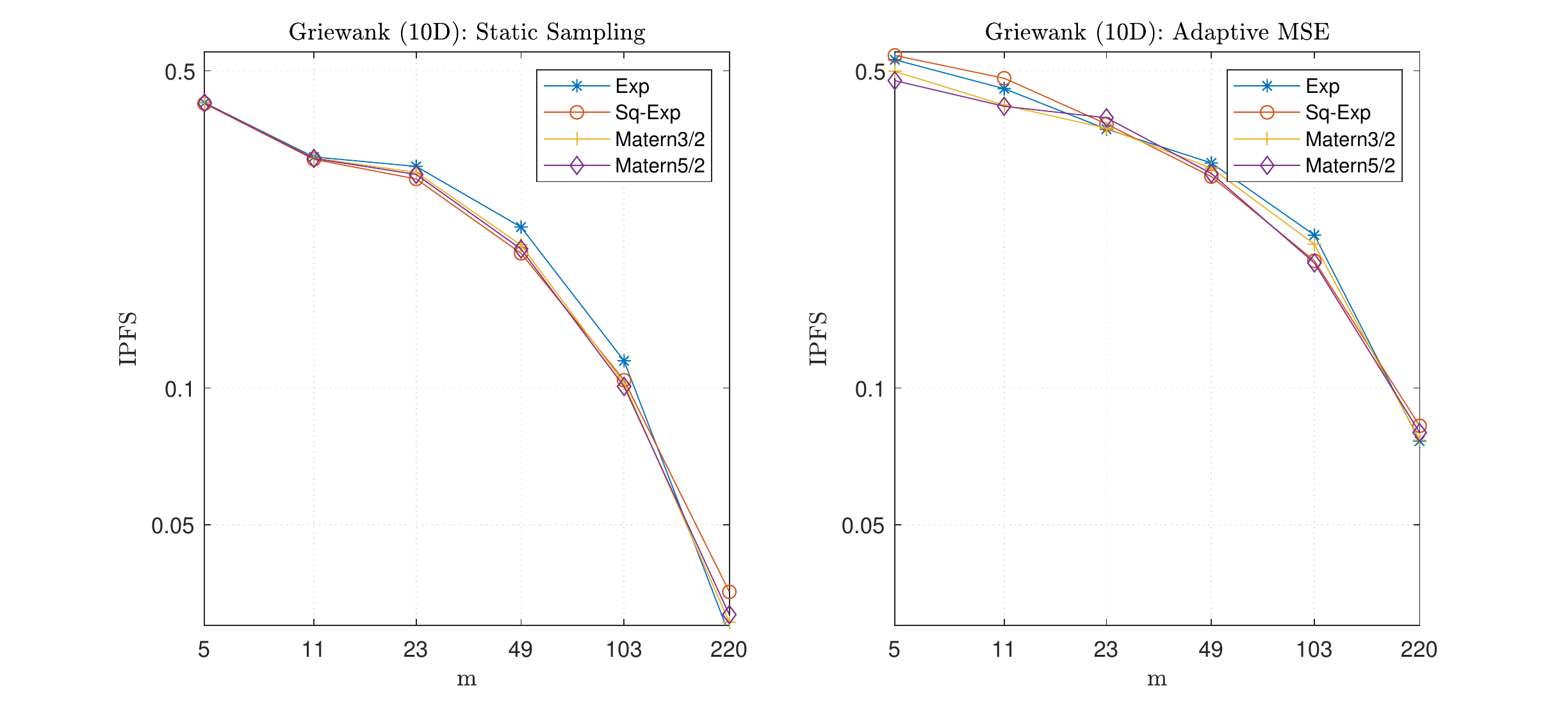}
	\end{center}
	\caption{Truncated $N(2.5, 0.75^2)$ on each dimension of $d=10$: The maximal IMSE and IPFS for the 10-dimensional Griewank's functions and four covariance kernels.}
	\label{fig:test2d10comp1}
\end{figure}

Figures \ref{fig:test1comp1}-\ref{fig:test2d10comp1} report the comparison results for our static sampling and the Adaptive MSE Procedure under the measures of the maximal IMSE and IPFS. We have the following observations:
\begin{itemize}
	\item For Case (i), where $d=1$ and the De Jong's functions are smooth enough, the Adaptive MSE Procedure has smaller maximal IMSE and IPFS than the static sampling from $\PP_{\X}$.
	\item For Cases (ii), (iii), and (iv) with the De Jong's functions in dimension $d=1,2,3$, where the normal variance becomes smaller, i.e., the sampling distribution $\PP_{\X}$ becomes more concentrated, the static sampling from $\PP_{\X}$ has slightly better performance than the Adaptive MSE Procedure, but overall their performances are similar.
	\item For Case (v), where $d=1$, the sampling distribution $\PP_{\X}$ is uniform, and the target is the Griewank's functions, we can see from Figure \ref{fig:test2d1unif} that the Adaptive MSE Procedure has slightly smaller maximal IMSE and IPFS than the static sampling, but overall their performances are similar.
	\item For Case (vi), where $d=1$, the sampling distribution $\PP_{\X}$ is truncated normal with a moderately large variance, and the target Griewank's functions have strong oscillation, we can see from Figure \ref{fig:test2comp1} that the static sampling almost always yields smaller maximal IMSE and IPFS than the Adaptive MSE Procedure.
	\item For Cases (vii) and (viii), where the dimension is high ($d=10$) and the target Griewank's functions have strong oscillation, we can see from Figures \ref{fig:test2d10unif} and \ref{fig:test2d10comp1} that the static sampling always yields much smaller maximal IMSE and IPFS than the Adaptive MSE Procedure, for both the uniform distribution and the truncated normal distribution.
\end{itemize}
In conclusion, the static sampling from $\PP_{\X}$ seems to yield comparable performance to the Adaptive MSE Procedure under the two measures in general. The static sampling tends to perform better than the Adaptive MSE Procedure when the target function has strong oscillation, the dimension becomes higher, and the covariate distribution $\PP_{\X}$ becomes more concentrated.

\subsection{Achieving a Target Precision of the Maximal IMSE} 
\label{sec:proc_targ}

Based on the linear decreasing trend of the maximal IMSE in Figure 5 of the main text, we propose a simple procedure to determine the sample size $m_0$ such that the maximal IMSE satisfies $\max_{i\in\{1,\ldots,k\}}{\E}_{\X_0}\left[\mse_{i,\opt}(\X_0)\right] = c_0$ for a target precision $c_0$. Suppose that we have already drawn $m$ covariate points $\X^{m}$ from $\PP_{\X}$ and each covariate point has $n$ simulation replications. Then, for an integer $L\geq 3$, we draw $L-1$ subsamples of sizes $m_1 < \ldots < m_{L-1} (<m_L\equiv m)$ from $\X^{m}$ without replacement. Denote these subsets as $\X^{m_1},\ldots,\X^{m_{L-1}}$. We then fit $(L-1)k$ SK models based on each dataset of $\X^{m_1},\ldots,\X^{m_{L-1}}$, and estimate the maximal IMSE for each subset using the Monte Carlo estimator described in Section 3 of the Online Supplement. We repeat this subsampling-fitting-estimating process for multiple times and take the average of the estimated maximal IMSE's at each size $m_1, \ldots, m_{L-1}, m_L$, denoted by $\max_{i\in\{1,\ldots,k\}} \widehat{\text{IMSE}}_i(m_l)$, $l=1,\ldots,L$. Finally, we fit the linear model $\log(\max_{i\in\{1,\ldots,k\}} \text{IMSE}_i) = c_1 + c_2 \log m + \text{error}$ using the pairs $\Big\{\big(\max_{i\in\{1,\ldots,k\}} \widehat{\text{IMSE}}_l(m_l), m_l\big): l=1,\ldots,L \Big\}$, and predict $m_0$ by $\widehat m_0 = \exp\left\{ (\log c_0-\widehat{c}_1)/\widehat{c}_2 \right\}$, where $\widehat{c}_1,\widehat{c}_2$ are the fitted linear coefficients.

Next, we apply this procedure to the M/M/1 queue example. We draw $m = 80$ covariate points from the sampling distribution $\PP_{\X}$ with $n=10$ replications, and estimate the maximal IMSE with subsample sizes $\{10,15,23,35,53,80\}$. For example, for the squared exponential kernel and uniform sampling distribution, we obtain the fitted linear regression model $\log( \max_{i\in\{1,\ldots,k\}} \text{IMSE}_i ) = -1.03\log(m) -4.58$ and the predicted $\widehat m_0\approx 119$ such that $\max_{i\in\{1,\ldots,k\}} \text{IMSE}_i = c_0= 7.5 \times 10^{-5}$. To numerically verify whether the true maximal IMSE is around $7.5 \times 10^{-5}$ at sample size $\widehat m_0=119$, we randomly draw another 39 covariate points from the uniform distribution, establish the SK models based on the union of the 39 new points and the 80 existing points, and compute the maximal IMSE. We repeat this process for 40 macro Monte Carlo replications. We find that the median maximal IMSE over the 40 macro replications is $7.37 \times 10^{-5}$. The numerical results for the two tested sampling distributions and four covariance kernels are summarized in Table \ref{tab:pred}. In almost all cases, the predicted $\widehat m_0$ values yield very similar or smaller maximal IMSE's compared to the target values. This demonstrates that our theory can help the decision makers determine the number of additional covariate points needed to achieve a target precision.

	\begin{table}[htbp]
		\caption{Prediction of sample size $m_0$ for a maximal IMSE precision $c_0$ based on $m=80$ covariate points.}
		\centering
		\begin{tabular}{c|c|c|cc|c|cc}	
			\hline
			& Kernels  & $c_0$ & $\widehat{c}_1$ &  $\widehat{c}_2$  & $\widehat m_0$ & Mean & Median  \\ \hline
			\multirow{4}{*}{\begin{tabular}[c]{@{}c@{}}uniform,   \\      $n=10$\end{tabular}} & \texttt{SqExp}      & $7.5 \times 10^{-5}$ & -1.03           & -4.58          & 119 & $7.37 \times 10^{-5}$ & $7.37 \times 10^{-5}$  \\
			& \texttt{Matern 5/2}  & $7.5 \times 10^{-5}$ & -1.12           & -4.06          & 130 & $7.07 \times 10^{-5}$ & $6.95 \times 10^{-5}$ \\
			& \texttt{Matern 3/2} & $7.5 \times 10^{-5}$ & -1.12           & -3.92          & 147 & $7.19 \times 10^{-5}$ & $6.95 \times 10^{-5}$  \\
			& \texttt{Exp}        & $2.0 \times 10^{-4}$ & -0.95           & -3.99          & 118 & $2.23 \times 10^{-4}$ & $2.09 \times 10^{-4}$  \\ \hline
			\multirow{4}{*}{\begin{tabular}[c]{@{}c@{}}truncated   \\ normal, \\      $n=10$\end{tabular}}  & \texttt{SqExp}      & $7.5 \times 10^{-5}$ & -1.00           & -4.70          & 122 & $7.44 \times 10^{-5}$ & $7.04 \times 10^{-5}$  \\
			& \texttt{Matern 5/2} & $7.5 \times 10^{-5}$ & -1.06           & -4.24          & 144 & $6.57 \times 10^{-5}$ & $6.45 \times 10^{-5}$  \\
			& \texttt{Matern 3/2} & $7.5 \times 10^{-5}$ & -1.08           & -4.04          & 160 & $6.67 \times 10^{-5}$ & $6.69 \times 10^{-5}$  \\
			& \texttt{Exp}        & $2.0 \times 10^{-4}$ & -0.95           & -4.00          & 117 & $2.34 \times 10^{-4}$ & $2.11 \times 10^{-4}$
			\\\hline
		\end{tabular}\label{tab:pred}
	\flushleft
	{\small Notes: \emph{``Mean" is the sample average of the maximal IMSE over 40 macro Monte Carlo replications. ``Median" is the sample median of the maximal IMSE over 40 macro Monte Carlo replications.}}
	\end{table}







\bibliographystyle{poms}
\bibliography{myrefs,papers}

\begin{thebibliography}{44}
\providecommand{\natexlab}[1]{#1}
\providecommand{\url}[1]{\texttt{#1}}
\providecommand{\urlprefix}{URL }
\providecommand{\bibAnnoteFile}[1]{%
  \IfFileExists{#1}{\begin{quotation}\noindent\textsc{Key:} #1\\
  \textsc{Annotation:}\ \input{#1}\end{quotation}}{}}
\providecommand{\bibAnnote}[2]{%
  \begin{quotation}\noindent\textsc{Key:} #1\\
  \textsc{Annotation:}\ #2\end{quotation}}
\providecommand{\bibinfo}[2]{#2}

\bibitem[{Ahmed \protect\BIBand{} Alkhamis(2009)}]{Ahmed2009}
\bibinfo{author}{Ahmed, M.~A.}, \bibinfo{author}{T.~M. Alkhamis}.
  \bibinfo{year}{2009}.
\newblock \bibinfo{title}{Simulation optimization for an emergency department
  healthcare unit in {K}uwait}.
\newblock \textit{\bibinfo{journal}{Eur J Oper Res}}, \bibinfo{volume}{198},
  \bibinfo{pages}{936--942}.
\bibAnnoteFile{Ahmed2009}

\bibitem[{Ankenman et~al.(2010)Ankenman, Nelson, \protect\BIBand{}
  Staum}]{ankenman2010}
\bibinfo{author}{Ankenman, B.~E.}, \bibinfo{author}{B.~L. Nelson},
  \bibinfo{author}{J.~Staum}. \bibinfo{year}{2010}.
\newblock \bibinfo{title}{Stochastic kriging for simulation metamodeling}.
\newblock \textit{\bibinfo{journal}{Oper Res}},
  \bibinfo{volume}{58}(\bibinfo{number}{2}), \bibinfo{pages}{371--382}.
\bibAnnoteFile{ankenman2010}

\bibitem[{Benini et~al.(1998)Benini, Hodgson, \protect\BIBand{}
  Siegel}]{Benini1998}
\bibinfo{author}{Benini, L.}, \bibinfo{author}{R.~Hodgson},
  \bibinfo{author}{P.~Siegel}. \bibinfo{year}{1998}.
\newblock \bibinfo{title}{System-level power estimation and optimization}.
\newblock \textit{\bibinfo{journal}{Proc 1998 International Symposium on Low
  Power Electronics and Design}}, \bibinfo{pages}{173--178}.
\bibAnnoteFile{Benini1998}

\bibitem[{Bertsimas et~al.(2017)Bertsimas, Kallus, Weinstein, \protect\BIBand{}
  Zhuo}]{bertsimas2017}
\bibinfo{author}{Bertsimas, D.}, \bibinfo{author}{N.~Kallus},
  \bibinfo{author}{A.~M. Weinstein}, \bibinfo{author}{Y.~D. Zhuo}.
  \bibinfo{year}{2017}.
\newblock \bibinfo{title}{Personalized diabetes management using electronic
  medical records}.
\newblock \textit{\bibinfo{journal}{Diabetes Care}},
  \bibinfo{volume}{40}(\bibinfo{number}{2}), \bibinfo{pages}{210--217}.
\bibAnnoteFile{bertsimas2017}

\bibitem[{Chen et~al.(2008)Chen, He, Fu, \protect\BIBand{} Lee}]{chen2008}
\bibinfo{author}{Chen, C.~H.}, \bibinfo{author}{D.~He},
  \bibinfo{author}{M.~Fu}, \bibinfo{author}{L.~H. Lee}. \bibinfo{year}{2008}.
\newblock \bibinfo{title}{Efficient simulation budget allocation for selecting
  an optimal subset}.
\newblock \textit{\bibinfo{journal}{INFORMS J Comput}},
  \bibinfo{volume}{20}(\bibinfo{number}{4}), \bibinfo{pages}{579--595}.
\bibAnnoteFile{chen2008}

\bibitem[{Chen \protect\BIBand{} Lee(2011)}]{chen2011}
\bibinfo{author}{Chen, C.~H.}, \bibinfo{author}{L.~H. Lee}.
  \bibinfo{year}{2011}.
\newblock \textit{\bibinfo{title}{Stochastic Simulation Optimization: An
  Optimal Computing Budget Allocation}}.
\newblock \bibinfo{publisher}{Singapore: World Scientific Publishing}.
\bibAnnoteFile{chen2011}

\bibitem[{Chen et~al.(2000)Chen, Lin, Y\"{u}cesan, \protect\BIBand{}
  Chick}]{chen2000}
\bibinfo{author}{Chen, C.~H.}, \bibinfo{author}{J.~Lin},
  \bibinfo{author}{E.~Y\"{u}cesan}, \bibinfo{author}{S.~E. Chick}.
  \bibinfo{year}{2000}.
\newblock \bibinfo{title}{Simulation budget allocation for further enhancing
  the efficiency of ordinal optimization}.
\newblock \textit{\bibinfo{journal}{Discrete Event Dyn S}},
  \bibinfo{volume}{10}, \bibinfo{pages}{251--270}.
\bibAnnoteFile{chen2000}

\bibitem[{Chen et~al.(2013)Chen, Ankenman, \protect\BIBand{}
  Nelson}]{chenx2013}
\bibinfo{author}{Chen, X.}, \bibinfo{author}{B.~E. Ankenman},
  \bibinfo{author}{B.~L. Nelson}. \bibinfo{year}{2013}.
\newblock \bibinfo{title}{Enhancing stochastic kriging metamodels with gradient
  estimators}.
\newblock \textit{\bibinfo{journal}{Oper Res}},
  \bibinfo{volume}{61}(\bibinfo{number}{2}), \bibinfo{pages}{512--528}.
\bibAnnoteFile{chenx2013}

\bibitem[{Dai(1996)}]{dai1996}
\bibinfo{author}{Dai, L.} \bibinfo{year}{1996}.
\newblock \bibinfo{title}{Convergence properties of ordinal comparison in the
  simulation of discrete event dynamic systems}.
\newblock \textit{\bibinfo{journal}{J Optimiz Theory App}},
  \bibinfo{volume}{91}(\bibinfo{number}{2}), \bibinfo{pages}{363--388}.
\bibAnnoteFile{dai1996}

\bibitem[{Ding et~al.(2005)Ding, Benyoucef, \protect\BIBand{} Xie}]{ding2005}
\bibinfo{author}{Ding, H.}, \bibinfo{author}{L.~Benyoucef},
  \bibinfo{author}{X.~Xie}. \bibinfo{year}{2005}.
\newblock \bibinfo{title}{A simulation optimization methodology for supplier
  selection problem}.
\newblock \textit{\bibinfo{journal}{Int J Comput Integ Manuf}},
  \bibinfo{volume}{18}, \bibinfo{pages}{210--224}.
\bibAnnoteFile{ding2005}

\bibitem[{Frazier et~al.(2008)Frazier, Powell, \protect\BIBand{}
  Dayanik}]{frazier2008}
\bibinfo{author}{Frazier, P.~I.}, \bibinfo{author}{W.~B. Powell},
  \bibinfo{author}{S.~Dayanik}. \bibinfo{year}{2008}.
\newblock \bibinfo{title}{A knowledge-gradient policy for sequential
  information collection}.
\newblock \textit{\bibinfo{journal}{SIAM J Contr Optim}},
  \bibinfo{volume}{47}(\bibinfo{number}{5}), \bibinfo{pages}{2410--2439}.
\bibAnnoteFile{frazier2008}

\bibitem[{Gao \protect\BIBand{} Chen(2017)}]{gao2017b}
\bibinfo{author}{Gao, S.}, \bibinfo{author}{W.~Chen}. \bibinfo{year}{2017}.
\newblock \bibinfo{title}{Efficient feasibility determination with multiple
  performance measure constraints}.
\newblock \textit{\bibinfo{journal}{IEEE Trans Automat Contr}},
  \bibinfo{volume}{62}, \bibinfo{pages}{113--122}.
\bibAnnoteFile{gao2017b}

\bibitem[{Gao et~al.(2017)Gao, Chen, \protect\BIBand{} Shi}]{gao2017a}
\bibinfo{author}{Gao, S.}, \bibinfo{author}{W.~Chen}, \bibinfo{author}{L.~Shi}.
  \bibinfo{year}{2017}.
\newblock \bibinfo{title}{A new budget allocation framework for the expected
  opportunity cost}.
\newblock \textit{\bibinfo{journal}{Oper Res}}, \bibinfo{volume}{65},
  \bibinfo{pages}{787--803}.
\bibAnnoteFile{gao2017a}

\bibitem[{Gao et~al.(2019{\natexlab{a}})Gao, Du, \protect\BIBand{}
  Chen}]{gao2019selecting}
\bibinfo{author}{Gao, S.}, \bibinfo{author}{J.~Du}, \bibinfo{author}{C.-H.
  Chen}. \bibinfo{year}{2019}{\natexlab{a}}.
\newblock \bibinfo{title}{Selecting the optimal system design under
  covariates}.
\newblock In \textit{\bibinfo{booktitle}{2019 ieee 15th international
  conference on automation science and engineering (case)}},
  \bibinfo{pages}{547--552}. \bibinfo{organization}{IEEE}.
\bibAnnoteFile{gao2019selecting}

\bibitem[{Gao et~al.(2019{\natexlab{b}})Gao, Li, \protect\BIBand{}
  Du}]{gao2019rate}
\bibinfo{author}{Gao, S.}, \bibinfo{author}{C.~Li}, \bibinfo{author}{J.~Du}.
  \bibinfo{year}{2019}{\natexlab{b}}.
\newblock \bibinfo{title}{Rate analysis for offline simulation online
  application}.
\newblock In \textit{\bibinfo{booktitle}{Proc. 2019 Winter Simulation Conf.}},
  \bibinfo{pages}{3468--3479}.
\bibAnnoteFile{gao2019rate}

\bibitem[{Garud et~al.(2017)Garud, Karimi, \protect\BIBand{} Kraft}]{Garud2017}
\bibinfo{author}{Garud, S.~S.}, \bibinfo{author}{I.~A. Karimi},
  \bibinfo{author}{M.~Kraft}. \bibinfo{year}{2017}.
\newblock \bibinfo{title}{Design of computer experiments: A review}.
\newblock \textit{\bibinfo{journal}{Computers and Chemical Engineering}},
  \bibinfo{volume}{106}, \bibinfo{pages}{71--95}.
\bibAnnoteFile{Garud2017}

\bibitem[{Glynn \protect\BIBand{} Juneja(2004)}]{glynn2004}
\bibinfo{author}{Glynn, P.}, \bibinfo{author}{S.~Juneja}. \bibinfo{year}{2004}.
\newblock \bibinfo{title}{A large deviations perspective on ordinal
  optimization}.
\newblock In \textit{\bibinfo{booktitle}{Proc. 2004 Winter Simulation Conf.}},
  \bibinfo{pages}{577--585}.
\bibAnnoteFile{glynn2004}

\bibitem[{Gu(2002)}]{Gu02}
\bibinfo{author}{Gu, C.} \bibinfo{year}{2002}.
\newblock \textit{\bibinfo{title}{{Smoothing Spline ANOVA Models}}}.
\newblock \bibinfo{publisher}{Springer, New York}.
\bibAnnoteFile{Gu02}

\bibitem[{Hensman et~al.(2014)Hensman, Fusi, \protect\BIBand{}
  Lawrence}]{hensman2013gaussian}
\bibinfo{author}{Hensman, J.}, \bibinfo{author}{N.~Fusi},
  \bibinfo{author}{N.~Lawrence}. \bibinfo{year}{2014}.
\newblock \bibinfo{title}{Gaussian processes for big data}.
\newblock In \textit{\bibinfo{booktitle}{Proc. 29th Conference on Uncertainty
  in Artificial Intelligence}}, \bibinfo{pages}{282--290}.
\bibAnnoteFile{hensman2013gaussian}

\bibitem[{Hong \protect\BIBand{} Jiang(2019)}]{hong2019}
\bibinfo{author}{Hong, L.~J.}, \bibinfo{author}{G.~Jiang}.
  \bibinfo{year}{2019}.
\newblock \bibinfo{title}{Offine simulation online application: a new framework
  of simulation-based decision making}.
\newblock \textit{\bibinfo{journal}{Asia Pac J Oper Res}},
  \bibinfo{volume}{36}(\bibinfo{number}{6}), \bibinfo{pages}{1940015}.
\bibAnnoteFile{hong2019}

\bibitem[{Hsing \protect\BIBand{} Eubank(2015)}]{HsiEub15}
\bibinfo{author}{Hsing, T.}, \bibinfo{author}{R.~Eubank}. \bibinfo{year}{2015}.
\newblock \textit{\bibinfo{title}{{Theoretical Foundations of Functional Data
  Analysis, with an Introduction to Linear operators}}}.
\newblock \bibinfo{publisher}{John Wiley \& Sons}.
\bibAnnoteFile{HsiEub15}

\bibitem[{Hsu et~al.(2012)Hsu, Kakade, \protect\BIBand{} Zhang}]{Hsuetal12}
\bibinfo{author}{Hsu, D.}, \bibinfo{author}{S.~M. Kakade},
  \bibinfo{author}{T.~Zhang}. \bibinfo{year}{2012}.
\newblock \bibinfo{title}{A tail inequality for quadratic forms of subgaussian
  random vectors}.
\newblock \textit{\bibinfo{journal}{Electronic Communications in Probability}},
  \bibinfo{volume}{17}(\bibinfo{number}{52}), \bibinfo{pages}{1--6}.
\bibAnnoteFile{Hsuetal12}

\bibitem[{Kim \protect\BIBand{} Nelson(2006)}]{kim2006}
\bibinfo{author}{Kim, S.~H.}, \bibinfo{author}{B.~L. Nelson}.
  \bibinfo{year}{2006}.
\newblock \bibinfo{title}{Selecting the best system}.
\newblock In \bibinfo{editor}{Henderson, S.~G.}, \bibinfo{editor}{B.~L.
  Nelson}, editors, \textit{\bibinfo{booktitle}{Simulation}}, Handbooks in
  Operations Research and Management Science, chapter~\bibinfo{chapter}{13},
  \bibinfo{pages}{501--534}. \bibinfo{publisher}{Elsevier},
  \bibinfo{address}{Amsterdam, Netherlands}.
\bibAnnoteFile{kim2006}

\bibitem[{Kleijnen(1993)}]{kleijnen1993}
\bibinfo{author}{Kleijnen, J. P.~C.} \bibinfo{year}{1993}.
\newblock \bibinfo{title}{Simulation and optimization in production planning: A
  case study}.
\newblock \textit{\bibinfo{journal}{Decis Support Syst}},
  \bibinfo{volume}{9}(\bibinfo{number}{3}), \bibinfo{pages}{269--280}.
\bibAnnoteFile{kleijnen1993}

\bibitem[{Kleijnen(2009)}]{kleijnen2009}
\bibinfo{author}{Kleijnen, J. P.~C.} \bibinfo{year}{2009}.
\newblock \bibinfo{title}{Kriging metamodeling in simulation: A review}.
\newblock \textit{\bibinfo{journal}{Eur J Oper Res}},
  \bibinfo{volume}{192}(\bibinfo{number}{3}), \bibinfo{pages}{707--716}.
\bibAnnoteFile{kleijnen2009}

\bibitem[{Kreh(2012)}]{Kre12}
\bibinfo{author}{Kreh, M.} \bibinfo{year}{2012}.
\newblock \textit{\bibinfo{title}{Bessel Functions. Lecture Notes, Penn State -
  G\"ottingen Summer School on Number Theory}}.
\bibAnnoteFile{Kre12}

\bibitem[{Law(2015)}]{law2015}
\bibinfo{author}{Law, A.~M.} \bibinfo{year}{2015}.
\newblock \textit{\bibinfo{title}{Simulation Modeling and Analysis}}.
\newblock \bibinfo{edition}{5th} edition. \bibinfo{publisher}{McGraw-Hill, New
  York}.
\bibAnnoteFile{law2015}

\bibitem[{Luo \protect\BIBand{} Duraiswami(2013)}]{luo2013}
\bibinfo{author}{Luo, Y.}, \bibinfo{author}{R.~Duraiswami}.
  \bibinfo{year}{2013}.
\newblock \bibinfo{title}{{Fast near-GRID Gaussian process regression}}.
\newblock In \textit{\bibinfo{booktitle}{Proc. 16th International Conference on
  Artificial Intelligence and Statistics}}, \bibinfo{pages}{424--432}.
\bibAnnoteFile{luo2013}

\bibitem[{Ni et~al.(2017)Ni, Ciocan, Henderson, \protect\BIBand{}
  Hunter}]{ni2017}
\bibinfo{author}{Ni, E.~C.}, \bibinfo{author}{D.~F. Ciocan},
  \bibinfo{author}{S.~G. Henderson}, \bibinfo{author}{S.~R. Hunter}.
  \bibinfo{year}{2017}.
\newblock \bibinfo{title}{Efficient ranking and selection in parallel computing
  environments}.
\newblock \textit{\bibinfo{journal}{Oper Res}},
  \bibinfo{volume}{65}(\bibinfo{number}{3}), \bibinfo{pages}{821--836}.
\bibAnnoteFile{ni2017}

\bibitem[{Qu \protect\BIBand{} Fu(2014)}]{qu2014}
\bibinfo{author}{Qu, H.}, \bibinfo{author}{M.~C. Fu}. \bibinfo{year}{2014}.
\newblock \bibinfo{title}{Gradient extrapolated stochastic kriging}.
\newblock \textit{\bibinfo{journal}{ACM Trans Model Comput Simul}},
  \bibinfo{volume}{24}(\bibinfo{number}{4}).
\newblock \bibinfo{note}{{A}rticle 3}.
\bibAnnoteFile{qu2014}

\bibitem[{Rasmussen \protect\BIBand{} Williams(2006)}]{RasWil06}
\bibinfo{author}{Rasmussen, C.~E.}, \bibinfo{author}{C.~K. Williams}.
  \bibinfo{year}{2006}.
\newblock \textit{\bibinfo{title}{{Gaussian Process for Machine Learning}}}.
\newblock \bibinfo{publisher}{MIT press}.
\bibAnnoteFile{RasWil06}

\bibitem[{Ryzhov(2016)}]{ryzhov2016}
\bibinfo{author}{Ryzhov, I.~O.} \bibinfo{year}{2016}.
\newblock \bibinfo{title}{On the convergence rates of expected improvement
  methods}.
\newblock \textit{\bibinfo{journal}{Oper Res}},
  \bibinfo{volume}{64}(\bibinfo{number}{6}), \bibinfo{pages}{1515--1528}.
\bibAnnoteFile{ryzhov2016}

\bibitem[{Sabuncuoglu \protect\BIBand{} Touhami(2002)}]{sabuncuoglu2002}
\bibinfo{author}{Sabuncuoglu, I.}, \bibinfo{author}{S.~Touhami}.
  \bibinfo{year}{2002}.
\newblock \bibinfo{title}{Simulation metamodeling with neural networks: an
  experimental investigation}.
\newblock \textit{\bibinfo{journal}{Internat J Production Res}},
  \bibinfo{volume}{40}, \bibinfo{pages}{2483--2505}.
\bibAnnoteFile{sabuncuoglu2002}

\bibitem[{Santin \protect\BIBand{} Schaback(2016)}]{SanSch16}
\bibinfo{author}{Santin, G.}, \bibinfo{author}{R.~Schaback}.
  \bibinfo{year}{2016}.
\newblock \bibinfo{title}{Approximation of eigenfunctions in kernel-based
  spaces}.
\newblock \textit{\bibinfo{journal}{Advances in Computational Mathematics}},
  \bibinfo{volume}{42}(\bibinfo{number}{4}), \bibinfo{pages}{973--993}.
\bibAnnoteFile{SanSch16}

\bibitem[{Shen et~al.(2021)Shen, Hong, \protect\BIBand{} Zhang}]{shen2019}
\bibinfo{author}{Shen, H.}, \bibinfo{author}{L.~J. Hong},
  \bibinfo{author}{X.~Zhang}. \bibinfo{year}{2021}.
\newblock \bibinfo{title}{Ranking and selection with covariates for
  personalized decision making}.
\newblock \textit{\bibinfo{journal}{INFORMS Journal on Computing}},
  \bibinfo{volume}{33}(\bibinfo{number}{4}), \bibinfo{pages}{1500--1519}.
\bibAnnoteFile{shen2019}

\bibitem[{Stein(1999)}]{Stein99}
\bibinfo{author}{Stein, M.~L.} \bibinfo{year}{1999}.
\newblock \textit{\bibinfo{title}{{Interpolation for Spatial Data: Some Theory
  for Kriging}}}.
\newblock \bibinfo{publisher}{Springer, New York}.
\bibAnnoteFile{Stein99}

\bibitem[{Steinwart et~al.(2009)Steinwart, Hush, \protect\BIBand{}
  Scovel}]{Steetal09}
\bibinfo{author}{Steinwart, I.}, \bibinfo{author}{D.~Hush},
  \bibinfo{author}{C.~Scovel}. \bibinfo{year}{2009}.
\newblock \bibinfo{title}{Optimal rates for regularized least squares
  regression}.
\newblock In \textit{\bibinfo{booktitle}{Proc. 22nd Annual Conference on
  Learning Theory}}, \bibinfo{pages}{79--93}.
\bibAnnoteFile{Steetal09}

\bibitem[{van~der Vaart \protect\BIBand{} van Zanten(2011)}]{VarZan11}
\bibinfo{author}{van~der Vaart, A.~W.}, \bibinfo{author}{J.~H. van Zanten}.
  \bibinfo{year}{2011}.
\newblock \bibinfo{title}{{Information rates of nonparametric Gaussian process
  methods}}.
\newblock \textit{\bibinfo{journal}{J Mach Learn Res}}, \bibinfo{volume}{12},
  \bibinfo{pages}{2095--2119}.
\bibAnnoteFile{VarZan11}

\bibitem[{Van~Trees(2001)}]{VT01}
\bibinfo{author}{Van~Trees, H.~L.} \bibinfo{year}{2001}.
\newblock \textit{\bibinfo{title}{{Detection, Estimation, and Modulation
  Theory}}}.
\newblock \bibinfo{publisher}{John Wiley \& Sons}.
\bibAnnoteFile{VT01}

\bibitem[{Wang \protect\BIBand{} Hu(2018)}]{wang2018}
\bibinfo{author}{Wang, B.}, \bibinfo{author}{J.~Hu}. \bibinfo{year}{2018}.
\newblock \bibinfo{title}{Some monotonicity results for stochastic kriging
  metamodels in sequential settings}.
\newblock \textit{\bibinfo{journal}{INFORMS J Comput}},
  \bibinfo{volume}{30}(\bibinfo{number}{2}), \bibinfo{pages}{278--294}.
\bibAnnoteFile{wang2018}

\bibitem[{Wilson \protect\BIBand{} Nickisch(2015)}]{wilson2015kernel}
\bibinfo{author}{Wilson, A.}, \bibinfo{author}{H.~Nickisch}.
  \bibinfo{year}{2015}.
\newblock \bibinfo{title}{{Kernel interpolation for scalable structured
  Gaussian processes (KISS-GP)}}.
\newblock In \textit{\bibinfo{booktitle}{International Conference on Machine
  Learning}}, \bibinfo{pages}{1775--1784}.
\bibAnnoteFile{wilson2015kernel}

\bibitem[{Zhang(2005)}]{Zha05}
\bibinfo{author}{Zhang, T.} \bibinfo{year}{2005}.
\newblock \bibinfo{title}{Learning bounds for kernel regression using effective
  data dimensionality}.
\newblock \textit{\bibinfo{journal}{Neural Comput}}, \bibinfo{volume}{17},
  \bibinfo{pages}{2077--2098}.
\bibAnnoteFile{Zha05}

\bibitem[{Zhang et~al.(2015)Zhang, Duchi, \protect\BIBand{}
  Wainwright}]{Zhaetal15}
\bibinfo{author}{Zhang, Y.}, \bibinfo{author}{J.~C. Duchi},
  \bibinfo{author}{M.~J. Wainwright}. \bibinfo{year}{2015}.
\newblock \bibinfo{title}{Divide and conquer kernel ridge regression: a
  distributed algorithm with minimax optimal rates}.
\newblock \textit{\bibinfo{journal}{J Mach Learn Res}}, \bibinfo{volume}{16},
  \bibinfo{pages}{3299--3340}.
\bibAnnoteFile{Zhaetal15}

\bibitem[{Zhou \protect\BIBand{} Xie(2015)}]{zhou2015}
\bibinfo{author}{Zhou, E.}, \bibinfo{author}{W.~Xie}. \bibinfo{year}{2015}.
\newblock \bibinfo{title}{Simulation optimization when facing input
  uncertainty}.
\newblock In \textit{\bibinfo{booktitle}{Proc. 2015 Winter Simulation Conf.}},
  \bibinfo{pages}{3714--3724}.
\bibAnnoteFile{zhou2015}

\end{thebibliography}

%

\end{document}